\documentclass[10pt,twocolumn,preprintnumbers,amsmath,amssymb,nofootinbib
,superscriptaddress]{revtex4-1}
\usepackage{graphicx,longtable,mathrsfs,color,array}
\usepackage[hidelinks]{hyperref}
\usepackage[usenames,dvipsnames]{xcolor} 
\usepackage{amssymb,amsmath,mathtools,mathrsfs,slashed} 
\usepackage{epsfig,subfigure,placeins,float} 
\usepackage{booktabs,longtable,ctable,multirow} 
\usepackage{exscale,relsize} 
\usepackage[normalem]{ulem} 
\usepackage{enumerate}
\usepackage{times, mathptmx} 
\usepackage[utf8]{inputenc}
\usepackage{color}
\usepackage{hyperref}
\usepackage{graphicx}
\usepackage{color}
\usepackage{graphicx,graphics}
\usepackage{bm}
\allowdisplaybreaks[1]

\begin{document}

\title{General theories of linear gravitational perturbations to a Schwarzschild Black Hole}
\author{Oliver J. Tattersall}
\email{oliver.tattersall@physics.ox.ac.uk}
\affiliation{Astrophysics, University of Oxford, DWB, Keble Road, Oxford OX1 3RH, UK}
\author{Pedro G. Ferreira}
\email{p.ferreira1@physics.ox.ac.uk}
\affiliation{Astrophysics, University of Oxford, DWB, Keble Road, Oxford OX1 3RH, UK}
\author{Macarena Lagos}
\email{mlagos@kicp.uchicago.edu}
\affiliation{Kavli Institute for Cosmological Physics, The University of Chicago, Chicago, IL 60637, USA}
\date{Received \today; published -- 00, 0000}

\begin{abstract}
We use the covariant formulation proposed in \cite{Tattersall:2017eav} to analyse the structure of linear perturbations about a spherically symmetric background in different families of gravity theories, and hence study how quasi-normal modes of perturbed black holes may be affected by modifications to General Relativity. We restrict ourselves to single-tensor, scalar-tensor and vector-tensor diffeomorphism-invariant gravity models in a Schwarzschild black hole background. We show explicitly the full covariant form of the quadratic actions in such cases, which allow us to then analyse odd parity (axial) and even parity (polar) perturbations simultaneously in a straightforward manner. 
\end{abstract}


\maketitle
\section{Introduction}

The recent detection of gravitational waves (GW) from the merger of black hole binaries by the advanced Laser Interferometric Gravitational Wave Observatory (LIGO) \cite{2016PhRvL.116f1102A,2016PhRvL.116x1103A,2017PhRvL.118v1101A} and advanced Virgo \cite{2017arXiv170909660T} has opened a new window of physics that will allow us to test gravity in completely new regimes \cite{2016PhRvL.116v1101A}. While General Relativity (GR) enjoys great success and accuracy in the weak field regime around the Solar System, the moderate and strong field regime has been previously unexplored and little observational data has been available so far \cite{Berti:2015itd}. This situation is changing and with the accumulation of data from new black hole mergers, in addition to the plans of future observatories such as eLISA, KAGRA, and the Einstein Telescope, we will be able to impose precise observational constraints in new regimes by analysing the evolution of GW signals.

The black hole remnant resulting from the merger of two black holes is, initially, highly deformed. It subsequently settles down to a quiescent state by emitting gravitational radiation - dubbed the `ringdown'. In GR, this process is described as the final black hole `shedding hair' and can be characterized by two parameters: the black hole's mass and angular momentum. The fact that black holes in GR have `no hair' has become one of the cornerstone results of modern gravity \cite{Dreyer:2003bv,Gossan:2011ha, Kamaretsos:2011um,Meidam:2014jpa, Nakano:2015uja, Cardoso:2016ryw, Thrane:2017lqn,  Glampedakis:2017dvb}. In extensions to GR the situation is different. The final state, the black hole remnant, may have additional structure or hair or it might not. But the structure of the GW signal could carry information about the underlying theory of gravity, even if the final equilibrium state is a black hole which is indistinguishable from those found in GR (for example Schwarzschild or Kerr). Thus, the characterisation of the ringdown might allow us to discriminate between GR and alternative theories of gravity \cite{Barausse:2008xv}.

Over the past decade, a set of numerical algorithms have been established to characterise the ringdown in terms of quasi-normal modes \cite{Berti:2009kk}.
A number of consistency checks have been proposed for testing the no-hair hypothesis by comparing the values of the dominant and sub-dominant quasinormal modes; as a by product, it should be possible to read off the spin and mass of the final black hole \cite{Berti:2005ys}. The errors and the associated signal-to-noise ratio of such procedures have been studied in detail \cite{Berti:2007zu,Yang:2017zxs,Baibhav:2017jhs} and it has been shown that it should be possible to find accurate constraints on black hole parameters from future data. 

There has been some attempts at exploring the ringdown process in specific extensions of GR. The evolution equations have been analysed for Jordan-Brans-Dicke gravity \cite{Harada:1997mr, PhysRevD.34.333}, for scalar-tensor theories with non-minimal derivative couplings \cite{Dong:2017toi}, for Einstein-Dilaton-Gauss-Bonnet gravity \cite{Blazquez-Salcedo:2016enn}, for TeVeS models \cite{Lasky:2010bd}, and for Dynamical Chern-Simons gravity \cite{Cardoso:2009pk,Molina:2010fb}. But it is fair to say that the literature is far from complete and comprehensive. With the advent of black hole spectroscopy it is timely to start exploring extensions of General Relativity more thoroughly with the hope that future data might allow us to place stringent constraints on such modifications. 

The study of the ringdown process through the quasi-normal modes involves the analysis of linear perturbations around a stationary black hole \cite{Berti:2009kk}. By studying the structure of the evolution equations, subject to a particular set of boundary conditions, one is able to determine frequencies and damping scales which contain a wealth of information. The problem is entirely analogous to that of analysing deviations from homogeneity on a cosmological spacetime (such as a Friedman-Robertson-Walker universe) 
\cite{Dodelson:2003ft}. There, one uses a set of basis functions tailored to homogeneous and isotropic spacetime and studies their evolution and spatial morphology. Comparing to cosmological observations one is then able to extract information about, for example, the expansion of the universe, the densities of the different energy components and the statistical properties of the initial conditions. 

Given the similarities between the study of quasi-normal modes and cosmological perturbations, it would make sense to explore whether techniques developed for cosmology might be applied to the study of black hole physics. The focus of this paper will be to show that a method developed for constructing general quadratic theories of gravity in the context of cosmological linear perturbation theory \cite{Lagos:2016wyv,2017JCAP...01..047L} can also be used to develop families of perturbed actions around black hole spacetimes. In doing so, it is possible to develop a formalism for quasi-normal modes in general theories of gravity.

While we will describe our method in more detail in Section \ref{SecFP}, it helps to briefly summarise our approach. We will construct general quadratic actions of the metric and any additional gravitational degrees of freedom around a black hole solution. With a judicious use of Noether's theorem, we will ensure that these quadratic actions are diffeomorphism invariant (or, to be more precise, gauge invariant on the specific background spacetime). These actions will depend on a small number of free functions that will affect the quasi-normal mode equations we derive. Thus, by constraining quasi-normal modes, we will be able to constrain the free functions and therefore extensions to General Relativity. Key to this construction is that these equations are built with a minimal set of assumptions and, as a result, should cover a wide range of models in the space of non-linear gravitational theories.

The focus of this paper will be on linear perturbations. For simplicity and clarity, we will restrict ourselves to a Schwarzschild background although the method we present should be applicable to Kerr or more exotic black holes arising in extensions of GR. This restriction merits a brief discussion. The most straightforward extension to GR is the addition of a non-minimally coupled scalar field -- scalar-tensor gravity theories. It is well established that a wide range of scalar-tensor theories have no hair and thus settle down to Schwarzschild or Kerr black holes \cite{2012PhRvL.108h1103S}. However it is possible to construct hairy black holes in scalar-tensor theories \cite{Sotiriou:2013qea}. The same can be said of theories where the extra gravitational degree of freedom is a 4-vector: for example generalized Proca theories \cite{Chagoya:2016aar,2017JCAP...08..024H}. In this paper, the extensions to GR we will consider involve either an extra scalar or vector field and given that these theories have regimes with a Schwarzschild solution, we are justified in restricting ourselves to having it as the background space time.

We structure the paper as follows: In Section \ref{SecFP} we summarise the method for constructing general quadratic actions in the covariant form and use it to derive the action of a free massless spin-2 field propagating on Minkowski space, which corresponds to linearised GR. In Sections \ref{secGR}-\ref{secVT}, we will derive the diffeomorphism-invariant quadratic actions of linear perturbations on a Schwarzschild background for three families of theories of gravity: containing a single-tensor field, a tensor field with a scalar field, and a tensor field with a vector field, respectively. In each case we will derive the equations of motion for odd parity (axial) and even parity (polar) type perturbations. In Section \ref{conclusion} we will discuss the results of our work and the method presented in this paper, as well as future work to be undertaken. 

Throughout this paper, indices using the greek alphabet ($\mu$, $\nu$, $\lambda$...) will denote space-time indices and run over coordinates 0-3. Capital Roman letters ($A$, $B$, $C$...) will denote angular indices and run over coordinates 2-3. The metric signature will be $(-,+,+,+)$, and we will use geometrised units in which $G=c=1$.


\section{Covariant action approach}\label{SecFP}
In this section we review the covariant method for constructing gauge invariant quadratic actions for linear perturbations, as first described in \cite{Tattersall:2017eav}, and illustrate it by recovering linear General Relativity in Minkowski space. We discuss the role of the global symmetry of the background and the local gauge symmetry of the perturbations in the method. The use of this method on cosmological backgrounds is detailed extensively in \cite{Tattersall:2017eav}.

We follow the same logic as in \cite{Lagos:2016wyv,2017JCAP...01..047L} but using a covariant approach. The main steps of the method are summarised as follows: 
\begin{enumerate}
\item For a given set of gravitational fields, choose a background and write a set of covariant projectors (a set of vectors and tensors) that foliate your spacetime following the global symmetries of the background. Then, consider linear perturbations for each gravitational (and matter) field.

\item Construct the most general quadratic action for the gravitational fields by writing all possible compatible contractions of the covariant background projectors and the linear perturbations. Introduce a free function of the background in front of each possible term and truncate the number of possible terms in the action by choosing a maximum number of derivatives.

\item Choose a desired gauge symmetry and impose local invariance of the quadratic action by solving a set of Noether constraints. The resulting action will be the most general quadratic gauge invariant action around a background with a given set of global symmetries. 

\end{enumerate} 

We now proceed to illustrate the method by following each one of the previous step in the case of a single tensor gravitational field $g_{\mu\nu}$ (or metric) in vacuum with a diffeomorphism invariant action. We start by following step 1. In this case, the background will correspond to Minkowski space: 
\begin{equation}
\bar{g}_{\mu\nu}=\eta_{\mu\nu},
\end{equation}
where the bar denotes the background value of the metric, and $\eta_{\mu\nu}=\text{diag}(-1,1,1,1)$ is the Minkowski metric. We know that this background has a global symmetry under the Poincare group, and thus we can describe the metric with only one projector, the tensor $\eta_{\mu\nu}$, that follows this symmetry. Hence, in this case, we do not need to make any particular foliation. Next, we consider linear perturbations and thus the full metric can be expressed as:
\begin{equation}
g_{\mu\nu}=\eta_{\mu\nu}+ h_{\mu\nu}; \quad | h_{\mu\nu}| \ll |\eta_{\mu\nu}|,
\end{equation}
where $h_{\mu\nu}$ is a linear perturbation, which in general is a function of space and time.

We now follow step 2 and write the most general covariant quadratic action leading to second-order derivative equations of motion. In this case, we can only have two different possible terms (modulo total derivatives):
\begin{equation}\label{GRgral}
S^{(2)}=\int d^4x \left[ \mathcal{A}^{\mu\alpha\beta\nu\gamma\delta}\bar{\nabla}_\mu h_{\alpha\beta} \bar{\nabla}_\nu h_{\gamma\delta} + \mathcal{B}^{\alpha\beta\gamma\delta}h_{\alpha\beta}h_{\gamma\delta}\right],
\end{equation} 
where $\bar{\nabla}_\mu$ are covariant derivatives with respect to the background metric, and the tensors $\mathcal{A}$ and $\mathcal{B}$ are, for consistency, arbitrary functions of the background. These tensors must respect the symmetries of the background and hence be constructed with the tensor $\eta_{\mu\nu}$. Explicitly, the most general form these tensors can take is the following:
\begin{align}\label{GRCoeffs}
\mathcal{A}^{\mu\alpha\beta\nu\gamma\delta} = & c_3 \eta^{\mu\nu}\eta^{\alpha\beta}\eta^{\gamma\delta}+ c_4 \eta^{\mu\alpha}\eta^{\nu\beta}\eta^{\gamma\delta} \nonumber \\ &+ c_5\eta^{\mu \nu}\eta^{\alpha\gamma}\eta^{\beta\delta} + c_6\eta^{\mu\gamma}\eta^{\nu\alpha}\eta^{\beta\delta}, \nonumber\\
\mathcal{B}^{\alpha\beta\gamma\delta} = & c_1 \eta^{\alpha\beta}\eta^{\gamma\delta}+ c_2 \eta^{\alpha\gamma}\eta^{\beta\delta},
\end{align}
where the scalars $c_n$ are free functions of the background, i.e.~constants in this case. We note that we have not actually written all the possible contractions in these tensors $\mathcal{A}$ and $\mathcal{B}$, but instead only those that are inequivalent after considering the contraction with the symmetric tensor perturbation $h_{\mu\nu}$ in the action in eq.~(\ref{GRgral}). 

By plugging in the expressions in eq.~(\ref{GRCoeffs}) into the quadratic action, and separating each term of the action explicitly, the resulting most general quadratic action takes the following form:
\begin{align}
 S^{(2)}= \int d^4x&\left[ c_1h^2+c_2h_{\mu\nu}h^{\mu\nu}+c_3\partial_\mu h \partial^\mu h +c_4 \partial_\mu h^{\mu\nu}\partial_\nu h \right.\nonumber \\ & \ \ \ \left.+c_5\partial_\mu h_{\nu\lambda}\partial^\mu h^{\nu\lambda}+c_6\partial_\mu h_{\nu\lambda}\partial^\nu h^{\mu\lambda}\right], \label{SFP}
\end{align}
where $h=\eta^{\mu\nu}h_{\mu\nu}$ and indices are lowered and raised with the background metric $\eta_{\mu\nu}$.

We now proceed to follow step 3, and we will impose symmetry under linear diffeomorphism invariance. Consider an infinitesimal coordinate transformation:
\begin{align}
 x^{\mu}\rightarrow x^\mu + \epsilon^\mu;\quad |\epsilon^\mu | \ll | x^\mu|, \label{coordtransformation}
\end{align}
where $\epsilon^\mu$ is a linear perturbation that depends on space and time. Under this transformation the background stays the same but the gravitational perturbation field changes as:
\begin{align}
 h_{\mu\nu}\rightarrow h_{\mu\nu}+\partial_\mu\epsilon_\nu+\partial_\nu\epsilon_\mu. \label{htransformation}
\end{align}
If we wish our theory to be invariant under this coordinate transformations, then the variation of the action in eq.~(\ref{SFP}) with respect to the transformation in eq.~(\ref{htransformation}) should vanish. After making suitable integrations by parts, we find that the variation of the action in this case gives:
\begin{align}
 \delta_\epsilon S^{(2)}=\int &d^4x\,\epsilon_\mu \left[-4c_2\partial_\nu h^{\mu\nu}-4c_1\partial^\mu h\right. \nonumber \\&+ 2\, (c_4+c_6)\partial^\mu\partial^\nu\partial^\lambda h_{\nu\lambda} +2\, (2c_5+c_6)\partial_\nu\Box h^{\mu\nu} \nonumber\\
 &\left.+2\, (2c_3+c_4)\partial^\mu\Box h\right], \label{deltaSFP}
\end{align}
where $\Box=\eta^{\mu\nu}\partial_\mu\partial_\nu$ is the d'Alembertian operator. For the action to be gauge invariant we need $\delta_\epsilon S^{(2)}$ to vanish for arbitrary $\epsilon^\mu$, and therefore the whole integrand to vanish. This leads to the following Noether identity:
\begin{eqnarray}
-4c_2\partial_\nu h^{\mu\nu}-4c_1\partial^\mu h+2\, (c_4+c_6)\partial^\mu\partial^\nu\partial^\lambda h_{\nu\lambda} &\nonumber \\ +2\, (2c_5+c_6)\partial_\nu\Box h^{\mu\nu} +2\, (2c_3+c_4)\partial^\mu\Box h &=0.
\end{eqnarray}
Since this identity must be satisfied off-shell, terms with different derivative structure must vanish independently, leading to the following set of Noether constraints:
\begin{align}
 c_1= & c_2=0,\nonumber\\
 c_4= & -c_6=-2c_3=2c_5.
\end{align}
These constraints are simple algebraic relations on the free coefficients $c_n$, and they ensure the action (\ref{SFP}) is linearly diffeomorphism invariant. Using our freedom to rescale the size of $h_{\mu\nu}$, we can set $-4c_4=M_{Pl}^2$, the reduced Planck mass (squared), and write the resulting quadratic action as:
\begin{align}
 S^{(2)}= \int d^4x\,\frac{M_{Pl}^2}{4}&\left[ \frac{1}{2}\partial_\mu h \partial^\mu h - \partial_\mu h^{\mu\nu}\partial_\nu h \right. \nonumber \\ & \ \ \left.-\frac{1}{2}\partial_\mu h_{\nu\lambda}\partial^\mu h^{\nu\lambda}+\partial_\mu h_{\nu\lambda}\partial^\nu h^{\mu\lambda}\right], \label{SFPfinal}
\end{align}
which we recognise as the quadratic expansion of the Einstein-Hilbert action about a Minkowski background \cite{2011PhRvD..83f4038S}. This is the most general quadratic action for a single metric, around Minkowski space, that is linearly diffeomorphism invariant and has second-order derivative equations of motion.


\section{Single-tensor theories on a Schwarzschild background}\label{secGR}
In this section we apply the covariant method for analysing perturbations around a spherically symmetric background. In particular, we consider the case when the gravitational field content is given by a single tensor field and construct the most general quadratic action around a stationary and static black hole background, that is invariant under linear coordinate transformations and has second-order derivative equations of motion. 

We start by following step 1. We assume that the background is given by the Schwarzschild metric:
\begin{equation}
ds^2=\bar{g}_{\mu\nu}dr^\mu dr^\nu=-fdt^2+\frac{1}{f}dr^2+r^2d\theta^2+r^2\sin^2\theta d\phi^2, \label{schmetric}
\end{equation}
where we have used spherical coordinates and defined $f=1-\frac{2m}{r}$, where $m$ is the mass of the central black hole. In order to describe this background in a covariant way, we foliate the spacetime according to the background symmetries. We make a 1+1+2 split and define a time-like unit vector $u^\mu$ and a space-like vector $r^\mu$, which induces orthogonal hypersurfaces with a spatial metric $\gamma_{\mu\nu}$ such that:
\begin{align}
\gamma_{\mu\nu}=\bar{g}_{\mu\nu}+u_\mu u_\nu-r_\mu r_\nu. \label{gammadef}
\end{align}
Thus , $u^\mu$, $r^\mu$ and $\gamma_{\mu\nu}$ act as the projectors for this spacetime. Specifically in this case, projectors are given by:
\begin{align}
u_\mu = & (-f^{\frac{1}{2}},\bf{0})_\mu,\\
r_\mu = & (0,f^{-\frac{1}{2}},0,0)_\mu,\\
\gamma_{AB} = &r^2\Omega_{AB},\\
\gamma_{\mu0} = & \gamma_{\mu 1}= 0,
\end{align}
where $\bf{0}$ is a 3D zero vector, and $\Omega_{AB}$ is the metric on the unit 2-sphere. These projectors are mutually orthogonal to one another:
\begin{equation}
 \gamma^{\mu\nu}u_\nu=0; \quad \gamma^{\mu\nu}r_\nu=0; \quad u^\mu r_\mu =0.
\end{equation}

We now move onto step 2 and construct the most general quadratic gravitational action. As in Section \ref{SecFP}, the most general action quadratic in $h_{\mu\nu}$ with up to second order equations of motion can be written as:
\begin{align}
S_G^{(2)}=\int d^4x\,r^2\sin\theta\;&\left[ \mathcal{A}^{\mu\nu\alpha\beta} h_{\mu\nu} h_{\alpha\beta}
 + \mathcal{B}^{\mu\nu\alpha\beta\delta}\bar{\nabla}_{\delta} h_{\mu\nu} h_{\alpha\beta}
\right. \nonumber \\ &\left. + \mathcal{C}^{\mu\nu\alpha\beta\kappa\delta}\bar{\nabla}_{\kappa} h_{\mu\nu} \bar{\nabla}_{\delta} h_{\alpha\beta}\right], \label{SgenGR}
\end{align}
where the coefficients $\mathcal{A}$, $\mathcal{B}$, and $\mathcal{C}$ are tensors depending on the background. Notice that here, unlike the action in Section \ref{SecFP}, we have a tensor with five indices $\mathcal{B}^{\mu\nu\alpha\beta\delta}$, which we previously ignored as it only contributes to the action as a boundary term in a Minkowski background. Also, for future convenience we have defined the tensors in action (\ref{SgenGR}) with a factor $r^2\sin\theta$ in front. 

We now write the most general form that the tensors $\mathcal{A}$, $\mathcal{B}$, and $\mathcal{C}$ can have respecting the symmetries of the background. In this case, they can be constructed using the three relevant projectors $u^\mu$, $r^\mu$, and $\gamma_{\mu\nu}$, in the following way:
\begin{widetext}
\begin{align}
\mathcal{A}^{\mu\nu\alpha\beta}= &A_1 \gamma^{\mu\nu} \gamma^{\alpha\beta}+\gamma^{\mu\nu} \left(A_2
 u^\alpha u^\beta+A_3 r^\alpha r^\beta+A_4 u^\alpha r^\beta\right)+\gamma^{\mu\alpha} \left(A_5 \gamma^{\nu\beta}+A_6 u^\nu u^\beta+A_7 r^\nu r^\beta+A_8
 u^\nu r^\beta\right)\nonumber\\
 &+u^\mu u^\nu \left(A_9 u^\alpha u^\beta+A_{10}
 r^\alpha r^\beta+A_{11} u^\alpha r^\beta\right)+r^\mu r^\nu \left(A_{12} r^\alpha r^\beta+A_{13} r^\alpha u^\beta\right)+A_{14}u^\mu r^\nu
 u^\alpha r^\beta\label{Atensor},\\
\mathcal{B}^{\mu\nu\alpha\beta\delta}=&\gamma^{\mu\nu}
 \gamma^{\alpha\delta} \left(u^\beta B_1+r^\beta B_2\right)+\gamma^{\mu\delta} \gamma^{\nu\alpha} \left(u^\beta B_3+r^\beta B_4\right)\nonumber\\
 &+\gamma^{\mu\nu}\left(u^\alpha u^\beta u^\delta
 B_5+r^\alpha r^\beta r^\delta B_6+u^\alpha r^\beta u^\delta B_7+u^\alpha r^\beta r^\delta B_8+r^\alpha r^\beta u^\delta B_9+u^\alpha
 u^\beta r^\delta B_{10}\right)\nonumber\\
 &+\gamma^{\mu\delta} \left(u^\nu u^\alpha u^\beta B_{11}+r^\nu r^\alpha r^\beta
 B_{12}+u^\nu u^\alpha r^\beta B_{13}+r^\nu u^\alpha r^\beta B_{14}+u^\nu
 r^\alpha r^\beta B_{15}+r^\nu u^\alpha u^\beta
 B_{16}\right)\nonumber\\
 &+\gamma^{\mu\alpha} \left(u^\nu r^\beta u^\delta
 B_{17}+u^\nu r^\beta r^\delta B_{18}\right)+r^\mu r^\nu u^\alpha u^\beta \left(r^\delta B_{19}+u^\delta
 B_{20}\right)+r^\mu r^\nu r^\alpha u^\beta r^\delta B_{21}+u^\mu u^\nu u^\alpha r^\beta u^\delta
 B_{22}\nonumber\\
 &+r^\mu r^\nu r^\alpha u^\beta u^\delta
 B_{23}+u^\mu u^\nu u^\alpha r^\beta r^\delta B_{24}\label{Btensor},\\
\mathcal{C}^{\mu\nu\alpha\beta\kappa\delta}=&C_1\gamma^{\mu\nu}\gamma^{\alpha\beta}\gamma^{\kappa\delta}+ C_2\gamma^{\mu\alpha}\gamma^{\nu\beta}\gamma^{\kappa\delta}+ C_3\gamma^{\mu\nu}\gamma^{\alpha\kappa}\gamma^{\beta\delta}+ C_4\gamma^{\mu\kappa}\gamma^{\alpha\beta}\gamma^{\nu\delta}
 +\left(C_5 \gamma^{\mu\nu}\gamma^{\alpha\beta}+ C_6\gamma^{\mu\alpha}\gamma^{\nu\beta}\right)u^\kappa u^\delta \nonumber\\
 &+ \left(C_7 \gamma^{\mu\nu}\gamma^{\kappa\delta}+ C_8\gamma^{\mu\kappa}\gamma^{\nu\delta}\right)u^\alpha u^\beta + C_{9}\gamma^{\alpha\beta} u^\mu u^\nu u^\kappa u^\delta+C_{10}\gamma^{\kappa\delta} u^\alpha u^\beta u^\mu u^\nu
+ \left(C_{11}\gamma^{\kappa\delta}\gamma^{\beta \nu}+ C_{12}\gamma^{\kappa\beta}\gamma^{\delta \nu}\right)u^\mu u^\alpha\nonumber\\
&+ \left(C_{13}\gamma^{\alpha\beta}\gamma^{\nu \delta}+ C_{14}\gamma^{\alpha\nu}\gamma^{\delta \beta}\right)u^\mu u^\kappa
+C_{15}\gamma^{\mu\alpha}u^\nu u^\beta u^\kappa u^\delta + C_{16}\gamma^{\mu\kappa}u^\nu u^\beta u^\alpha u^\delta+ C_{17} u^\mu u^\alpha u^\nu u^\beta u^\kappa u^\delta\nonumber\\
&+\left(C_{18} \gamma^{\mu\nu}\gamma^{\alpha\beta}+ C_{19}\gamma^{\mu\alpha}\gamma^{\nu\beta}\right)r^\kappa r^\delta + \left(C_{20} \gamma^{\mu\nu}\gamma^{\kappa\delta}+ C_{21}\gamma^{\mu\kappa}\gamma^{\nu\delta}\right)r^\alpha r^\beta + C_{22}\gamma^{\alpha\beta} r^\mu r^\nu r^\kappa r^\delta+C_{23}\gamma^{\kappa\delta} r^\alpha r^\beta r^\mu r^\nu
\nonumber\\
&+ \left(C_{24}\gamma^{\kappa\delta}\gamma^{\beta \nu}+ C_{25}\gamma^{\kappa\beta}\gamma^{\delta \nu}\right)r^\mu r^\alpha+ \left(C_{26}\gamma^{\alpha\beta}\gamma^{\nu \delta}+ C_{27}\gamma^{\alpha\nu}\gamma^{\delta \beta}\right)r^\mu r^\kappa
+C_{28}\gamma^{\mu\alpha}r^\nu r^\beta r^\kappa r^\delta + C_{29}\gamma^{\mu\kappa}r^\nu r^\beta r^\alpha r^\delta\nonumber\\
&+ C_{30} r^\mu r^\alpha r^\nu r^\beta r^\kappa r^\delta+\gamma^{\mu\nu}\left(C_{31}\gamma^{\alpha\beta}r^\kappa u^\delta +C_{32}\gamma^{\alpha\kappa}u^\beta r^\delta+C_{33}\gamma^{\alpha\kappa}u^\delta r^\beta+C_{34}\gamma^{\kappa\delta}r^\alpha u^\beta\right)\nonumber\\
&+\gamma^{\mu\alpha}\left(C_{35}\gamma^{\nu\beta}u^\kappa r^\delta +C_{36}\gamma^{\kappa\delta}r^\nu u^\beta\right)+\gamma^{\mu\kappa}\left(C_{37}\gamma^{\alpha\delta}r^\nu u^\beta+C_{38}\gamma^{\nu\delta}r^\alpha u^\beta +C_{39}\gamma^{\nu\alpha}r^\beta u^\delta +C_{40}\gamma^{\nu\alpha}u^\beta r^\delta\right)\nonumber\\
&+\gamma^{\mu\nu} \left(r^\alpha r^\beta u^\kappa u^\delta C_{41}+u^\alpha u^\beta r^\kappa r^\delta C_{42}+r^\alpha u^\beta
 r^\kappa u^\delta C_{43}\right)+\gamma^{\kappa\delta} \left(u^\mu u^\nu r^\alpha r^\beta C_{44}+r^\mu u^\nu
 r^\alpha u^\beta C_{45}\right)\nonumber\\
 &+\gamma^{\mu\kappa} \left(u^\nu
 r^\alpha r^\beta u^\delta C_{46}+r^\nu
 u^\alpha u^\beta r^\delta C_{47}+u^\nu u^\alpha r^\beta r^\delta C_{48}+r^\nu u^\alpha r^\beta u^\delta C_{49}\right)\nonumber\\
 &+\gamma^{\mu\alpha} \left(r^\nu r^\beta u^\kappa u^\delta
 C_{50}+u^\nu u^\beta r^\kappa
 r^\delta C_{51}+u^\nu r^\beta u^\kappa r^\delta C_{52}\right)+u^\mu
 u^\nu u^\alpha r^\beta \gamma^{\kappa\delta}C_{53}\nonumber\\
 &+ \gamma^{\mu\nu} \left(u^\alpha u^\beta u^\kappa r^\delta
 C_{54}+u^\alpha r^\beta u^\kappa u^\delta C_{55}\right)+\gamma^{\mu\kappa} \left(u^\nu u^\alpha r^\beta u^\delta C_{56}+u^\nu u^\alpha u^\beta
 r^\delta C_{57}+r^\nu u^\alpha u^\beta u^\delta
 C_{58}\right)\nonumber\\
 &+\gamma^{\mu\alpha} \left(u^\nu u^\beta
 u^\kappa r^\delta C_{59}+u^\nu r^\beta u^\kappa u^\delta C_{60}\right)+r^\mu r^\nu r^\alpha u^\beta \gamma^{\kappa\delta}C_{61} +\gamma^{\mu\nu}
 \left(r^\alpha r^\beta r^\kappa u^\delta C_{62}+r^\alpha u^\beta r^\kappa r^\delta
 C_{63}\right)\nonumber\\
 &+\gamma^{\mu\kappa} \left(r^\nu r^\alpha u^\beta
 r^\delta C_{64}+r^\nu r^\alpha r^\beta u^\delta
 C_{65}+u^\nu r^\alpha r^\beta r^\delta C_{66}\right)+\gamma^{\mu\alpha} \left(r^\nu r^\beta r^\kappa u^\delta C_{67}+r^\nu
 u^\beta r^\kappa r^\delta C_{68}\right)\nonumber\\
 &+u^\mu u^\nu u^\alpha u^\beta r^\kappa u^\delta
 C_{69}+r^\mu u^\nu u^\alpha u^\beta u^\kappa u^\delta C_{70}+r^\mu r^\nu r^\alpha r^\beta r^\kappa u^\delta C_{71}+u^\mu r^\nu r^\alpha
 r^\beta r^\kappa r^\delta C_{72}+u^\mu u^\nu u^\alpha u^\beta r^\kappa r^\delta C_{73}\nonumber\\
 &+r^\mu u^\nu u^\alpha u^\beta r^\kappa u^\delta
 C_{74}+r^\mu r^\nu u^\alpha u^\beta u^\kappa
 u^\delta C_{75}+r^\mu u^\nu r^\alpha u^\beta u^\kappa u^\delta C_{76}+r^\mu r^\nu
 r^\alpha r^\beta u^\kappa u^\delta C_{77}+u^\mu r^\nu r^\alpha r^\beta u^\kappa r^\delta
 C_{78}\nonumber\\
 &+u^\mu u^\nu r^\alpha r^\beta r^\kappa r^\delta C_{79}+u^\mu r^\nu u^\alpha r^\beta r^\kappa r^\delta
 C_{80}+u^\mu r^\nu r^\alpha
 r^\beta u^\kappa u^\delta C_{81}+u^\mu u^\nu r^\alpha
 r^\beta u^\kappa r^\delta C_{82}+u^\mu r^\nu u^\alpha r^\beta u^\kappa r^\delta C_{83}\nonumber\\
 &+u^\mu u^\nu u^\alpha
 r^\beta r^\kappa r^\delta C_{84},\label{Ctensor}
\end{align}
\end{widetext}
where, as in the previous section, we have only defined the set of tensors that lead to distinct terms in the quadratic action\footnote{Whilst in principle one should symmetrise over the indices of $\mathcal{A}$, $\mathcal{B}$, and $\mathcal{C}$ in order to obtain the most general tensors, the additional symmetrised terms do not contribute any new terms to the action so they have been ommited.}. Here, the coefficients $A_n$, $B_n$, and $C_n$ are arbitrary scalar functions of the background, and hence of radius. We note that the tensors $\mathcal{A}$, $\mathcal{B}$, and $\mathcal{C}$ could come from the background metric $\bar{g}_{\mu\nu}$ and its derivatives to arbitrary order. Hence, we are restricting the number of derivatives allowed for the perturbations $h_{\mu\nu}$, but not for the background. We comment here that, in using only the projectors $u^\mu, r^\mu,$ and $\gamma^{\mu\nu}$, we have implicitly restricted ourselves to studying theories that do not include parity violation. To study such theories, for example Chern-Simons theories of gravity \cite{Alexander:2009tp}, we would also have to use the four dimensional Levi-Civita tensor $\epsilon^{\mu\nu\alpha\beta}$ when constructing our background tensors. 

From equations (\ref{Atensor})-(\ref{Ctensor}) we can see how less symmetric backgrounds can lead to a larger number of free parameters in the gravitational action. Whereas in Minkowski the action in step 2 had only 6 free constant parameters, in a spherically symmetric background we find 122 free functions of radius. As we shall see later, we will also find more Noether constraints in this section, and so the total gauge invariant action will have only one extra free parameter compared to the Minkowski case.

Having obtained an explicit expression for the coefficients in eq.~(\ref{SgenGR}), we proceed to step 3. We want the total quadratic action to be linearly diffeomorphism invariant. In this case, the metric perturbation will transform as the Lie derivative of the background metric along an infinitesimal coordinate transformation vector $\epsilon^\mu$. That is,
\begin{align}
 h_{\mu\nu}\rightarrow h_{\mu\nu}+\bar{\nabla}_\mu\epsilon_\nu+\bar{\nabla}_\nu\epsilon_\mu, \label{hgaugetransformation}
\end{align}
where again $\epsilon_\mu$ is an arbitrary gauge parameter.
 The action given by eq.~(\ref{SgenGR}) can now be varied to find the Noether identities. Schematically, an infinitesimal variation of the total action can be written as:
\begin{equation}
\delta S^{(2)}_G=\int d^4x \left[ \mathcal{E}^{\mu\nu} \delta h_{\mu\nu}\right],
\end{equation}
where $\delta$ denotes a functional variation, and $\mathcal{E}^{\mu\nu}$ is the equations of motion of the perturbation field $h_{\mu\nu}$. We now consider the functional variation of the action when the perturbation field transform as in eq.~(\ref{hgaugetransformation}). After making suitable integrations by parts we find:
\begin{equation}\label{GRActionVariation}
\delta_\epsilon S^{(2)}_T= \int d^4x\left[ -2\bar{\nabla}_\nu\left(\mathcal{E}^{\mu\nu}\right)\right]\epsilon_\mu,
\end{equation}
where we have used the fact that $\mathcal{E}^{\mu\nu}$ is a symmetric tensor. For the total action to be gauge invariant we impose $\delta_\epsilon S^{(2)}_T=0$, which leads to four Noether identities given by each one of the components of the bracket in eq.~(\ref{GRActionVariation}). From these Noether identities we can read a number of Noether constraints that will relate the values of the free parameters $A_n$, $B_n$ and $C_n$ of the quadratic gravitational action. In order to read off the Noether constraints easily, we rewrite the Noether identities solely in terms of the projectors $u^\mu$, $r^\mu$ and $\gamma_{\mu\nu}$, by eliminating all covariant derivatives of the background using the equations in Appendix \ref{appendixS1}. For instance, we will rewrite the covariant derivative of a function $G$ as:
\begin{align}
\bar{\nabla}_\mu G=f^{\frac{1}{2}}r_\mu\frac{\partial G}{\partial r}.
\end{align}
In this way, due to the fact that the projectors are mutually orthogonal, any perturbation field contracted with different projectors or different index structure must vanish independently. Through this process, we obtain 120 Noether constraints for the coefficients $A_n$, $B_n$, and $C_n$ (see Appendix \ref{appendixT}). Thus, we are left with only two free parameters: a free function of $r$, $C_1$, and a constant, $C_{41}$. In fact, we find that all terms which depend on the parameter $C_1$ cancel in the final action, thus leaving the action dependent only on the constant $C_{41}$. We can thus write the total \textit{gauge invariant} action as:
\begin{align}
S_G^{(2)}=\int d^4x\,r^2\sin\theta\,M_{Pl}^2\,\mathcal{L}_{EH}\label{LfinalGR},
\end{align}
where we have chosen $C_{41}=-\frac{1}{4}M_{Pl}^2$, with $M_{Pl}$ being the reduced Planck mass, in order to describe modifications from GR. The Lagrangian $\mathcal{L}_{EH}$ is the quadratic expansion of the Einstein-Hilbert action, i.e. $\frac{1}{2}\sqrt{-g}R$, and is given by:
\begin{align}
 \mathcal{L}_{EH}= &\frac{1}{8}\bar{\nabla}_\mu h \bar{\nabla}^\mu h - \frac{1}{4}\bar{\nabla}_\mu h^{\mu\nu}\bar{\nabla}_\nu h-\frac{1}{8}\bar{\nabla}_\mu h_{\nu\lambda}\bar{\nabla}^\mu h^{\nu\lambda}\nonumber \\ &+\frac{1}{4}\bar{\nabla}^\mu h_{\mu\lambda}\bar{\nabla}_\nu h^{\nu\lambda}+\frac{1}{4}h^{\mu\rho}(h^{\nu\sigma} \bar{R}_{\rho\nu\mu\sigma}-h^{\nu}_{\rho}\bar{R}_{\mu\nu})\nonumber\\
 &+\frac{1}{16}\bar{R}(h^2-2h_{\mu\nu}h^{\mu\nu})+\frac{1}{4}\bar{R}_{\mu\nu}(2h^\mu_{\,\sigma}h^{\sigma\nu}-hh^{\mu\nu}),
 \label{EHaction}
 \end{align}
where $\bar{R}$, $\bar{R}_{\mu\nu}$ and $\bar{R}_{\rho\nu\mu\sigma}$ are the Ricci scalar, Ricci tensor and Riemann tensor for the background metric, respectively. 
 
Having found the most general gauge invariant quadratic action for a single tensor field on a Schwarzschild background, we can now find the equations of motion for different types of perturbations. Due to the spherical symmetry of the background, perturbations can be decomposed into tensor spherical harmonics and classified in terms of their parity: either odd (axial) or even (polar) \cite{Regge:1957td,Rezzolla:2003ua}. As our action is gauge invariant, we are free to choose a convenient gauge for our calculations. We will work in the Regge-Wheeler gauge \cite{Regge:1957td}, in which our odd and even perturbations take the following form \cite{Regge:1957td,Rezzolla:2003ua}:
 \begin{align}
 h_{\mu\nu,lm}^{odd}=&
 \begin{pmatrix}
 0&0&h_0(r)B^{lm}_\theta&h_0(r)B^{lm}_\phi\\
 0&0&h_1(r)B^{lm}_\theta&h_1(r)B^{lm}_\phi\\
 sym&sym&0&0\\
 sym&sym&0&0
 \end{pmatrix}e^{-i\omega t},
 \label{hodd}\\
 h_{\mu\nu,lm}^{even}=&
 \begin{pmatrix}
 H_0(r)f&H_1(r)&0&0\\
 sym&\frac{H_2(r)}{f}&0&0\\
 0&0&K(r)r^2&0\\
 0&0&0&K(r)r^2\sin\theta
 \end{pmatrix}Y^{lm}e^{-i\omega t},
 \label{heven}
\end{align}
where $sym$ indicates a symmetric entry, $B^{lm}_\mu$ is the odd parity vector spherical harmonic and $Y^{lm}$ is the standard scalar spherical harmonic, as described in \cite{Martel:2005ir,Ripley:2017kqg} (note there are slight differences in convention between the definitions of tensorial spherical harmonics used in \cite{Martel:2005ir} and \cite{Ripley:2017kqg}). Here, the amplitude of linear perturbations is described by the functions $h_i$, $H_i$ and $K$. The properties of tensor spherical harmonics and of the Schwarzschild spacetime are explored in great length in \cite{Martel:2005ir,Ripley:2017kqg}; the calculations of those papers were used throughout the calculations made here. We have also assumed a time dependence of $e^{-i\omega t}$ for our perturbations, due to the static nature of the background spacetime. Furthermore, spherical harmonic indices will be omitted from now on, with each equation assumed to hold for a given $l$ (we will find that the equations of motion are independent of $m$, which is unsurprising due to the spherical symmetry of the background). In general, the metric perturbation will be represented by a sum over $l$, $m$, and $\omega$ of the modes.

Clearly eq.~(\ref{LfinalGR}) shows that we have recovered the correct quadratic expansion of GR for single tensor theories of gravity. It will, however, be instructive for later sections to proceed with the full analysis of the equations of motion derived from the action given by eq.~(\ref{LfinalGR}). 

\subsection{Odd parity perturbations}
We will first consider odd parity perturbations, where $h_{\mu\nu}$ is given by eq.~(\ref{hodd}). We find the following two Euler-Lagrange equations upon varying eq.~(\ref{LfinalGR}) with respect to $h_0$ and $h_1$, respectively:
\begin{align}
\frac{d^2h_0}{dr^2}+i\omega\frac{dh_1}{d r}+i\omega\frac{2h_1}{r}-\frac{h_0}{r^2}f^{-1}\left(l(l+1)-\frac{4m}{r}\right)&=0,\label{hodd1}\\
f^{-1}\left(2i\omega\frac{h_0}{r}-i\omega\frac{d h_0}{d r}+\omega^2h_1\right)-\frac{h_1}{r^2}(l+2)(l-1)&=0\label{hodd2}.
\end{align}
Multiplying eq.~(\ref{hodd1}) by $-i\omega$ and taking the $r$ derivative of eq.~(\ref{hodd2}) and substituting, we find:
\begin{align}
-i\omega h_0=f\frac{d}{d r}\left(h_1f\right)\label{hodd3}.
\end{align}
Using eq.~(\ref{hodd3}) to eliminate $h_0$ from eq.~(\ref{hodd2}), we arrive at the famous Regge-Wheeler equation \cite{Regge:1957td}:
\begin{align}
\frac{d^2 Q}{d r_\ast^2}+\left[\omega^2-V_{RW}(r)\right]Q=0,\label{reggewheeler}
\end{align}
where we have introduced the Regge-Wheeler function $Q$ and the tortoise coordinate $r_\ast$ \cite{Regge:1957td} such that:
\begin{align}
 Q=&h_1\frac{f}{r}\label{rwfunction},\\
 dr_\ast=&f^{-1}dr\label{tortoise},
\end{align} 
whilst the potential $V_{RW}(r)$ is given by
\begin{align}
V_{RW}=\left(1-\frac{2m}{r}\right)\left(\frac{1}{r^2}l(l+1)-\frac{6m}{r^3}\right).\label{Vreggewheeler}
\end{align}

\subsection{Even parity perturbations}
For even parity perturbations, where $h_{\mu\nu}$ is given by eq.~(\ref{heven}), four Euler-Lagrange equations are found upon varying eq.~(\ref{LfinalGR}) with respect to $H_0$, $H_1$, $H_2$, and $K$. After a series of straightforward, but lengthy, manipulations, the following set of equations is found:
\medskip
\begin{widetext}
\begin{align}
&\frac{d K}{d r}+\frac{r-3m}{r(r-2m)}K-\frac{1}{r}H_0+\frac{1}{2}\frac{l(l+1)}{i\omega r^2}H_1=0\label{heven1},\\
&\frac{d H_0}{d r}+\frac{r-3m}{r(r-2m)}K-\frac{r-4m}{r(r-2m)}H_0+\left[\frac{i\omega r}{r-2m}++\frac{1}{2}\frac{l(l+1)}{i\omega r^2}\right]H_1=0,\\
&\frac{d H_1}{d r}+\frac{i\omega r}{r-2m}K+\frac{i\omega r}{r-2m}H_0+\frac{2m}{r(r-2m)}H_1=0,
\end{align}
which satisfy the following algebraic identity:
\begin{equation}
\left[\frac{6m}{r}+(l+2)(l-1)\right]H_0-\left[(l+2)(l-1)-\frac{2\omega^2r^3}{r-2m}+\frac{2m(r-3m)}{r(r-2m)}\right]K-\left[2i\omega r+\frac{l(l+1)m}{i\omega r^2}\right]H_1=0\label{heven4},
\end{equation}
\end{widetext}
and the relation $H_0=H_2$ is also found. We can make the following field redefinitions, as described in \cite{Gleiser:2006yz}, in terms of the Zerilli function $\psi(r)$
\begin{align}
K=&g_1(r)\psi+\left(1-\frac{2m}{r}\right)\frac{\partial \psi}{\partial r},\nonumber\\
H_1=&-i\omega \left(g_2(r)\psi+r\frac{\partial \psi}{\partial r}\right),\nonumber\\
H_0=&\frac{\partial}{\partial r}\left[\left(1-\frac{2m}{r}\right)\left(g_2(r)\psi+r\frac{\partial \psi}{\partial r}\right)\right]-K,\label{zerillidef}
\end{align}
where we have introduced:
\begin{align}
g_1(r)=&\frac{L(L+1)r^2+2Lmr+6m^2}{r^2(Lr+3m)},\nonumber\\
g_2(r)=&\frac{Lr^2-3Lmr-3m^2}{(r-2m)(Lr+3m)},\nonumber\\
2L=&(l+2)(l-1).\label{evenauxfunctions}
\end{align}
After making the substitutions given by eq.~(\ref{zerillidef}) in eq.~(\ref{heven4}), we find a single equation determining the evolution of perturbations, the familiar Zerilli equation \cite{Zerilli:1970se}:
\begin{align}
\frac{d^2 \psi}{d r_\ast^2} + \left[\omega^2-V_Z(r)\right]\psi=0\label{zerilli},
\end{align}
where the potential $V_Z(r)$ is given by
\begin{align}
V_Z(r)=2\left(1-\frac{2m}{r}\right)\frac{L^2r^2\left[(L+1)r+3m\right]+9m^2(Lr+m)}{r^3(Lr+3m)^2}.\label{Vzerilli}
\end{align}

For both odd and even parity perturbations, we recover the Regge-Wheeler and Zerilli equations as in GR. This is the expected result for a theory containing a single tensor perturbation about a Schwarzschild background. Having found that $C_1$ vanishes from the final gauge invariant action, and setting $C_{41}=-\frac{1}{4}M_{Pl}^2$, as explained above, there is no further parameter freedom in our theory. This result may seem to be in contrast to a similar calculation performed on a cosmological background \cite{Lagos:2016wyv,Tattersall:2017eav}, where it was found that a time-dependent Planck mass was allowed, and the running of this generalised Planck mass induced modifications in the equations for linear cosmological perturbations. However, in such a case the background evolution of the metric was left free, but if the background evolution was fixed to be that of GR (as in this paper), then the generalised Planck mass would have to be constant and thus no modified evolution for perturbations would be found.


\section{Scalar-Tensor theories on a Schwarzschild background}\label{SecSTBH}

Having studied the case of a single-tensor perturbation on a Schwarzschild background, we now construct the most general gravitational action for perturbations of a tensor and a scalar field, that leads to second order equations of motion and is linearly diffeomorphism invariant. The stability of stationary black holes under perturbations in scalar-tensor theories has been studied in \cite{Ganguly:2017ort,Kobayashi:2012kh,Kobayashi:2014wsa}. We follow the covariant procedure as in the previous section, but with the addition of a gravitational scalar field $\chi$:
\begin{align}
 \chi=\bar{\chi}(r)+\delta\chi; \quad |\delta\chi|\ll |\bar{\chi}|,
\end{align}
where $\bar{\chi}$ is the background value of the scalar field and $\delta\chi$ is a linear perturbation non-minimally coupled to the metric $g_{\mu\nu}$ and its perturbation, $h_{\mu\nu}$. We will assume that we are considering scalar-tensor theories of gravity where a no-hair theorem exists, such that the background spacetime is still described by the Schwarzschild solution given by eq.~(\ref{schmetric}) \cite{2012PhRvL.108h1103S,Sotiriou:2013qea,Creminelli:2017sry}. In fact, due to constraints on the speed excess of gravitational waves \cite{Baker:2017hug,2017arXiv171005901M,2017arXiv171005893S} from multi-messenger measurements of binary neutron star mergers \cite{2017arXiv171005832T,2017arXiv171005834L,2017arXiv171005833L}, many scalar-tensor theories that might have supported black hole hair have now been strongly constrained and, barring any fine tuning, effectively ruled out in favour of models that do not support scalar hair. Furthermore, due to our implicit assumption that no parity violation occurs in the theories studied, theories such as Chern-Simons gravity \cite{Alexander:2009tp} are not covered by the following analysis of scalar-tensor theories. Perturbations about a Schwarzschild background in Chern-Simons gravity have been studied in \cite{Cardoso:2009pk,Molina:2010fb}.

In the case that the background is given by eq.~(\ref{schmetric}), the background value of the scalar field, $\bar{\chi}$ must correspond to the trivial solution of a constant \cite{2012PhRvL.108h1103S,Sotiriou:2013qea}:
\begin{align}
\bar{\nabla}_\mu\bar{\chi}=0.
\end{align}
Note that the perturbation to the scalar, $\delta\chi$, is non-trivial and still depends on the space-time coordinates. 

Since we have the same background as in the previous section, we continue to use the 1+1+2 split of spacetime with the projectors $u^\mu$, $r^\mu$, and $\gamma_{\mu\nu}$. 

We proceed to step 2 and write down the most general scalar-tensor gravitational action as:
\begin{align}
 S_G^{(2)}=\int d^4x&\,r^2\sin\theta\; \left[\mathcal{A}^{\mu\nu\alpha\beta} h_{\mu\nu} h_{\alpha\beta}
 + \mathcal{B}^{\mu\nu\alpha\beta\delta}\bar{\nabla}_{\delta} h_{\mu\nu} h_{\alpha\beta}
\right. \nonumber\\
&\left.+ \mathcal{C}^{\mu\nu\alpha\beta\kappa\delta}\bar{\nabla}_{\kappa} h_{\mu\nu} \bar{\nabla}_{\delta} h_{\alpha\beta} +A_{\chi}(\delta\chi)^2 \right. \nonumber\\
&\left.+\mathcal{A}_\chi^{\mu\nu}\delta\chi h_{\mu\nu}+\mathcal{B}_\chi^{\mu\nu\delta}h_{\mu\nu}\bar{\nabla}_\delta\delta\chi \right. \nonumber\\
&\left.+\mathcal{C}_\chi^{\mu\nu}\bar{\nabla}_\mu\delta\chi\bar{\nabla}_\nu\delta\chi+\mathcal{D}_\chi^{\mu\nu\delta\kappa}\bar{\nabla}_\kappa\delta\chi\bar{\nabla}_\delta h_{\mu\nu}\right], \label{SgenSTBH}
\end{align}
where the $\mathcal{A}$, $\mathcal{B}$, and $\mathcal{C}$ are the same as those given by (\ref{Atensor})-(\ref{Ctensor}). We see that we also have two new tensors describing the self-interactions of the scalar field and three for the interactions between the scalar and tensor fields. These new tensors are arbitrary functions of the background, and hence must follow the background symmetry and can be constructed solely from the projectors $u^\mu$, $r^\mu$, and $\gamma_{\mu\nu}$. Similarly as in the previous section, we proceed to write down the most general forms these five new tensors can take: 
\begin{widetext}
\begin{align}
 \mathcal{A}_\chi^{\mu\nu}=& \ A_{\chi1}u^\mu u^\nu +A_{\chi2}\gamma^{\mu\nu}+A_{\chi3}r^\mu r^\nu+A_{\chi4}r^\mu u^\nu, \\
 \mathcal{B}_\chi^{\mu\nu\delta}=& \ B_{\chi1}u^\mu u^\nu u^\delta +B_{\chi2}u^\delta\gamma^{\mu\nu}+B_{\chi3}u^\mu\gamma^{\delta\nu}+B_{\chi4}r^\mu r^\nu r^\delta +B_{\chi5} r^\delta \gamma^{\mu\nu}+B_{\chi6}r^\mu \gamma^{\nu\delta}+B_{\chi7}r^\delta u^\mu u^\nu +B_{\chi8}u^\delta r^\mu r^\nu\nonumber\\
 & +B_{\chi9}u^\delta u^\mu r^\nu+B_{\chi10}r^\delta r^\mu u^\nu,\\
 \mathcal{C}_\chi^{\mu\nu}=& \ C_{\chi1}u^\mu u^\nu +C_{\chi2}\gamma^{\mu\nu}+C_{\chi3}r^\mu r^\nu +C_{\chi4} u^\mu r^\nu,\\
 \mathcal{D}_\chi^{\mu\nu\delta\kappa}=&\ D_{\chi1}u^\mu u^\nu u^\delta u^\kappa +D_{\chi2}u^\mu u^\nu \gamma^{\kappa\delta} + D_{\chi3}u^\kappa u^\delta \gamma^{\mu\nu} \ + D_{\chi4} u^\mu u^\kappa \gamma^{\delta\nu} + D_{\chi5}\gamma^{\mu\nu}\gamma^{\kappa\delta} + D_{\chi6}\gamma^{\mu\kappa}\gamma^{\nu\delta}+ D_{\chi7}r^\mu r^\nu r^\delta r^\kappa \nonumber\\
 &+D_{\chi8}r^\mu r^\nu \gamma^{\kappa\delta} + D_{\chi9}r^\kappa r^\delta \gamma^{\mu\nu}+ D_{\chi10} r^\mu r^\kappa \gamma^{\delta\nu}+D_{\chi11}\gamma^{\mu\nu}r^\kappa u^\delta +D_{\chi12}\gamma^{\mu\delta}u^\mu r^\kappa+D_{\chi13}\gamma^{\mu\delta}r^\mu u^\kappa +D_{\chi14}u^\mu r^\nu \gamma^{\kappa\delta}\nonumber \\
 & +D_{\chi15}r^\mu r^\nu u^\kappa u^\delta+D_{\chi16}r^\mu u^\nu r^\delta u^\kappa +D_{\chi17}u^\mu u^\nu r^\kappa r^\delta +D_{\chi18}r^\mu r^\nu r^\delta u^\kappa+D_{\chi19}u^\mu r^\nu r^\kappa r^\delta +D_{\chi20}u^\mu u^\nu u^\delta r^\kappa\nonumber \\
 & +D_{\chi21} r^\mu u^\nu u^\kappa u^\delta,
\end{align}
\end{widetext}
while $A_{\chi}$ is a scalar and hence simply considered to be free function of $r$. Here, each of the coefficients $A_{\chi\,n}$, $B_{\chi\,n}$, $C_{\chi\,n}$, and $D_{\chi\,n}$ are free functions of $r$ also. We see that we have 30 additional free functions due to the inclusion of the scalar field $\chi$. 

We now proceed to step 3. As before, we impose linear diffeomorphism invariance of the total action given by eq.~(\ref{SgenSTBH}). While the metric transforms as in eq.~(\ref{hgaugetransformation}) under an infinitesimal coordinate transformation, the new scalar field transforms as:
\begin{align}
\delta\chi \rightarrow \delta\chi + \epsilon^\mu\bar{\nabla}_\mu \bar{\chi}. \label{scalargaugetransformation}
\end{align}
Note that as we are assuming that our background is Schwarzschild, and as such has no scalar `hair', $\bar{\nabla}_\mu\bar{\chi}$ vanishes leaving $\delta\chi$ gauge invariant. 

The total action given by eq.~(\ref{SgenSTBH}) can now be varied under the gauge transformation. As in the previous sections, we obtain a number of Noether constraints by enforcing independent terms in the Noether identities to vanish. Due to $\delta\chi$ being gauge invariant, the Noether constraints that are obtained in Section \ref{secGR} are also valid for the analysis of the action given by eq.~(\ref{SgenSTBH}). The additional Noether constraints found for the $A_{\chi\,n}$, $B_{\chi\,n}$, $C_{\chi\,n}$, and $D_{\chi\,n}$ are given in Appendix \ref{appendixSTNC}. We find that the final action depends on 10 free parameters from the original action given by eq.~(\ref{SgenSTBH}):
\begin{align}
C_{41}, \; A_{\chi0}, \; C_{\chi1-4}, \; D_{\chi5}, \; D_{\chi8}, \; D_{\chi11}, \; D_{\chi15},\label{STparameters}
\end{align}
where once again $C_{41}$ is a constant whilst the other 9 parameters are free to be functions of $r$. Note that $A_{\chi0}$ and $C_{\chi1-4}$ are unconstrained due to $\delta\chi$ being gauge invariant on a Schwarzschild background. The final quadratic gauge-invariant action for scalar-tensor theories on a Schwarzschild background can be written as
\begin{align}
S^{(2)}_G=\int d^4x\,r^2\sin\theta\,M_{Pl}^2\left[\mathcal{L}_{EH}+\mathcal{L}_{\chi}\right],\label{LfinalST}
\end{align}
where $\mathcal{L}_{EH}$ is given by eq.~(\ref{EHaction}), and again we have chosen $M_{Pl}^2=-4C_{41}$ in order to describe modifications from GR. Thus, the entire action depends on 9 free parameters. The additional Lagrangian due to the inclusion of the scalar field $\chi$ is given by
\begin{widetext}
\begin{align}
M_{Pl}^2\mathcal{L}_{\chi}=&A_{\chi0}\left(\delta\chi\right)^2+C_{\chi1}u^\mu u^\nu \bar{\nabla}_\mu\delta\chi\bar{\nabla}_\nu\delta\chi +C_{\chi2}\gamma^{\mu\nu} \bar{\nabla}_\mu\delta\chi\bar{\nabla}_\nu\delta\chi+C_{\chi3}r^\mu \bar{\nabla}_\mu\delta\chi\bar{\nabla}_\nu\delta\chi r^\nu +C_{\chi4} u^\mu r^\nu \bar{\nabla}_\mu\delta\chi\bar{\nabla}_\nu\delta\chi\nonumber\\
&-\frac{1}{4m^2}\left(2D_{\chi5}(f-1)^2-2D_{\chi8}(f-1)^2+m\left(4f\left(\frac{dD_{\chi5}}{dr}+\frac{d^2D_{\chi15}}{dr^2}m+\frac{dD_{\chi8}}{dr}(f-1)-\frac{dD_{\chi5}}{dr}f\right)\right.\right.\nonumber\\
 &\left.\left.+\frac{dD_{\chi15}}{dr}\left(1+2f-3f^2\right)\right)\right)u^\mu u^\nu h_{\mu\nu}\delta\chi+\frac{\sqrt{f}}{2m}D_{\chi5}(f-1)r^\mu \gamma^{\nu\delta} h_{\mu\nu}\bar{\nabla}_\delta\delta\chi\nonumber\\
 &-\frac{1}{4m}\left(\frac{dD_{\chi15}}{dr}(f-1)^2+4\left(\frac{dD_{\chi8}}{dr}(1-f)+\frac{d^2D_{\chi8}}{dr^2}mf\right)+\frac{dD_{\chi5}}{dr}\left(f^2-1\right)\right)\gamma^{\mu\nu}h_{\mu\nu}\delta\chi\nonumber\\
 &+\frac{1}{4m^2}\left(f-1\right)\left(2D_{\chi5}(f-1)-2D_{\chi8}(f-1)+m\left(\frac{dD_{\chi15}}{dr}(f-1)+4\frac{dD_{\chi8}}{dr}f\right)\right)r^\mu r^\nu\nonumber\\
 &+\frac{1}{4m\sqrt{f}}\left(D_{\chi11}\left(-1-2f+3f^2\right)-4\frac{dD_{\chi11}}{dr}mf\right)u^\delta\gamma^{\mu\nu} h_{\mu\nu}\bar{\nabla}_\delta\delta\chi\nonumber\\
 &+\frac{1}{2m\sqrt{f}}\left(D_{\chi11}(1-f)+2\frac{dD_{\chi11}}{dr}mf\right)u^\mu\gamma^{\delta\nu} h_{\mu\nu}\bar{\nabla}_\delta\delta\chi+\frac{1}{4m\sqrt{f}}\left(D_{\chi15}(f-1)^2+4D_{\chi8}f(f-1)\right)r^\mu r^\nu r^\delta h_{\mu\nu}\bar{\nabla}_\delta\delta\chi\nonumber\\
 & +\frac{1}{4m\sqrt{f}}\left(-D_{\chi15}-2D_{\chi8}(f-1)^2-4\frac{dD_{\chi8}}{dr}mf+D_{\chi15}f(2-f)-D_{\chi5}\left(f^2-1\right)\right) r^\delta \gamma^{\mu\nu} h_{\mu\nu}\bar{\nabla}_\delta\delta\chi\nonumber\\
 &+\frac{1}{4m\sqrt{f}}\left(D_{\chi15}(f-1)^2-4f\left(\frac{dD_{\chi15}}{dr}m-D_{\chi5}(f-1)+D_{\chi8}(f-1)\right)\right)r^\delta u^\mu u^\nu h_{\mu\nu}\bar{\nabla}_\delta\delta\chi\nonumber\\
 & -\frac{2\sqrt{f}(f-1)}{m}\left(D_{\chi5}-D_{\chi8}-D_{\chi15}\right)u^\delta u^\mu r^\nu h_{\mu\nu}\bar{\nabla}_\delta\delta\chi-\frac{\sqrt{f}(f-1)}{m}D_{\chi11}r^\delta r^\mu r^\nu h_{\mu\nu} h_{\mu\nu}\bar{\nabla}_\delta\delta\chi\nonumber\\
 &+\left(-D_{\chi5}+D_{\chi8}+D_{\chi15}\right)u^\mu u^\nu \gamma^{\kappa\delta} \bar{\nabla}_\kappa\delta\chi\bar{\nabla}_\delta h_{\mu\nu} + \left(-D_{\chi5}+D_{\chi8}+D_{\chi15}\right)u^\kappa u^\delta \gamma^{\mu\nu} \bar{\nabla}_\kappa\delta\chi\bar{\nabla}_\delta h_{\mu\nu} \nonumber\\
 & -2\left(-D_{\chi5}+D_{\chi8}+D_{\chi15}\right) u^\mu u^\kappa \gamma^{\delta\nu} \bar{\nabla}_\kappa\delta\chi\bar{\nabla}_\delta h_{\mu\nu} + D_{\chi5}\gamma^{\mu\nu}\gamma^{\kappa\delta} \bar{\nabla}_\kappa\delta\chi\bar{\nabla}_\delta h_{\mu\nu} - D_{\chi5}\gamma^{\mu\kappa}\gamma^{\nu\delta} \bar{\nabla}_\kappa\delta\chi\bar{\nabla}_\delta h_{\mu\nu}\nonumber\\
 &+D_{\chi8}r^\mu r^\nu \gamma^{\kappa\delta} \bar{\nabla}_\kappa\delta\chi\bar{\nabla}_\delta h_{\mu\nu} + D_{\chi8}r^\kappa r^\delta \gamma^{\mu\nu} \bar{\nabla}_\kappa\delta\chi\bar{\nabla}_\delta h_{\mu\nu}-2 D_{\chi8} r^\mu r^\kappa \gamma^{\delta\nu} \bar{\nabla}_\kappa\delta\chi\bar{\nabla}_\delta h_{\mu\nu}+D_{\chi11}\gamma^{\mu\nu}r^\kappa u^\delta \bar{\nabla}_\kappa\delta\chi\bar{\nabla}_\delta h_{\mu\nu}\nonumber\\
 & -D_{\chi11}\gamma^{\mu\delta}u^\mu r^\kappa \bar{\nabla}_\kappa\delta\chi\bar{\nabla}_\delta h_{\mu\nu}-D_{\chi11}\gamma^{\mu\delta}r^\mu u^\kappa \bar{\nabla}_\kappa\delta\chi\bar{\nabla}_\delta h_{\mu\nu} +D_{\chi11}u^\mu r^\nu \gamma^{\kappa\delta} \bar{\nabla}_\kappa\delta\chi\bar{\nabla}_\delta h_{\mu\nu}\nonumber \\
 & +D_{\chi15}r^\mu r^\nu u^\kappa u^\delta \bar{\nabla}_\kappa\delta\chi\bar{\nabla}_\delta h_{\mu\nu}-2D_{\chi15}r^\mu u^\nu r^\delta u^\kappa \bar{\nabla}_\kappa\delta\chi\bar{\nabla}_\delta h_{\mu\nu} +D_{\chi15}u^\mu u^\nu r^\kappa r^\delta \bar{\nabla}_\kappa\delta\chi\bar{\nabla}_\delta h_{\mu\nu}.\label{Lchiplus}
 \end{align}
\end{widetext}
As in Section \ref{secGR}, having obtained a form for the fully covariant diffeomorphism invariant action, we can study the odd and even parity perturbations separately. 

\subsection{Odd parity perturbations}\label{STodd}
We will first consider odd parity perturbations, where $h_{\mu\nu}$ is given by eq.~(\ref{hodd}). Since $\delta\chi$ has no contribution to the odd parity sector due to being a scalar, and is also gauge invariant (hence the gravitational self-interactions are the same as in those in the previous section), the odd parity gravitational perturbations are again governed by the Regge-Wheeler equation given by eq.~(\ref{reggewheeler}).

\subsection{Even parity perturbations}\label{STeven}
For even parity perturbations $h_{\mu\nu}$ is given by eq.~({\ref{heven}), whilst we decompose $\delta\chi$ into spherical harmonics like so (following the convention of \cite{PhysRevD.34.333})
\begin{align}
\delta\chi^{lm}=\frac{2\varphi(r)}{r}Y^{lm}e^{-i\omega t},
\end{align}
where $Y^{lm}$ is again the standard scalar spherical harmonic \cite{Martel:2005ir,Ripley:2017kqg}.

We again vary the gauge invariant scalar tensor action with respect to $H_0$, $H_1$, $H_2$, $K$, and $\varphi$ and combine the metric perturbations into a single function $\tilde{\psi}$ using the following substitutions
\begin{align}
K=&g_1(r)\tilde{\psi}+\left(1-\frac{2m}{r}\right)\frac{\partial \tilde{\psi}}{\partial r}+\beta(r)\frac{2\varphi}{r},\nonumber\\
H_1=&-i\omega \left(g_2(r)\tilde{\psi}+r\frac{\partial \tilde{\psi}}{\partial r}\right),\nonumber\\
H_0=&\frac{\partial}{\partial r}\left[\left(1-\frac{2m}{r}\right)\left(g_2(r)\tilde{\psi}+r\frac{\partial \tilde{\psi}}{\partial r}\right)\right]-K,\label{zerillidef2}
\end{align}
where $g_1$, $g_2$, and $L$ are again given by eqs.~(\ref{evenauxfunctions}). Note the difference between the substitutions given above to those previously given in eq.~(\ref{zerillidef}), where we have introduced the parameter $\beta$ in eq.~(\ref{zerillidef2}), which is a dimensionless function of $r$ that we are free to choose, reflecting the freedom to make a further field redefinition by mixing the metric and scalar perturbations. The relation between $H_0$, $H_2$, and $\varphi$ for scalar-tensor theories is given in Appendix \ref{appendixSTRel}. Note that, as $H_2$ is given by in terms of $H_0$ \textit{and} $\varphi$, the metric and scalar perturbations will in general be mixed. The following equations of motion are found
\begin{align}
\frac{d^2\tilde{\psi}}{dr_\ast^2}+\left(\omega^2-V_Z\right)\tilde{\psi}+a_1\varphi=&0,\label{STeq1}\\
b_1\frac{d^2\varphi}{dr_\ast^2}+b_2\frac{d\varphi}{dr_\ast}+b_3\varphi+b_4\frac{d^2\tilde{\psi}}{dr_\ast^2}+b_5\frac{d\tilde{\psi}}{dr_\ast}+b_6\tilde{\psi}=&0\label{STeq2},
\end{align}
where the $a_n$ and $b_n$ are functions of $r$ and of the 9 remaining free parameters of the theory, as well as functions of $\omega$ and $l$ (see Appendix \ref{appendixSTEven}). $V_Z$ is the Zerilli potential given by eq.~(\ref{Vzerilli}). It is interesting to note that eq.~(\ref{STeq1}) is not in the most general form that any second order equation for $\tilde{\psi}$ and $\varphi$ could take, e.g. there are no terms proportional to $\frac{d\varphi}{dr_\ast}$. This is due to the fact that these equations come from an action principle and thus they must be integrable.
In addition, as we will discuss later, the function $a_1$ in eq.~(\ref{STeq1}) will depend on the free function $\beta(r)$ and in virtue of an appropriate choice for $\beta$, we will always be able to make $a_1=0$. 

Eqs.~(\ref{STeq1})-(\ref{STeq2}) form a pair of homogeneous coupled ordinary differential equations with non-constant coefficients. By introducing the following fields
\begin{align}
\tilde{\Psi}=\frac{d\tilde{\psi}}{dr_\ast},\;\Phi=\frac{d\varphi}{dr_\ast},\label{STPhi}
\end{align}
we can write eqs.~(\ref{STeq1})-(\ref{STPhi}) as the first order matrix equation
\begin{align}
\frac{d}{dr_\ast}\bm{\Lambda}=-\bm{M}\bm{\Lambda},
\end{align}
where
\begin{align}
\bm{\Lambda}=\begin{bmatrix}
\tilde{\Psi}\\
\tilde{\psi}\\
\Phi\\
\varphi\\
\end{bmatrix},\;
\bm{M}=\begin{bmatrix}
0&\omega^2-V_Z&0&a_1\\
-1&0&0&0\\
\frac{b_5}{b_1}&\frac{b_6-b_4\left(\omega^2-V_z\right)}{b_1}&\frac{b_2}{b_1}&\frac{b_3-b_4a_1}{b_1}\\
0&0&-1&0
\end{bmatrix}.
\end{align}
We have then found that, as expected, these scalar-tensor theories propagate one degree of freedom, in addition to the two metric perturbations. Furthermore, we can see that, in general, even though the background black hole has no hair and it is identical to GR, at the level of perturbations the scalar field can be excited and generate hair. This means that the evolution of metric perturbations will be generically modified and hence the detection of quasi normal modes in gravitational wave experiments would allow us to test and distinguish scalar-tensor models from GR. 

Next, we proceed to work out two specific examples of scalar-tensor theories and show explicitly how the equations of motion can be modified. In particular, we will consider two examples: one in which the evolution of even perturbations is different to GR, and another example where both odd and even perturbations evolve exactly in the same way as GR because the terms that modify gravity vanish in a Schwarzschild background. 

\subsection{Examples}

As explained in Section \ref{STodd}, the equation of motion for odd parity metric perturbations, i.e. the Regge-Wheeler equation, is unaffected by the presence of scalar field perturbations. This is due to the trivial background profile of the scalar field, and the fact that $\delta\chi$ is purely of even parity. Thus, in the following examples only the equations for even parity perturbations will be shown in detail.
\subsubsection{Brans-Dicke}
Let us take our test action to take the form of a simple Brans-Dicke model with scalar field mass $\mu$ \cite{PhysRevD.34.333}:
\begin{align}
S=\int d^4x\;\sqrt{-g}\;\frac{M_{Pl}^2}{2}\;\left[\chi R-\frac{\Omega}{\chi}\nabla_\mu \chi \nabla^\mu \chi -\mu^2\chi^2\right], \label{SBD}
\end{align}
where $R$ is the Ricci scalar and $\Omega$ is a constant. Perturbing eq.~(\ref{SBD}) to quadratic order in linear perturbations we find the following values for the free parameters listed in eq.~(\ref{STparameters}):
\begin{align}
A_{\chi0}=&\;-\frac{1}{2}\mu^2,\quad C_{\chi1}=-C_{\chi2}=-C_{\chi3}=-\frac{\Omega}{2},\nonumber\\
D_{\chi15}=&-D_{\chi5}=-D_{\chi8}=\frac{1}{4},\label{BDparams}
\end{align}
with $C_{\chi4}$ and $D_{\chi11}$ vanishing. Here we have ignored the overall scaling of $M_{Pl}^2$. With this parameter choice, we find the following equations of motion:
\begin{widetext}
\begin{align}
\frac{d^2\tilde{\psi}}{dr_\ast^2}+\left(\omega^2-V_Z\right)\tilde{\psi}+4(2\beta-1)\frac{9m^2+\left(l^2+l-4\right)mr+r^2\left(2-l-l^2+r^2\omega^2\right)}{r^3\left(6m+(l+2)(l-1)r\right)}\varphi=&0,\label{zerilliBD}\\
\frac{d^2\varphi}{dr_\ast^2}+\left[\omega^2-\left(1-\frac{2m}{r}\right)\left(\frac{1}{r^2}l(l+1)+\frac{2 m}{r^3}+\frac{8\mu^2}{3-8\Omega}\right)\right]\varphi=&0,
\end{align}
\end{widetext}
Note that we can use our freedom to make a field redefinition to set $\beta=\frac{1}{2}$ and mix the even parity metric and scalar perturbations. This field redefinition removes the $\varphi$ contribution to eq.~(\ref{zerilliBD}) and leaves it in an identical form to the GR Zerilli equation. The field $\tilde{\psi}$ obeying this Zerilli equation is now, however, a mixture of the even parity metric and scalar perturbations, rather than a pure metric perturbation as in the GR case. Furthermore, with the Brans-Dicke parameter choice given by eq.~(\ref{BDparams}), the relation between the metric perturbations $H_2$ and the other perturbations is not simply $H_2=H_0$ as with GR (see Appendix \ref{appendixSTRel}), but rather involves the scalar perturbation $\varphi$ as well.

We can define a generalised Regge-Wheeler potential
\begin{align}
\hat{V}_{RW}=\left(1-\frac{2m}{r}\right)\left(\frac{1}{r^2}l(l+1)+\frac{2\sigma m}{r^3}\right),\label{Vhat}
\end{align}
where $\sigma=1-s^2$, and with $s$ being the spin of the field being perturbed. We see that in the case of a massless scalar field (i.e. $\mu=0$) the scalar perturbation obeys an equation of motion of the form
\begin{align}
\frac{d^2P}{dr_\ast^2}+\left(\omega^2-\hat{V}_{RW}\right)P=0,\label{RWgeneral}
\end{align}
where $P$ is some perturbed field of spin $s$. $\hat{V}_{RW}$ would, for example, be evaluated with $s=2$ for metric perturbations, and with $s=0$ for scalar perturbations. Thus for a massless scalar field, both the odd parity metric perturbation and the scalar perturbation obey the generalised Regge-Wheeler equation given by eq.~(\ref{RWgeneral}). This is the result shown in \cite{PhysRevD.34.333}, where an analysis on the stability of these perturbations was performed. In \cite{Saijo:1996iz} it was shown that, in Kerr spacetime, whilst gravitational waves (i.e.~the metric perturbations) might dominate over the Brans-Dicke scalar waves, an observation of the polarisation of the gravitational waves (a now realistic prospect \cite{2017arXiv170909660T}) could divide the tensor and scalar parts. 

\subsubsection{Cubic Galilieon}

For the Cubic Galileon model, that is cosmologically relevant, the action takes the form \cite{Barreira:2013eea}
\begin{align}
S=\int d^4x \sqrt{-g}\;&\left[\frac{M_{Pl}^2}{2}R-\frac{1}{2}c_2\nabla_\mu \chi \nabla^\mu \chi\right.\nonumber\\
&\left.-\frac{c_3}{M_{Pl}H_0^2}\Box\chi\nabla_\mu \chi \nabla^\mu \chi\right],\label{Sgal}
\end{align}
where $c_2$ and $c_3$ are dimensionless constants, and $H_0$ is the value of the Hubble parameter today. Again perturbing eq.~(\ref{Sgal}) to quadratic order in linear perturbations (about a constant background scalar field), we find the following values for the free parameters listed in eq.~(\ref{STparameters}):
\begin{align}
C_{\chi1}=-C_{\chi2}=-C_{\chi3}=c_2, 
\end{align}
with the rest of the parameters vanishing.

With this parameter choice we find the following equations of motion:
\begin{widetext}
\begin{align}
\frac{d^2\tilde{\psi}}{dr_\ast^2}+\left(\omega^2-V_Z\right)\tilde{\psi}+4\beta\frac{9m^2+\left(l^2+l-4\right)mr+r^2\left(2-l-l^2+r^2\omega^2\right)}{r^3\left(6m+(l+2)(l-1)r\right)}\varphi=&0, \label{zerilliCG}\\
\frac{d^2\varphi}{dr_\ast^2}+\left(\omega^2-\hat{V}_{RW}\right)\varphi=&0,
\end{align}
\end{widetext}
where $\hat{V}_{RW}$ is given by eq.~(\ref{Vhat}) and is evaluated with $s=0$. We can now use our freedom to choose $\beta$ to remove the $\varphi$ contribution from eq.~(\ref{zerilliCG}). By choosing $\beta=0$ we recover the GR Zerilli equation for the metric perturbation, whilst the scalar perturbation obeys the scalar Regge-Wheeler equation. We emphasise that in this case, with $\beta=0$, the field redefinitions made in eq.~(\ref{zerillidef2}) are equivalent to those in eq.~(\ref{zerillidef}), and hence $\tilde{\psi}=\psi$. Therefore, in this model we find that both even and odd perturbations are unaffected by the presence of the scalar field. Indeed, as seen from Appendix \ref{appendixSTRel}, we find that the metric perturbation $H_2$ is related to the other fields through $H_2=H_0$, as in GR.

This is the expected result for a minimally coupled massless scalar field such as the Cubic Galileon, considering that the higher order derivative term in the action given by eq.~(\ref{Sgal}) (parameterised by $c_3$) vanishes at quadratic order for a trivial background solution for $\varphi$.

\subsection{Field redefinitions}\label{fieldsST}

In the above example of Brans-Dicke gravity, we find that the simple field redefinition given by setting $\beta(r)=\frac{1}{2}$ in eq.~(\ref{zerillidef2}) allows to find a combination of the scalar and metric perturbations that obeys the Zerilli equation, as in GR. In fact, we find that it is \textit{always} possible to set $a_1=0$ in eq.~(\ref{STeq1}) through such a field redefinition by making the following choice for $\beta$:
\begin{widetext}
\begin{align}
M_{Pl}^2\beta=&\frac{1}{l (l+1) r^2 \omega \left(-2 \left(l^2+l-4\right)m r+r^2 \left(l^2+l-r^2 \omega ^2-2\right)-9 m^2\right)}\left(-2 \omega \left(l (l+1) r^2 D_{\chi15} \left(-3 m^2+2m r+r^4 \omega ^2\right)\right.\right.\nonumber\\
 &+2 m (2 m-r) \left(\frac{dD_{\chi15}}{dr} m \left(l (l+1)r^2+8 m^2-4 m r\right)+2 (2 m-r) \left(\frac{dD_{\chi8}}{dr}\left(\left(l^2+l-2\right) r^2+8 m^2-4 m r\right)\right.\right.\nonumber\\
 &\left.\left.\left.+r \left(2 r \left(\frac{dD_{\chi5}}{dr}+i \frac{dD_{\chi11}}{dr} r \omega \right)+\frac{d^2D_{\chi15}}{dr^2} m (r-2 m)-2\frac{d^2D_{\chi8}}{dr^2} (r-2 m)^2\right)\right)\right)\right)\nonumber\\
 & \left.+2 l (l+1) r^2 \omega\frac{dD_{\chi8}}{dr} \left(2 \left(l^2+l-5\right) mr-\left(l^2+l-2\right) r^2+12 m^2\right)-i l (l+1) rD_{\chi11} (2 m-r) \left(l^2 m+l m-2 r^3 \omega^2\right)\right).
\end{align}
\end{widetext}
Thus a field $\tilde{\psi}$ that obeys the standard GR Zerilli equation can always be found. Therefore, in order to solve the evolution of perturbations in scalar-tensor theories, we would have to solve the standard Zerilli equation first and then separately solve the additional scalar field equation. This is an extremely useful tool given the amount of study already devoted to the solutions and quasi-normal modes of the Zerilli equation \cite{1975RSPSA.343..289C,1975RSPSA.344..441C}. Note, however, that in general $\tilde{\psi}$ will represent a mixture of metric and scalar perturbations, and not the pure metric perturbation of GR. Furthermore, the scalar field perturbation may be excited by a second family of quasi-normal modes, different to the GR spectrum calculated from the Zerilli equation, by solving eq.~(\ref{STeq2}) with $\tilde{\psi}=0$.


\allowdisplaybreaks[1]
\section{Vector-Tensor theories on a Schwarzschild background}\label{secVT}

We now study the case of vector-tensor theories of gravity, and construct the most general gravitational action for linear perturbations of a tensor and a vector field that leads to second order equations of motion and is linearly diffeomorphism invariant. We follow the covariant procedure as in the previous sections, but with the addition of a gravitational vector field $\zeta^\mu$:
\begin{align}
 \zeta^\mu=\bar{\zeta}_r(r)r^\mu+\bar{\zeta}_t(r)u^\mu+\delta\zeta^\mu; \quad |\delta\zeta^\mu|\ll |\bar{\zeta}^\mu|,
\end{align}
where $\bar{\zeta}_r$ and $\bar{\zeta}_t$ are the background values of the field in the $r^\mu$ and $u^\mu$ directions, respectively. We assume the background value of the vector field to be radius-dependent, and to only have components parallel to $u^\mu$ and $r^\mu$, in order to comply with the global symmetries of the background. The vector perturbation $\delta\zeta^\mu$ is a linear perturbation non-minimally coupled to the metric $g_{\mu\nu}$ and its perturbation, $h_{\mu\nu}$. 

We again choose to use the `hair-less' Schwarzschild solution as our background spacetime, as in previous sections. For consistency, in this case, we must impose that the background vector field vanishes:
\begin{align}
\bar{\zeta}_r=\bar{\zeta}_t=0\label{vectorbackground}.
\end{align}
The perturbed vector field $\delta\zeta^\mu$ is, however, non-zero. Note that this is slightly different to the case of the scalar-tensor theories discussed in Section \ref{SecSTBH}, where the requirement of having no scalar hair simply imposed that the background value of the scalar field be constant (rather than vanishing). This is because in scalar-tensor theories, a constant non-zero background scalar field would only alter the action through the addition of an overall constant that has no physical effect, and hence the background metric solution is the same as the one in GR. However, in vector-tensor theories, a constant background vector field would generically couple to the metric through covariant derivatives (i.e.~to the Christoffel symbols), forcing the background metric away from a Schwarzschild solution. 

As we are using the same background spacetime as in previous sections, we continue to use a 1+1+2 split of the background with the projectors $u^\mu$, $r^\mu$, and $\gamma_{\mu\nu}$. We now proceed to step 2 and write down the most general vector-tensor gravitational action as:
\begin{align}
 S_G^{(2)}=\int d^4x\,&r^2\sin\theta\; \left[\mathcal{A}^{\mu\nu\alpha\beta} h_{\mu\nu} h_{\alpha\beta}
 + \mathcal{B}^{\mu\nu\alpha\beta\delta}\bar{\nabla}_{\delta} h_{\mu\nu} h_{\alpha\beta}
\right. \nonumber\\
&\left.+ \mathcal{C}^{\mu\nu\alpha\beta\kappa\delta}\bar{\nabla}_{\kappa} h_{\mu\nu} \bar{\nabla}_{\delta} h_{\alpha\beta}+\mathcal{A}_{\zeta^2}^{\mu\nu}\delta\zeta_\mu \delta\zeta_\nu\right. \nonumber\\
&\left.+\mathcal{A}_{\zeta h}^{\mu\nu\lambda}\delta\zeta_\lambda h_{\mu\nu}+\mathcal{B}_{\zeta h}^{\mu\nu\lambda\kappa}h_{\mu\nu}\bar{\nabla}_\kappa\delta\zeta_\lambda \right. \nonumber\\
&\left.+\mathcal{B}_{\zeta^2}^{\mu\nu\kappa}\delta\zeta_\mu\bar{\nabla}_\kappa\delta\zeta_\nu+\mathcal{C}_\zeta^{\mu\nu\kappa\delta}\bar{\nabla}_\kappa\delta\zeta_\mu\bar{\nabla}_\delta\delta\zeta_\nu\right.\nonumber\\
&\left.+\mathcal{D}_\zeta^{\mu\nu\lambda\delta\kappa}\bar{\nabla}_\kappa\delta\zeta_\lambda\bar{\nabla}_\delta h_{\mu\nu}\right], \label{SgenVT}
\end{align}
where the $\mathcal{A}$, $\mathcal{B}$, and $\mathcal{C}$ are the same as those given by (\ref{Atensor})-(\ref{Ctensor}). We see that we also have three new tensors describing the self-interactions of the vector field and three for the interactions between the vector and tensor fields. These new tensors are arbitrary functions of the background, and hence must follow the background symmetry and can be constructed solely from the projectors $u^\mu$, $r^\mu$, and $\gamma_{\mu\nu}$. Similarly as in the previous section, we proceed to write down the most general forms these six new tensors can take: 
\begin{widetext}
\begin{align}
\mathcal{A}_{\zeta^2}^{\mu\nu}=&A_{\zeta1}u^\mu u^\nu + A_{\zeta2}\gamma^{\mu\nu} + A_{\zeta3}r^\mu r^\nu + A_{\zeta4}u^\mu r^\nu, \\
\mathcal{A}_{\zeta h}^{\mu\nu\lambda}=&A_{\zeta5}u^\mu u^\nu u^\lambda + A_{\zeta6} \gamma^{\mu\nu}u^\lambda +A_{\zeta7}u^\mu \gamma^{\nu\lambda}+A_{\zeta8}r^\mu r^\nu r^\lambda + A_{\zeta9} \gamma^{\mu\nu}r^\lambda +A_{\zeta10}r^\mu \gamma^{\nu\lambda} +A_{\zeta11} r^\mu r^\nu u^\lambda\nonumber\\
& + A_{\zeta12} u^\mu u^\nu r^\lambda+A_{\zeta13} r^\mu u^\nu u^\lambda +A_{\zeta14}u^\mu r^\nu r^\lambda,\\
 \mathcal{B}_{\zeta h}^{\mu\nu\lambda\kappa}=&B_{\zeta1}u^\mu u^\nu u^\lambda u^\kappa +B_{\zeta2}u^\mu u^\nu\gamma^{\lambda\kappa}+B_{\zeta3}u^\kappa u^\lambda \gamma^{\mu\nu}+B_{\zeta4}u^\mu u^\lambda\gamma^{\kappa\nu}+B_{\zeta5}u^\mu u^\kappa\gamma^{\nu\lambda}+B_{\zeta6}\gamma^{\mu\nu}\gamma^{\lambda\kappa} +B_{\zeta7}\gamma^{\mu\kappa}\gamma^{\nu\lambda}\nonumber\\
 &+B_{\zeta8}r^\mu r^\nu r^\lambda r^\kappa +B_{\zeta9}r^\mu r^\nu\gamma^{\lambda\kappa}+B_{\zeta10}r^\kappa r^\lambda \gamma^{\mu\nu}+B_{\zeta11}r^\mu r^\lambda\gamma^{\kappa\nu}+B_{\zeta12}r^\mu r^\kappa\gamma^{\nu\lambda}+B_{\zeta13}u^\mu u^\nu r^\kappa r^\lambda +B_{\zeta14}r^\mu r^\nu u^\kappa u^\lambda\nonumber \\
 &+B_{\zeta15}u^\mu u^\nu u^\kappa r^\lambda +B_{\zeta16}u^\mu u^\nu r^\kappa u^\lambda +B_{\zeta17} r^\mu r^\nu r^\kappa u^\lambda +B_{\zeta18} r^\mu r^\nu u^\kappa r^\lambda+B_{\zeta19}r^\mu u^\nu r^\kappa u^\lambda+B_{\zeta20}r^\mu u^\nu r^\kappa r^\lambda \nonumber\\
 &+B_{\zeta21}r^\mu u^\nu u^\kappa r^\lambda +B_{\zeta22}r^\mu u^\nu u^\kappa u^\lambda+B_{\zeta23}\gamma^{\mu\nu}u^\kappa r^\lambda+B_{\zeta24}\gamma^{\mu\nu}r^\kappa u^\lambda+B_{\zeta25}\gamma^{\kappa\lambda}u^\mu r^\nu +B_{\zeta26}\gamma^{\mu\kappa}u^\nu r^\lambda+B_{\zeta27}\gamma^{\mu\kappa}r^\nu u^\lambda\nonumber\\
 &+B_{\zeta28}\gamma^{\nu\lambda}u^\mu r^\kappa+B_{\zeta29}\gamma^{\nu\lambda}r^\mu u^\kappa,\\
 \mathcal{B}_{\zeta^2}^{\mu\nu\kappa}=&B_{\zeta30}u^\mu \gamma^{\kappa\nu} +B_{\zeta31}r^\mu \gamma^{\kappa\nu} +B_{\zeta32}u^\mu u^\kappa r^\nu +B_{\zeta33}u^\mu r^\kappa r^\nu,\\
 \mathcal{C}_{\zeta}^{\mu\nu\kappa\delta}=&C_{\zeta1}u^\mu u^\nu u^\kappa u^\delta +C_{\zeta2}u^\mu u^\nu \gamma^{\kappa\delta} +C_{\zeta3}u^\kappa u^\delta\gamma^{\mu\nu} +C_{\zeta4}u^\mu u^\delta\gamma^{\nu\kappa}+C_{\zeta5}\gamma^{\mu\nu}\gamma^{\kappa\delta}+C_{\zeta6}\gamma^{\mu\delta}\gamma^{\nu\kappa}+C_{\zeta7}r^\mu r^\nu r^\kappa r^\delta\nonumber\\
 & +C_{\zeta8}r^\mu r^\nu \gamma^{\kappa\delta} +C_{\zeta9}r^\kappa r^\delta\gamma^{\mu\nu} +C_{\zeta10}r^\mu r^\delta\gamma^{\nu\kappa}+C_{\zeta11}r^\mu u^\nu \gamma^{\kappa\delta} +C_{\zeta12}r^\mu u^\kappa \gamma^{\delta\nu} + C_{\zeta13}r^\kappa u^\delta \gamma^{\mu\nu} + C_{\zeta14} r^\mu r^\nu u^\kappa u^\delta \nonumber\\
 &+C_{\zeta15}r^\mu r^\nu u^\kappa r^\delta +C_{\zeta16} u^\mu u^\nu r^\kappa r^\delta +C_{\zeta17}u^\mu u^\nu u^\kappa r^\delta +C_{\zeta18} r^\mu u^\nu u^\kappa u^\delta +C_{\zeta19} r^\mu u^\nu r^\kappa r^\delta +C_{\zeta20} r^\mu u^\nu u^\kappa r^\delta,\\
 \mathcal{D}_{\zeta}^{\mu\nu\lambda\kappa\delta}=&D_{\zeta1}u^\mu u^\nu u^\lambda u^\kappa u^\delta+D_{\zeta2}u^\lambda u^\kappa u^\delta \gamma^{\mu\nu} + D_{\zeta3}u^\lambda u^\mu u^\nu \gamma^{\kappa\delta} +D_{\zeta4}u^\lambda u^\mu u^\kappa \gamma^{\delta\nu} +D_{\zeta5}u^\mu u^\nu u^\delta \gamma^{\kappa\lambda} +D_{\zeta6}u^\mu u^\kappa u^\delta \gamma^{\nu\lambda} \nonumber\\
 & + D_{\zeta7}u^\lambda\gamma^{\mu\nu}\gamma^{\kappa\delta} + D_{\zeta8}u^\lambda\gamma^{\mu\kappa}\gamma^{\delta\nu} +D_{\zeta9}u^\mu\gamma^{\nu\kappa}\gamma^{\delta\lambda}+D_{\zeta10}u^\kappa\gamma^{\mu\nu}\gamma^{\delta\lambda} +D_{\zeta11}u^\kappa\gamma^{\mu\delta}\gamma^{\nu\lambda}+D_{\zeta12}u^\mu\gamma^{\nu\lambda}\gamma^{\kappa\delta}\nonumber\\
 &+D_{\zeta13}r^\mu r^\nu r^\lambda r^\kappa r^\delta+D_{\zeta14}r^\lambda r^\kappa r^\delta \gamma^{\mu\nu} + D_{\zeta15}r^\lambda r^\mu r^\nu \gamma^{\kappa\delta} +D_{\zeta16}r^\lambda r^\mu r^\kappa \gamma^{\delta\nu} +D_{\zeta17}r^\mu r^\nu r^\delta \gamma^{\kappa\lambda} +D_{\zeta18}r^\mu r^\kappa r^\delta \gamma^{\nu\lambda} \nonumber\\
 & + D_{\zeta19}r^\lambda\gamma^{\mu\nu}\gamma^{\kappa\delta} + D_{\zeta20}r^\lambda\gamma^{\mu\kappa}\gamma^{\delta\nu} +D_{\zeta21}r^\mu\gamma^{\nu\kappa}\gamma^{\delta\lambda}+D_{\zeta22}r^\kappa\gamma^{\mu\nu}\gamma^{\delta\lambda} +D_{\zeta23}r^\kappa\gamma^{\mu\delta}\gamma^{\nu\lambda}+D_{\zeta24}r^\mu\gamma^{\nu\lambda}\gamma^{\kappa\delta}\nonumber\\
 &+D_{\zeta25}\gamma^{\mu\nu}r^\kappa r^\delta u^\lambda+D_{\zeta26}\gamma^{\mu\nu}u^\kappa u^\delta r^\lambda +D_{\zeta27}\gamma^{\mu\nu} u^\kappa r^\delta u^\lambda +D_{\zeta28}\gamma^{\mu\nu} u^\kappa r^\delta r^\lambda +D_{\zeta29}\gamma^{\lambda\kappa}r^\mu r^\nu u^\delta +D_{\zeta30}\gamma^{\lambda\kappa}u^\mu u^\nu r^\delta \nonumber\\
 &+D_{\zeta31}\gamma^{\lambda\kappa}r^\mu u^\nu u^\delta+D_{\zeta32}\gamma^{\lambda\kappa}r^\mu u^\nu r^\delta +D_{\zeta33}\gamma^{\mu\lambda}u^\nu r^\kappa r^\delta +D_{\zeta34}\gamma^{\mu\lambda}u^\nu u^\kappa r^\delta +D_{\zeta35}\gamma^{\mu\lambda}r^\nu u^\kappa u^\delta +D_{\zeta36}\gamma^{\mu\lambda}r^\nu u^\kappa r^\delta\nonumber\\
 &+D_{\zeta37}\gamma^{\kappa\delta}u^\mu u^\nu r^\lambda+D_{\zeta38}\gamma^{\kappa\delta}r^\mu r^\nu u^\lambda +D_{\zeta39}\gamma^{\kappa\delta}u^\mu r^\nu u^\lambda+D_{\zeta40}\gamma^{\kappa\delta}u^\mu r^\nu r^\lambda +D_{\zeta41}r^\mu r^\nu r^\lambda u^\kappa u^\delta +D_{\zeta42}r^\mu r^\nu u^\lambda r^\kappa u^\delta\nonumber\\
 & +D_{\zeta43}r^\mu r^\nu u^\lambda r^\kappa r^\delta +D_{\zeta44}r^\mu r^\nu r^\lambda u^\kappa r^\delta +D_{\zeta45}u^\mu u^\nu r^\lambda u^\kappa u^\delta +D_{\zeta46}u^\mu u^\nu u^\lambda r^\kappa u^\delta +D_{\zeta47}u^\mu u^\nu u^\lambda r^\kappa r^\delta \nonumber\\
 &+D_{\zeta48}u^\mu u^\nu r^\lambda u^\kappa r^\delta +D_{\zeta49}u^\mu r^\nu u^\lambda u^\kappa u^\delta +D_{\zeta50}u^\mu r^\nu r^\lambda r^\kappa r^\delta +D_{\zeta51}u^\mu r^\nu u^\lambda r^\kappa u^\delta +D_{\zeta52}u^\mu r^\nu r^\lambda u^\kappa u^\delta \nonumber\\
 &+D_{\zeta53}u^\mu r^\nu u^\lambda r^\kappa r^\delta +D_{\zeta54}u^\mu r^\nu r^\lambda u^\kappa r^\delta +D_{\zeta55}\gamma^{\mu\delta}r^\nu u^\lambda r^\kappa +D_{\zeta56}\gamma^{\mu\delta}r^\nu r^\lambda u^\kappa +D_{\zeta57}\gamma^{\mu\delta} r^\nu u^\lambda u^\kappa+D_{\zeta58}\gamma^{\mu\delta}u^\nu u^\lambda r^\kappa\nonumber\\
 &+D_{\zeta59}\gamma^{\mu\delta}u^\nu r^\lambda u^\kappa+D_{\zeta60}\gamma^{\mu\delta}u^\nu r^\lambda r^\kappa +D_{\zeta61}r^\mu r^\nu u^\lambda u^\kappa u^\delta +D_{\zeta62}u^\mu u^\nu r^\lambda r^\kappa r^\delta.
\end{align}
\end{widetext}
Each of the coefficients $A_{\zeta\,n}$, $B_{\zeta\,n}$, $C_{\zeta\,n}$, and $D_{\zeta\,n}$ are free functions of $r$, giving an additional 130 free functions in the most general action (given by eq.~(\ref{SgenVT})) due to the inclusion of the vector field $\zeta^\mu$.

We now proceed to step 3. As before, we impose linear diffeomorphism invariance of the total action given by eq.~(\ref{SgenVT}). While the metric transforms as in eq.~(\ref{hgaugetransformation}) under an infinitesimal coordinate transformation, a vector field perturbations generically transforms as
\begin{align}
\delta\zeta^\mu \rightarrow \delta\zeta^\mu +& \epsilon^\nu\bar{\nabla}_\nu\left(\bar{\zeta}_r(r)r^\mu+\bar{\zeta}_t(r)u^\mu\right)\nonumber\\
&-\left(\bar{\zeta}_r(r)r^\nu+\bar{\zeta}_t(r)u^\nu\right)\bar{\nabla}_\nu \epsilon^\mu, \label{vectorgaugetransformation}
\end{align}
which means that the vector perturbation $\delta\zeta^\mu$ is diffeomorphism invariant in the background given by eq.~(\ref{vectorbackground}).

The total action given by eq.~(\ref{SgenVT}) can now be varied under the gauge transformation. As in the previous sections, we obtain a number of Noether constraints by enforcing independent terms in the Noether identities to vanish. As in Section \ref{SecSTBH}, due to $\delta\zeta^\mu$ being gauge invariant, the Noether constraints that are obtained in Section \ref{secGR} are also valid for the analysis of the action given by eq.~(\ref{SgenVT}). The additional Noether constraints for the coefficients $A_{\zeta\,n}$, $B_{\zeta\,n}$, $C_{\zeta\,n}$, and $D_{\zeta\,n}$ are given in Appendix \ref{appendixVTNC}.

We find that the final action depends on the following 39 free parameters from the original action given by eq.~(\ref{SgenVT}):
\begin{widetext}
\begin{align}
C_{41},\;A_{\zeta 1-4},\;B_{\zeta 30-33}, \;C_{\zeta 1-20}, \;D_{\zeta2},\; D_{\zeta7}, \;D_{\zeta11},\; D_{\zeta15}, \;D_{\zeta19}, \;D_{\zeta21},\; D_{\zeta26}, \;D_{\zeta27}, \;D_{\zeta38},\; D_{\zeta60},\label{VTparameters}
\end{align}
\end{widetext}
where again $C_{41}$ is a constant, whereas all the other parameters are free functions of radius. We note that 28 of these free parameters, namely $A_{\zeta 1-4},\;B_{\zeta 30-33}, \;C_{\zeta 1-20}$, describe vector self-interaction terms, and as such are left unconstrained due to the gauge invariant nature of $\delta\zeta^\mu$ in the background we are considering. The remaining 11 free parameters are those that are left after solving the Noether constraints generated by imposing diffeomorphism invariance.

The final quadratic gauge-invariant action for vector-tensor theories on a pure Schwarzschild background can thus be written as
\begin{align}
S^{(2)}_G=\int d^4x\,r^2\sin\theta\,M_{Pl}^2\left[\mathcal{L}_{EH}+\mathcal{L}_{\zeta}\right],\label{LfinalVT}
\end{align}
where $\mathcal{L}_{EH}$ is given by eq.~(\ref{EHaction}), and we have chosen $M_{Pl}^2=-4C_{41}$. Thus we find that the whole action depends on 38 free parameters. The additional Lagrangian $\mathcal{L}_{\zeta}$ due to the addition of the vector field $\zeta^\mu$ is not presented here for brevity's sake, however the Noether constraints presented in Appendix \ref{appendixVTNC} can simply be substituted into eq.~(\ref{SgenVT}) to find the full covariant action.

As in the previous sections, having obtained a form for the fully covariant diffeomorphism invariant action, we proceed to study the odd and even parity perturbations separately. In general, vector-tensor theories can propagate a massive or massless spin-1 particle and hence at most three different polarisations (or degrees of freedom). As we will see next, one of these polarisations couples to the odd parity metric perturbations, and thus modifies the evolution of odd perturbations, contrary to scalar-tensor theories. This suggests that the odd parity sector might be used to test and distinguish vector-tensor and scalar-tensor modified gravity theories.

\subsection{Odd parity perturbations}\label{VTodd}
We will first consider odd parity perturbations, where $h_{\mu\nu}$ is given by eq.~(\ref{hodd}), whilst $\delta\zeta^\mu$ is given by
\begin{align}
\delta\zeta_\mu^{lm}=z_0(r)e^{-i\omega t}B_\mu^{lm},
\end{align}
with $B_\mu^{lm}$ being the odd parity vector spherical harmonic as described in \cite{Martel:2005ir,Ripley:2017kqg}. After varying the action given by eq.~(\ref{LfinalVT}) with respect to $h_0$, $h_1$, and $z_0$, we find the following system of second order ODEs:
\begin{align}
\frac{d^2 Q}{dr_\ast^2}+\left(\omega^2-V_{RW}\right)Q+c_1\frac{d^2 z_0}{dr_\ast^2}+c_2\frac{d z_0}{dr_\ast}+c_3z_0&=0,\label{VToddeq1}\\
d_4\frac{d Q}{dr_\ast}+d_5Q+d_1\frac{d^2 z_0}{dr_\ast^2}+d_2\frac{d z_0}{dr_\ast}+d_3z_0&=0\label{VToddeq2},
\end{align}
where $Q$ is the Regge-Wheeler function given by eq.~(\ref{rwfunction}), $r_\ast$ is the tortoise coordinate given by eq.~(\ref{tortoise}), and $V_{RW}$ is the Regge-Wheeler potential as given by eq.~(\ref{Vreggewheeler}). The $c_n$ and $d_n$ are functions of $r$, $l$, $\omega$, and 10 of the 38 free functions of the theory (see Appendix \ref{appendixVTOddEq}). The relation linking the metric perturbation $h_0$ to $h_1$ (and thus to $Q$ through eq.~(\ref{rwfunction})) and $z_0$ is given in Appendix \ref{appendixVTOdd}. 

Eqs.~(\ref{VToddeq1})-(\ref{VToddeq2}) form a pair of homogeneous coupled ordinary differential equations with non-constant coefficients. By introducing the following fields
\begin{align}
\mathcal{Q}=\frac{dQ}{dr_\ast},\;\mathcal{Z}_0=\frac{dz_0}{dr_\ast},\label{VTZ0}
\end{align}
we can write eqs.~(\ref{VToddeq1})-(\ref{VToddeq2}) and (\ref{VTZ0}) as the first order matrix equation
\begin{align}
\frac{d}{dr_\ast}\bm{\Lambda}=-\bm{M}\bm{\Lambda},
\end{align}
where
\begin{widetext}
\begin{align}
\bm{\Lambda}=\begin{bmatrix}
\mathcal{Q}\\
Q\\
\mathcal{Z}_0\\
z_0\\
\end{bmatrix},\;
\bm{M}=\begin{bmatrix}
-c_1\frac{d_4}{d_1}&\omega^2-V_{RW}-c_1\frac{d_5}{d_1}&c_2-c_1\frac{d_2}{d_1}&c_3-c_1\frac{d_3}{d_1}\\
-1&0&0&0\\
\frac{d_4}{d_1}&\frac{d_5}{d_1}&\frac{d_2}{d_1}&\frac{d_3}{d_1}\\
0&0&-1&0
\end{bmatrix}.
\end{align}
\end{widetext}

\subsection{Even parity perturbations}\label{VTeven}
For even parity perturbations $h_{\mu\nu}$ is given by eq.~({\ref{heven}), whilst we decompose $\delta\zeta^\mu$ as:

\begin{align}
\delta\zeta_\mu^{lm}=\left(-\frac{z_1(r)}{\sqrt{f}}Y^{lm}u_\mu + z_2(r)\sqrt{f}Y^{lm}r_\mu + z_3(r)E^{lm}_\mu\right)e^{-i\omega t},
\end{align}
where $E^{lm}_\mu$ is the even parity vector spherical harmonic \cite{Martel:2005ir,Ripley:2017kqg}. 

Next, we vary the gauge-invariant vector-tensor action given by eq.~(\ref{LfinalVT}) with respect to $H_0$, $H_1$, $H_2$, $K$, $z_1$, $z_2$, and $z_3$ in order to obtain the relevant set of equations of motion. We combine the metric perturbations into a single `Zerilli function' $\psi$ using the substitutions given by eq.~(\ref{zerillidef}). The relation between $H_0$ and $H_2$ for vector-tensor theories is given in Appendix \ref{appendixVTEven}. The following coupled equations of motion are found for the even parity perturbations:
\begin{align}
&\frac{d^2\psi}{dr_\ast^2}+\left(\omega^2-V_Z\right)\psi+e_1\frac{dz_3}{dr\ast}+e_2z_1+e_3z_2+e_4z_3=0, \label{VTeveneq1}\\
&f_1\frac{d^2z_3}{dr_\ast^2}+f_2\frac{dz_3}{dr_\ast}+f_3z_3+f_4\frac{d^2\psi}{dr_\ast^2}+f_5\frac{d\psi}{dr_\ast}+f_6\psi+f_7\frac{dz_1}{dr_\ast}\nonumber\\
&+f_8z_1+f_9\frac{dz_2}{dr_\ast}+f_{10}z_2=0, \label{VTeveneq2}\\
&j_1\frac{d^2z_1}{dr_\ast^2}+j_2\frac{dz_1}{dr_\ast}+j_3z_1+j_4\frac{d^2\psi}{dr_\ast^2}+j_5\frac{d\psi}{dr_\ast}+j_6\psi+j_7\frac{d^2z_2}{dr_\ast^2}\nonumber\\
&+j_8\frac{dz_2}{dr_\ast}+j_9z_2+j_{10}\frac{dz_3}{dr_\ast}+j_{11}z_3=0, \label{VTeveneq3}\\
&k_1\frac{d^2z_1}{dr_\ast^2}+k_2\frac{dz_1}{dr_\ast}+k_3z_1+k_4\frac{d^2\psi}{dr_\ast^2}+k_5\frac{d\psi}{dr_\ast}+k_6\psi+k_7\frac{d^2z_2}{dr_\ast^2}\nonumber\\
&+k_8\frac{dz_2}{dr_\ast}+k_9z_2+k_{10}\frac{dz_3}{dr_\ast}+k_{11}z_3=0\label{VTeveneq4},
\end{align}
where the $e_n$, $f_n$, $j_n$, and $k_n$ are functions of of $r$, $l$, $\omega$, and of all 38 the remaining free parameters of the theory (see Appendix \ref{appendixVTEvenEq}); $V_Z$ is the Zerilli potential given by eq.~(\ref{Vzerilli}). Note that, similarly to eq.~({\ref{STeq1}) describing the equation of motion for $\tilde{\psi}$ in scalar-tensor theories of gravity, eq.~(\ref{VTeveneq1}) does not include terms that could in general be present in the most general second order equation for $\tilde{\psi}$. For example, there are no terms proportional to $\frac{dz_1}{dr_\ast}$ in eq.~(\ref{VTeveneq1}). Thus even the most general equation of motion for $\tilde{\psi}$ in vector-tensor theories is a subset of the most general second order equation of motion for $\tilde{\psi}$ imaginable due to their integrability.

Eqs.~(\ref{VTeveneq1})-(\ref{VTeveneq4}) form a set of homogeneous coupled ordinary differential equations with non-constant coefficients. By introducing the following fields
\begin{align}
\Psi=\frac{d\psi}{dr_\ast},\;\mathcal{Z}_1=\frac{dz_1}{dr_\ast},\;\mathcal{Z}_2=\frac{dz_2}{dr_\ast},\;\mathcal{Z}_3=\frac{dz_3}{dr_\ast},\label{VTZ3}
\end{align}
we can write eqs.~(\ref{VTeveneq1})-(\ref{VTZ3}) as the first order matrix equation
\begin{align}
\frac{d}{dr_\ast}\bm{\Lambda}=-\bm{M}\bm{\Lambda},
\end{align}
where
\begin{widetext}
\begin{align}
\bm{\Lambda}=\begin{bmatrix}\label{VTmatrix}
\Psi\\
\psi\\
\mathcal{Z}_1\\
z_1\\
\mathcal{Z}_2\\
z_2\\
\mathcal{Z}_3\\
z_3\\
\end{bmatrix},\;
\bm{M}=\begin{bmatrix}
0&\omega^2-V_Z&0&e_2&0&e_3&e_1&e_4\\
-1&0&0&0&0&0&0&0\\
J_1&J_2&J_3&J_4&J_5&J_6&J_7&J_8\\
0&0&-1&0&0&0&0&0\\
K_1&K_2&K_3&K_4&K_5&K_6&K_7&K_8\\
0&0&0&0&-1&0&0&0\\
\frac{f_5}{f_1}&\frac{f_6+f_4\left(V_Z-\omega^2\right)}{f_1}&\frac{f_7}{f_1}&\frac{f_8-e_2f_4}{f_1}&\frac{f_9}{f_1}&\frac{f_{10}-e_3f_4}{f_1}&\frac{f_2-e_1f_4}{f_1}&\frac{f_3-e_4f_4}{f_1}\\
0&0&0&0&0&0&-1&0
\end{bmatrix}.
\end{align}
\end{widetext}
The $J_n$ and $K_n$ are combinations of the $e_n$, $f_n$, $j_n$, and $k_n$, given in Appendix \ref{appendixVTEvenEq}. Here we can see that there can be three dynamical vector degrees of freedom contributing to the even parity sector -- namely $z_1$, $z_2$ and $z_3$ -- which gives a total of four when counting the odd parity perturbation $z_0$ as well. As previously mentioned, we might have naively expected at most three vector degrees of freedom, corresponding to the three polarisations of a massive spin-1 particle. However, general vector-tensor theories can be unhealthy and propagate an additional ghostly mode. Indeed, in \cite{Lagos:2016wyv,Tattersall:2017eav} the same result was found for linear perturbations around a cosmological background in vector-tensor theories. This ghostly mode can be recast as a scalar field with negative kinetic energy that makes the physical system unstable. Usually, specific conditions must be imposed in vector-tensor theories (and modified gravity theories, more generally) in order to avoid such an unstable mode. In the case presented in this paper, we can fix some of the free parameters appropriately and reduce the number of vector dynamical degrees of freedom from four to three and therefore describe healthy vector-tensor theories only. For instance, we can choose the free parameters such that $f_1=f_2=f_4=0$ so that the field $z_3$ becomes an auxiliary variable that can simply be worked out from eq.~(\ref{VTeveneq2}) in terms of the other dynamical fields in order to reduce the whole even-parity system to a set of three second-order coupled ODEs (for two dynamical vector and one dynamical tensor degrees of freedom). 

Next, similarly to the previous section, we proceed to work out two specific examples of vector-tensor theories and show explicitly the equations of motion for odd and parity sectors. In both cases we consider healthy theories and, as a result, we find that $z_3$ becomes an auxiliary field, as previously discussed, so these models propagate at most three vector degrees of freedom, as expected. In addition, in both examples we find that while vector perturbations evolve in a non-trivial way, metric perturbations evolve exactly in the same as in GR. Unfortunately, we have been unable to find non-linear vector-tensor models that lead to non-trivial metric perturbations. In particular, we looked at the currently most general fully diffeomorphism-invariant vector-tensor theory, known as Generalised Proca \cite{2017JCAP...08..024H}, which seems to be lacking second-order derivative couplings between the metric and vector perturbations for our chosen black hole background. Our results on the general parametrised vector-tensor action show that modified metric perturbations are allowed though, and therefore it will be interesting to explore in the future what non-linear interactions can be constructed to obtain such modifications.

\subsection{Examples}
\subsubsection{Standard Proca field}\label{procaexample}
For the case of a Proca field with constant mass $\mu$ the action is given by \cite{2017JCAP...08..024H}:
\begin{align}
S=\int d^4x\;\sqrt{-g}\;\left[\frac{M_{Pl}^2}{2}R-\frac{1}{4}F_{\alpha\beta}F^{\alpha\beta}-\frac{1}{2}\mu^2\zeta_\alpha \zeta^\alpha\right],
\end{align}
where $F_{\alpha\beta}=\nabla_\alpha \zeta_\beta-\nabla_\beta \zeta_\alpha$ is the field strength. Perturbing the fields, about a vanishing background for the case of the vector field $\zeta^\alpha$, and expanding to quadratic order, we find the following values of the parameters given in eq.~(\ref{VTparameters})
\begin{align}
A_{\zeta1}=&-A_{\zeta2}=-A_{\zeta3}=\frac{1}{2}\mu^2,\nonumber\\
C_{\zeta2}=&C_{\zeta3}=C_{\zeta6}=C_{\zeta14}=C_{\zeta16}=-C_{\zeta5}=-C_{\zeta8}=\frac{1}{2},\nonumber\\
C_{\zeta4}=&C_{\zeta20}=-C_{\zeta10}=2C_{\zeta9}=-1,
\end{align}
with the rest of 24 parameters vanishing.
With this set of parameters, we find that for odd parity perturbations $Q$ and $z_0$ obey the following set of equations:
\begin{align}
&\frac{d^2Q}{dr_\ast^2}+\left(\omega^2-\hat{V}_{RW}\right)Q=0,\nonumber\\
&\frac{d^2z_0}{dr_\ast^2}+\left(\omega^2-\hat{V}_{RW}-\left(1-\frac{2m}{r}\right)\mu^2\right)z_0=0,\label{EOModdproca}
\end{align}
where $\hat{V}_{RW}$ is given by eq.~(\ref{Vhat}) and is evaluated with $s=1,2$ for the vector and metric perturbations respectively. 

For even parity perturbations, we find the following set of equations:
\begin{widetext}
\begin{align}
\frac{d^2\psi}{dr_\ast^2}+\left(\omega^2-V_Z\right)\psi=&\;0, \nonumber\\
\frac{d^2Z}{dr_\ast^2}+\left(\omega^2-\mu^2+\frac{2m\mu}{r}-\hat{V}_{RW}\right)Z-\frac{2\mu^2r}{i\omega}\left(1-\frac{2m}{r}\right)z_1=&\;0, \nonumber\\
\mu^2\left[\frac{d^2z_1}{dr_\ast^2}+\frac{2}{r}\left(1-\frac{2m}{r}\right)\frac{dz_1}{dr_\ast}+\left(\omega^2-\mu^2+\frac{2m\mu}{r}-\hat{V}_{RW}\right)z_1-\frac{2im\omega}{r^4}\left(1-\frac{2m}{r}\right)Z\right]=&\;0,\label{EOMevenproca}
\end{align}
\end{widetext}
where $Z$ is given by (following the convention of \cite{Zhang:2006hh}):
\begin{align}
Z=r^2\left(z_2+\frac{1}{i\omega}\frac{dz_1}{dr}\right),\label{Zdefproca}
\end{align}
and $z_3$ is related to the other fields through
\begin{align}
z_3=\frac{1}{l(l+1)}\left[\left(1-\frac{2m}{r}\right)\frac{dZ}{dr}-\frac{1}{i\omega}\left(l(l+1)+r^2\mu^2\right)z_1\right],\label{z3defproca}
\end{align}
$V_Z$ is given by eq.~(\ref{Vzerilli}) while $\hat{V}_{RW}$, given by eq.~(\ref{Vhat}), is again evaluated with $s=1$ for the vector perturbations. 

From these equations, we can make a number of remarks. We can see that, as expected, the vector perturbations propagate only three degrees of freedom in total ($z_0$, $z_1$ and $Z$), instead of four, because $z_3$ has become an auxiliary variable given by eq.~(\ref{z3defproca}). This is because the Proca action is constructed in such a way that it is healthy and does not propagate an additional ghostly mode. We also note that for a generic mass $\mu$, the metric perturbations $Q$ and $\psi$ obey the usual Regge-Wheeler and Zerilli equations respectively, as in GR. Furthermore, we find that the metric perturbations $h_0$ and $H_2$ (of odd and even parity, respectively) are related to the other perturbed fields through their usual GR relations. Thus the metric perturbations evolve exactly as in GR, as expected for a minimally coupled Proca field. In addition, the odd parity vector perturbation $z_0$ obeys a Regge-Wheeler style equation with a modified potential due to the mass of the Proca field, whilst the even parity vector perturbations $z_1$ and $Z$ are governed by a pair of coupled second order differential equations.

In the case of a pure massless Maxwell field (i.e. for $\mu=0$), we see that we are left with $z_0$ and $Z$ as the only two degrees of freedom for odd and even parity vector perturbations, respectively, which would correspond to the two polarisations of a massless spin-1 particle. Both of these fields now obey the standard Regge-Wheeler equation (with the potential $\hat{V}_{RW}$ evaluated for $s=1$) for $\mu=0$ \cite{Zhang:2006hh,Toshmatov:2015wga,Medved:2003rga}. 

\subsubsection{Sixth order coupling to Proca field}
\label{procadual}
In \cite{2017JCAP...08..024H}, it is shown that in generalised Proca theories we can achieve a Schwarzschild black hole with a sixth order coupling between the metric and the Proca field like so
\begin{align}
S=\int d^4x\;\sqrt{-g}&\;\left[\frac{M_{Pl}^2}{2}R-\frac{1}{4}F_{\alpha\beta}F^{\alpha\beta}+G_6\;L^{\mu\nu\alpha\beta}\nabla_\mu\zeta_\nu\nabla_\alpha\zeta_\beta\right],\label{Sproca6}
\end{align}
where
\begin{align}
L^{\mu\nu\alpha\beta}=-\frac{1}{4}\epsilon^{\mu\nu\rho\sigma}\epsilon^{\alpha\beta\gamma\delta}R_{\rho\sigma\gamma\delta},
\end{align}
and where $\epsilon^{\mu\nu\rho\sigma}$ is the Levi-Civita tensor, normalised such that $\epsilon^{\mu\nu\rho\sigma}\epsilon_{\mu\nu\rho\sigma}=-4!$. In the case that the background vector field vanishes, $G_6$ is a constant with dimensions $mass^2$. 

Perturbing the action given by eq.~(\ref{Sproca6}) about a Schwarzschild background for the $g_{\mu\nu}$ and a vanishing background for $\zeta^\mu$, and expanding to quadratic order, we find most of the parameters in eq.~(\ref{VTparameters}) to vanish except the following ones: 

\begin{align}
C_{\zeta2}=&C_{\zeta3}=-C_{\zeta8}=\frac{1}{2}-G_6\frac{m}{r^3},\nonumber\\
C_{\zeta4}=&2C_{\zeta9}=-1+ G_6\frac{2m}{r^3},\nonumber\\
C_{\zeta6}=&C_{\zeta14}=C_{\zeta16}=-C_{\zeta5}=\frac{1}{2}+G_6\frac{2m}{r^3},\nonumber\\
C_{\zeta10}=&1-G_6\frac{2m}{r^3},\nonumber\\
C_{\zeta20}=&-1-G_6\frac{4m}{r^3}.
\end{align}

With this parameter choice, we find the following equations of motion for odd parity perturbations:
\begin{widetext}
\begin{align}
\frac{d^2Q}{dr_\ast^2}+\left(\omega^2-\hat{V}_{RW(s=2)}\right)Q=&\;0,\nonumber\\
\frac{d^2z_0}{dr_\ast^2}+\frac{6G_6m}{r}\left(1-\frac{2m}{r}\right)\frac{1}{r^3-2G_6m}\frac{dz_0}{dr_\ast}+\left[\omega^2-\frac{1}{r^3-2G_6m}\left(r^3\hat{V}_{RW(s=1)}-2G_6m\hat{V}_{RW(s=\pm\sqrt{\frac{5}{2}})}\right)\right]z_0=&\;0.\label{EOModdproca6}
\end{align}
\end{widetext}
We see that the metric perturbation $Q$ obeys the usual Regge-Wheeler equation as in GR, with $\hat{V}_{RW}$ given by eq.~(\ref{Vhat}) evaluated for $s=2$. In addition, $h_0$ is related to the other perturbed fields as in GR (c.f.~eq.~(\ref{hodd3})). Thus, the odd parity metric perturbations evolve exactly as in GR. The odd parity vector perturbation $z_0$, however, obeys a second order equation of motion with a friction-like term proportional to $\frac{dz_0}{dr_\ast}$ and a modified potential where the contribution from the sixth order coupling term is such that a Regge-Wheeler potential with $s^2=\frac{5}{2}$ arises. We see that in the case that $G_6=0$, i.e. with no sixth order term, the usual odd parity equation for a massless Proca field (given by eq.~(\ref{EOModdproca}) for the vector mode $z_0$ is recovered.

For even parity perturbations, we find:
\begin{widetext}
\begin{align}
\frac{d^2\psi}{dr_\ast^2}+\left(\omega^2-V_Z\right)\psi=&\;0,\nonumber\\
G_6\left[\frac{d^2z_1}{dr_\ast^2}-i\omega\frac{G_6}{r^2}\left(1-\frac{2m}{r}\right)\frac{dZ}{dr_\ast}-\frac{G_6}{r}\frac{dz_1}{dr_\ast}+i\omega\frac{G_6}{r^3}\left(1-\frac{2m}{r}\right)\left(3-\frac{7m}{r}\right)Z+\frac{G_6m}{r^3}\left(3-\frac{5m}{r}\right)z_1\right]=&\;0\label{procaeveneom1}\\
 \frac{d^2Z}{dr_\ast^2} -\frac{dZ}{dr_\ast}\frac{6     G_6  m (2 m-r) \left(3  G_6  m-4 r^3\right)}{r^2 \left(r^3-2  G_6  m\right)
   \left(3  G_6  m+r^3\right)}+\frac{dz_1}{dr_\ast}\frac{2 i   G_6  m \left(2  G_6  m \left(3 m^2-m r+r^4 \omega
   ^2\right)-r^3 \left(-3 m^2+2 m r+r^4 \omega ^2\right)\right)}{r \omega  (2 m-r) \left(2  G_6  m-r^3\right) \left(3
    G_6  m+r^3\right)}&\nonumber\\
 +\frac{1}{r^4 \left(r^3-2  G_6  m\right) \left(3  G_6 
   m+r^3\right)} Z  \left(-2  G_6 ^2 m^2 \left(-\left(4 l^2+4 l+53\right) m r+r^2 \left(2 l^2+2 l+2 r^2
   \omega ^2+9\right)+69 m^2\right)\right.&\nonumber\\
  \left.+ G_6  m r^3 \left(-4 \left(2 l^2+2 l+41\right) m r+4 \left(l^2+l+9\right) r^2+183
   m^2\right)+r^7 \left(l^2 (2 m-r)+l (2 m-r)+r^3 \omega ^2\right)\right)&\nonumber\\
   -z_1\frac{2 i  G_6  m^2  \left(2  G_6  m \left(3 m^2-m r+r^4 \omega ^2\right)-r^3 \left(-3 m^2+2
   m r+r^4 \omega ^2\right)\right)}{r^3 \omega  (2 m-r) \left(2  G_6  m-r^3\right) \left(3  G_6  m+r^3\right)}=&\;0\label{procaeveneom2}
\end{align}
\end{widetext}
The metric perturbation $\psi$ obeys the usual Zerilli equation as in GR, with $V_Z$ given by eq.~(\ref{Vzerilli}), and with $H_2=H_0$ as in GR, whilst the even parity vector perturbations $Z$ given by eq.~(\ref{Zdefproca}). 

We see in eq.~(\ref{procaeveneom1}) and eq.~(\ref{procaeveneom2}) that the even parity vector perturbations $Z$ and $z_1$ obey a set of coupled second order equations of motion. Through a field redefinition of the type $Z\rightarrow Z+\beta_1(r)z_1+\beta_2(r)\frac{dz_1}{dr_\ast}$, and by making appropriate choices of $\beta_1$ and $\beta_2$, eq.~(\ref{procaeveneom2}) can be made into a single second order equation for $Z$. Such a choice is not presented here due to the complexity of the expressions but the salient point is that such a field redefinition can be made. Unlike the case of a massless Maxwell field, the example of which is given above in Section \ref{procaexample}, the odd and even parity vector perturbations do not appear to obey the same equations of motion.

It is interesting to note that in this example, the even parity vector perturbations are governed by a single equation for $Z$ (as discussed above), whilst this is not the case for a standard Proca field with a non-zero mass (eq.~(\ref{EOMevenproca})). It is perhaps more enlightening to rewrite the action given by eq.~(\ref{Sproca6}) in the following way
\begin{align}
S=\int d^4x\;\sqrt{-g}&\;\left[\frac{M_{Pl}^2}{2}R-\frac{1}{4}F_{\alpha\beta}F^{\alpha\beta} -\frac{G_6}{4}\;R^{\mu\nu\alpha\beta}\star F_{\mu\nu}\star F_{\alpha\beta}\right],\label{Sproca6dual}
\end{align}
where $\star F_{\mu\nu}$ is the \textit{dual} field strength tensor given by
\begin{align}
\star F^{\mu\nu}=\frac{1}{2}\epsilon^{\mu\nu\alpha\beta}F_{\alpha\beta}.
\end{align}
Here we can see that the action given by eq.~(\ref{Sproca6dual}) represents a $U(1)$ symmetry respecting massless vector field \cite{2017JCAP...08..024H}. Thus it is unsurprising that we find just two vector degrees of freedom, $z_0$ and $Z$, in eqs.~(\ref{EOModdproca6})-(\ref{procaeveneom2}) (after making a suitable field redefinition of $Z$ as mentioned above), similarly to the case of a massless Maxwell field. The equations for metric perturbations are unmodified with respect to GR because the third term in the action (\ref{Sproca6dual}) does not contribute with linear or quadratic metric perturbations in the specific background we have considered here. 

\subsection{Field redefinitions}

As in Section \ref{fieldsST}, we find that it is in general possible to write eq.~(\ref{VTeveneq1}) in the form of the standard Zerilli equation by making a field redefinition. If, instead of using the substitutions given by eq.~(\ref{zerillidef}) to combine the metric perturbations into the standard GR Zerilli function $\psi$, we use the field $\tilde{\psi}$ given by:
\begin{widetext}
\begin{align}
K=&g_1(r)\tilde{\psi}+\left(1-\frac{2m}{r}\right)\frac{\partial \tilde{\psi}}{\partial r}-\frac{r^2 \left(\left(l^2+l-2\right) r+6 m\right)\left(e_1\frac{dz_3}{dr\ast}+e_2z_1+e_3z_2+e_4z_3\right)}{2 \left(2 \left(l^2+l-4\right) m r-r^2
 \left(l^2+l-r^2 \omega ^2-2\right)+9 m^2\right)},\nonumber\\
H_1=&-i\omega \left(g_2(r)\tilde{\psi}+r\frac{\partial \tilde{\psi}}{\partial r}\right),\nonumber\\
H_0=&\frac{\partial}{\partial r}\left[\left(1-\frac{2m}{r}\right)\left(g_2(r)\tilde{\psi}+r\frac{\partial \tilde{\psi}}{\partial r}\right)\right]-K,\label{zerillidef}
\end{align}
\end{widetext}
then $\tilde{\psi}$ will obey the Zerilli equation as in GR. Note, however, that the function $\tilde{\psi}$ will now in general be a mixture of even parity metric and vector perturbations, and thus the metric perturbations will evolve differently than in GR. Furthermore, the vector field perturbations may be excited by further families of quasi-normal modes, different to the GR spectrum calculated from the Zerilli equation, by solving eq.~(\ref{VTeveneq2})-(\ref{VTeveneq4}) with $\tilde{\psi}=0$.

In the case of odd parity perturbations, we find that it is always possible to partially decouple $Q$ and $z_0$ eq.~(\ref{VToddeq1}) through a field redefinition, and hence obtain a single equation for a new field $\tilde{Q}$ with an additional equation that mixes $\tilde{Q}$ and $z_0$. However, the potential sourcing the equation for $\tilde{Q}$ will be different to the Regge-Wheeler potential, and hence the equation will be different to that of GR. 


\section{Conclusion}\label{conclusion}
In this paper we have analysed the structure of linear perturbations around black holes in modified gravity theories. In particular, we applied the covariant approach developed in \cite{Tattersall:2017eav} to construct the most general diffeomorphism-invariant quadratic actions for linear perturbations around a Schwarzschild black hole for three families of gravity theories: single-tensor, scalar-tensor, and vector-tensor theories. These actions contain a number of free parameters -- functions of the background -- that describe all the possible modifications to GR that are compatible with the given field content and symmetries. Therefore, these actions allow us to study, in a unified manner, a number of scalar-tensor models such as Covariant Galileons and Brans-Dicke, as well as vector-tensor models such as Maxwell and Proca. A particularly interesting and novel vector-tensor theory was discussed in Subsection \ref{procadual}, which involves the coupling of the dual Maxwell tensor to the Riemann tensor, preserving $U(1)$ gauge invariance. Our focus has been on perturbations of Schwarzschild spacetimes but the method used here is general and systematic and can thus be straightforwardly applied to other spherically symmetric backgrounds with non-trivial solutions for the additional gravity field. Such an extension would allow us to study the dynamics of linear perturbations in modified gravity with hairy solutions such as Einstein-Aether \cite{Eling:2006ec}. Furthermore, the method presented here is readily generalisable to non-spherically symmetric backgrounds, for example rotating black holes. For slowly rotating black holes, various no-hair theorems for scalar and vector fields (with non-minimal coupling or otherwise) are presented in \cite{Sotiriou:2013qea,PhysRevD.5.2403}, however perturbations to hairy rotating black holes could also be analysed in the manner presented in this paper. Such an analysis could lead to a generalisation of the Teukolsky equation \cite{Teukolsky:2014vca} for perturbations about rotating black holes in modified theories of gravity.

For each of the three families of modified gravity theories, we have found the equations of motion governing odd and even parity perturbations, in terms of the free parameters. In general, we found that even though at the level of the background all models considered have no hair (a Schwarzschild metric) and behave as GR, at the level of perturbations additional degrees of freedom are indeed excited and thus there is a dynamical hair that gives a modified evolution for linear perturbations \cite{Barausse:2008xv}. Nevertheless, we also find specific examples in which the additional degrees of freedom are not excited and thus perturbations evolve as in GR. In particular, we find that general single-tensor models behave exactly as GR at the level of linear perturbations. For scalar-tensor theories, we find the most general action to have 9 free parameters (functions of radius). All of these parameters affect the evolution of even perturbations, while odd perturbations evolve as in GR. For vector-tensor theories, the most general action depends on 38 free parameters (all functions of radius) and generically they will modify the evolution of odd and even perturbations. More specifically, we find that 10 free parameters modify the evolution of odd perturbations, whilst all 38 affect even perturbations. 

As a comparison, we mention that in the corresponding calculations of diffeomorphism-invariant quadratic actions about a cosmological Friedmann-Robertson-Walker (FRW) background presented in \cite{Tattersall:2017eav,Lagos:2016wyv}, fewer free parameters were found. For instance, there are four free parameters for scalar-tensor theories about an FRW background compared to 9 free parameters about a Schwarzschild background. As discussed in \cite{Tattersall:2017eav} the global symmetries of the background play a crucial role in determining the number of free parameters. In general, the less symmetric the background, the more free parameters are needed to describe general linear perturbations. Therefore, the larger number of free parameters found in this paper is not surprising. Furthermore, in the case of the pure Schwarzschild background studied here, the scalar self interactions are unconstrained because the scalar field perturbation is gauge invariant, contrary to the FRW case. Similarly, a large number of free parameters in the vector-tensor action are left unconstrained due to the vector field perturbation being gauge invariant.

The equations of motion derived in this paper are the most general ones for each family theory, and they provide a valuable tool for exploring modified theories of gravity with gravitational waves, and also for exotic test fields. This provides a new tool to the usual approach to quasi-normal mode analysis of black holes. Given an equation of motion, one can calculate the quasi-normal modes of the system, for example through the methods of \cite{Molina:2010fb,Pani:2012zz}. With future improved observations of quasi-normal modes from binary black hole events one could constrain the free parameters presented in this paper by constraining the effect these terms would have on the waveform. Whereas in practice it may not be possible to constrain 9 or 38 arbitrary functions of radius, these free parameters can be reduced by adding theoretical stability constraints, or they can be chosen to, for example, correspond to a particular non-linear theory, or they can be fitted with some specific functional forms.

An interesting feature that arose in the specific examples we considered here is that it was possible, in all cases, to write the evolution equations as GR-like Zerilli, or Regge-Wheeler equations in addition to a sourced evolution equation for the extra degrees of freedom. While one might expect that for minimally coupled theories, we showed that this was also true in the case of non-minimal coupling: for JBD gravity, we showed that a combination of the Zerilli function with the extra degree of freedom {\it also} satisfied the standard Zerilli equation of GR. In fact, we have shown that it is {\it always} possible to find such a combination of the even parity metric perturbations (i.e.~the Zerilli function) and the extra degrees of freedom such that this new combination satisfies the standard Zerilli equation of GR. A by-product of the fact that we are able to reduce the even parity evolution equations to a GR-like Zerilli equation is that we can already claim that a subset of the quasi-normal modes, in the cases considered here, will be exactly as in GR. There will be additional modes arising from the, sourced, extra degree of freedom. The perturbations will then, in general, be represented by linear combinations of the different families of quasi-normal modes. An important exercise, for future work, will be to determine how the lowest order modes -- i.e.~the modes which have highest signal to noise in current and future observations of ringdown -- will be affected by these extra modes, beyond those of GR.

An interesting recent development is the detection of the binary neutron star merger with gravitational wave signal GW170817 \cite{PhysRevLett.119.161101} and an electromagnetic counterpart GRB 170817A \cite{2041-8205-848-2-L12,2041-8205-848-2-L13,2041-8205-848-2-L14,2041-8205-848-2-L15}. The fact that the gravitational and electromagnetic waves are effectively coincident was subsequently used to place tight constraints on the difference in their velocities and, as a result, to place strong constraints on the range of possible extensions to General Relativity. In particular it was found that, in some sense, the simplest forms of non-minimal coupling were allowed in scalar-tensor and vector-tensor theories \cite{Baker:2017hug,Creminelli:2017sry,Sakstein:2017xjx,Ezquiaga:2017ekz}, severely limiting the allowed range of cosmological models. Given how restrictive the constraints are, it would make sense to focus on how it restricts the allowed families of black hole solutions to the classes of theories being considered in this paper. For a start, and more generally, it would be interesting to identify how many theories still allow for hairy black holes. But more specifically, it would be useful to check if the constraints on the speed of gravitational waves greatly restrict the number (or form) of the free parameters that appear in our actions for a perturbed Schwarzschild spacetime. 

Finally, and to emphasize our main motivation for pursuing this research, with the advent of black hole spectroscopy, it makes sense to explore methods which can be used to not only test the consistency of data with GR but also explore alternatives. In particular, and as in cosmology, it should be possible to use linear perturbations around the final state to constrain extensions to GR in a systematic way. In this paper we have proposed such an approach. The next step is to extend this approach beyond spherical symmetry and explore the general structure of the quasi-normal modes that arise in solutions to these equations. Only then will we be able to reap the benefits of analysing the ringdown from the data from aLIGO, its sister experiments, and their successors.


\textit{Acknowledgements ---} We thank A.~Buonnano, R.~Brito, and all of the members of the Astrophysical and Cosmological Relativity Division of the AEI Golm for useful conversations and advice. We are also grateful to V.~Cardoso, L.~Heisenberg, T.~Sotiriou, A.~Starinets, L.~Hui, and R.~Penco for further useful discussions. The $x$\textit{Tras} package for Mathematica \cite{Nutma:2013zea} was used in the computation of some of the results presented here. OJT was supported by the Science and Technology Facilities Council (STFC) Project Reference 1804725. OJT also thanks the AEI Golm for hosting him whilst part of this work was completed, funded by an STFC LTA grant. PGF acknowledges support from Leverhulme, STFC, BIPAC and the ERC. ML was supported by the Kavli Institute for Cosmological Physics through an endowment from the Kavli Foundation and its founder Fred Kavli.

\appendix

\section{Covariant quantities for the Schwarzschild background}\label{appendixS1}

For the Schwarzschild background, the background spacetime is not flat. Thus we need expressions for the Christoffel symbols and curvature tensors of the background \textit{in terms of the background quantities} to properly evaluate the Noether constraints arising from the variation of (\ref{SgenGR}). The relevant expressions can be shown to be:
\begin{align}
\bar{\nabla}_\mu u_\nu=&\;-\frac{\left(1-f\right)^2}{4m\sqrt{f}}u_\mu r_\nu, \\
\bar{\nabla}_\mu r_\nu=&\;-\frac{\left(1-f\right)^2}{4m\sqrt{f}}u_\mu u_\nu+\frac{\left(1-f\right)\sqrt{f}}{2m}\gamma_{\mu\nu}, \\
\bar{\nabla}_\mu\gamma_{\alpha\beta}=&\;u_\alpha\bar{\nabla}_\mu u_\beta + u_\beta\bar{\nabla}_\mu u_\alpha - r_\alpha\bar{\nabla}r_\beta - r_\beta\bar{\nabla}r_\alpha, \\
\bar{R}^\rho_{\,\sigma\mu\nu}=&\frac{\left(1-f\right)^3}{8m^2}\left(-2\left(u^\rho x_\sigma u_\mu x_\nu-x^\rho u_\sigma u_\mu x_\nu - u^\rho x_\sigma x_\mu u _\nu\right.\right. \nonumber\\
&\left.\left. + x^\rho u_\sigma x_\mu u _\nu\right)+\left(u^\rho u_\mu \gamma_{\sigma\nu} - \gamma^\rho_\nu u_\sigma u_\mu - u^\rho u_\nu \gamma_{\sigma\mu} \right.\right.  \nonumber\\
&\left.\left.+ \gamma^\rho_\mu u_\sigma u_\nu\right)-\left(r^\rho r_\mu \gamma_{\sigma\nu} - \gamma^\rho_\nu r_\sigma r_\mu - r^\rho r_\nu \gamma_{\sigma\mu} \right.\right.\nonumber\\
&\left.\left.+ \gamma^\rho_\mu r_\sigma r_\nu\right)+2\left(\gamma^\rho_\mu\gamma_{\sigma\nu}-\gamma^\rho_\nu\gamma_{\sigma\mu}\right)\right)\\
\bar{R}_{\mu\nu}=&\;0,\\
\bar{R}=&\;0,
\end{align}
where $f=1-\frac{2m}{r}$, $\bar{R}^\rho_{\,\sigma\mu\nu}$ is the background Riemann curvature tensor, $\bar{R}_{\mu\nu}=\bar{R}^\rho_{\mu\rho\nu}$ is the background Ricci tensor, and $\bar{R}=\bar{g}^{\mu\nu}\bar{R}_{\mu\nu}$ is the background Ricci scalar.

\section{Single-Tensor theories}\label{appendixT}
The Noether constraints for the coefficients $A_i$, $B_i$ and $C_i$ are the following:
\begin{widetext}
	\begin{align}
	-C_2=&-\frac{1}{2}C_3=\frac{1}{2}C_4=C_1, \; 2C_5=-2C_6=C_7=-C_8=-C_{11}=C_{12}=-\frac{1}{2}C_{13}=\frac{1}{2}C_{14}=-2C_{18}=2C_{19}=-C_{20}=C_{41},\nonumber\\
	C_{21}=&C_{24}=-C_{25}=\frac{1}{2}C_{26}=-\frac{1}{2}C_{27}=C_{42}=-\frac{1}{2}C_{43}=C_{44}=-C_{45}=-\frac{1}{2}C_{46}=-\frac{1}{2}C_{47}=\frac{1}{2}C_{48}=\frac{1}{2}C_{49}=-C_{50}=C_{41},\nonumber\\ 
	-C_{51}=& \frac{1}{2}C_{52}=C_{41}, \; B_{19}=-\frac{1}{2}B_{23}=2B_{12}=-B_{13}=B_{16}=-\frac{1}{2}B_{17}=-\frac{C_{41}}{m}\sqrt{f}(f-1), \; B_4=-\frac{1}{2}B_2-\frac{-2C_1}{m}\sqrt{f}(f-1),\nonumber\\
	 B_6=&B_{10}=\frac{C_{41}}{4m\sqrt{f}}(f-1)(3f-1),\; A_1=-A_5=-\frac{C_1}{4m^2}(f-1)^2(2f-1),\; A_2=A_3=\frac{C_{41}}{4m^2}(f-1)^3,\; A_6=-\frac{C_{41}}{4m^2}(f-1)^2f,\nonumber\\
	A_7=&\frac{(f-1)^2}{4m^2}(-4C_1+C_{41}(3f-2)), \; A_{12}=-\frac{C_{41}}{4m^2}(f-1)^2,\; 	A_{14}=-\frac{C_{41}}{4m^2}(f-1)^2(2f-1),\label{GRnoether}
	\end{align}
\end{widetext}
with all other remaining coefficients vanishing. In addition, we find that $C_{41}$ must be a constant.

\section{Scalar-Tensor theories}\label{appendixST}
\subsection{Noether Constraints}\label{appendixSTNC}
The Noether constraints for the $A_{\chi\,n}$, $B_{\chi\,n}$, $C_{\chi\,n}$, and $D_{\chi\,n}$ are given by:
\begin{widetext}
\begin{align}
 A_{\chi1}=&-\frac{1}{4m^2}\left(2D_{\chi5}(f-1)^2-2D_{\chi8}(f-1)^2+m\left(4f\left(\frac{dD_{\chi5}}{dr}+\frac{d^2D_{\chi15}}{dr^2}m+\frac{dD_{\chi8}}{dr}(f-1)-\frac{dD_{\chi5}}{dr}f\right)\right.\right.\nonumber\\
 &\left.\left.+\frac{dD_{\chi15}}{dr}\left(1+2f-3f^2\right)\right)\right), \nonumber\\
 A_{\chi2}=&-\frac{1}{4m}\left(\frac{dD_{\chi15}}{dr}(f-1)^2+4\left(\frac{dD_{\chi8}}{dr}(1-f)+\frac{d^2D_{\chi8}}{dr^2}mf\right)+\frac{dD_{\chi5}}{dr}\left(f^2-1\right)\right), \nonumber\\
 A_{\chi3}=&\frac{1}{4m^2}\left(f-1\right)\left(2D_{\chi5}(f-1)-2D_{\chi8}(f-1)+m\left(\frac{dD_{\chi15}}{dr}(f-1)+4\frac{dD_{\chi8}}{dr}f\right)\right), \nonumber\\
 B_{\chi2}=&\frac{1}{4m\sqrt{f}}\left(D_{\chi11}\left(-1-2f+3f^2\right)-4\frac{dD_{\chi11}}{dr}mf\right), \; B_{\chi3}=\frac{1}{2m\sqrt{f}}\left(D_{\chi11}(1-f)+2\frac{dD_{\chi11}}{dr}mf\right), \nonumber\\
 B_{\chi4}=&\frac{1}{4m\sqrt{f}}\left(D_{\chi15}(f-1)^2+4D_{\chi8}f(f-1)\right), \nonumber\\
B_{\chi5}=&\frac{1}{4m\sqrt{f}}\left(-D_{\chi15}-2D_{\chi8}(f-1)^2-4\frac{dD_{\chi8}}{dr}mf+D_{\chi15}f(2-f)-D_{\chi5}\left(f^2-1\right)\right), \; B_{\chi6}=\frac{\sqrt{f}}{2m}D_{\chi5}(f-1), \nonumber\\
 B_{\chi7}=&\frac{1}{4m\sqrt{f}}\left(D_{\chi15}(f-1)^2-4f\left(\frac{dD_{\chi15}}{dr}m-D_{\chi5}(f-1)+D_{\chi8}(f-1)\right)\right), \; B_{\chi9}=-\frac{2\sqrt{f}(f-1)}{m}\left(D_{\chi5}-D_{\chi8}-D_{\chi15}\right), \nonumber\\
 B_{\chi10}=&-\frac{\sqrt{f}(f-1)}{m}D_{\chi11}, \; D_{\chi2}=D_{\chi3}=-\frac{1}{2}D_{\chi4}=-D_{\chi5}+D_{\chi8}+D_{\chi15}, \; D_{\chi6}=-D_{\chi5}, \nonumber\\
 D_{\chi9}=&-\frac{1}{2}D_{\chi10}=D_{\chi8}, \; D_{\chi12}=D_{\chi13}=-D_{\chi14}=-D_{\chi11}, \; -\frac{1}{2}D_{\chi16}=D_{\chi17}=D_{\chi15},
\end{align}
\end{widetext}
with all other remaining coefficients vanishing. We also mention that $C_1$ is left unconstrained but does not appear in the final action. 

\subsection{Even Parity Perturbations: Relation between $H_0$, $H_2$, and $\varphi$}\label{appendixSTRel}
In Section \ref{STeven} the following relation is found between $H_0$, $H_2$, and $\varphi$:
\begin{widetext}
\begin{align}
H_2=&\;H_0+\varphi\frac{2}{l \left(l^3+2 l^2-l-2\right) r^4 C_{41} (r-2
 m)} \left(-l \left(l^3+2 l^2-l-2\right) r^3 D_{\chi5} (2
 m-r)\right.\nonumber\\
 &\left.-4 \left(-i l (l+1) m r^4 \omega D_{\chi11} +(2 m-r) \left(2
 \left(\frac{dD_{\chi8}}{dr}\right) m \left(-2 \left(l^2+l+8\right) m^2 r+3 l (l+1) m r^2+24 m^3+r^5 \omega
 ^2\right)\right.\right.\right.\nonumber\\
 &\left.\left.\left.+r \left(r \left(-i \left(\frac{dD_{\chi11}}{dr}\right) l^2 m r^2 \omega +i \left(\frac{dD_{\chi11}}{dr}\right) l^2 r^3
 \omega -i \left(\frac{dD_{\chi11}}{dr}\right) l m r^2 \omega +i \left(\frac{dD_{\chi11}}{dr}\right) l r^3 \omega +2 i
 \left(\frac{dD_{\chi11}}{dr}\right) m r^2 \omega\right.\right.\right.\right.\right.\nonumber\\
 &\left.\left.\left.\left.\left. +\left(\frac{dD_{\chi5}}{dr}\right) m \left(\left(l^2+l+4\right) m-r
 \left(l^2+l+2 r^2 \omega ^2\right)\right)-4 i \left(\frac{d^2D_{\chi11}}{dr^2}\right) m^2 r^2 \omega +2 i
 \left(\frac{d^2D_{\chi11}}{dr^2}\right) m r^3 \omega -4 \left(\frac{d^2D_{\chi5}}{dr^2}\right) m^2 r\right.\right.\right.\right.\right.\nonumber\\
 &\left.\left.\left.\left.\left.+2 \left(\frac{d^2D_{\chi5}}{dr^2}\right) m
 r^2+\left(\frac{d^3D_{\chi15}}{dr^3}\right) m^2 (r-2 m)^2+16 \left(\frac{d^3D_{\chi8}}{dr^3}\right) m^4-24 \left(\frac{d^3D_{\chi8}}{dr^3}\right) m^3 r+12
 \left(\frac{d^3D_{\chi8}}{dr^3}\right) m^2 r^2\right.\right.\right.\right.\right.\nonumber\\
 &\left.\left.\left.\left.\left.-2 \left(\frac{d^3D_{\chi8}}{dr^3}\right) m r^3\right)-2 \left(\frac{d^2D_{\chi15}}{dr^2}\right) m \left(6
 m^3-3 m^2 r-m r^4 \omega ^2+r^5 \omega ^2\right)\right.\right.\right.\right.\nonumber\\
 &\left.\left.\left.\left.-2 \left(\frac{d^2D_{\chi8}}{dr^2}\right) m (2 m-r)
 \left(-\left(l^2+l+4\right) m r+\left(l^2+l-2\right) r^2+12
 m^2\right)\right)\right)\right.\right.\nonumber\\
 &\left.\left.+\left(\frac{dD_{\chi15}}{dr}\right) m^2 \left(2 m r^2 \left(l^2+l-2 r^2 \omega
 ^2+2\right)+r^3 \left(-l^2-l+3 r^2 \omega ^2\right)+24 m^3-20 m^2
 r\right)\right)\right)\nonumber\\
 &+\frac{d\varphi}{dr}\frac{8}{l \left(l^3+2 l^2-l-2\right) r^3
 C_{41}} \left(i l (l+1) r^4 \omega D_{\chi11} +m
 \left(-\left(\frac{dD_{\chi15}}{dr}\right) \left(4 m^3-2 m^2 r+r^5 \omega ^2\right)\right.\right.\nonumber\\
 &\left.\left.-(2 m-r)
 \left(\left(\frac{dD_{\chi8}}{dr}\right) \left(\left(l^2+l-2\right) r^2+8 m^2-4 m r\right)+r \left(2 r
 \left(\left(\frac{dD_{\chi5}}{dr}\right)+i \left(\frac{dD_{\chi11}}{dr}\right) r \omega \right)\right.\right.\right.\right.\nonumber\\
 &\left.\left.\left.\left.+\left(\frac{d^2D_{\chi15}}{dr^2}\right) m (r-2 m)-2 \left(\frac{d^2D_{\chi8}}{dr^2}\right)
 (r-2 m)^2\right)\right)\right)\right).
\end{align}
\end{widetext}

\subsection{Even Parity Perturbations: Equations of motion coefficients}\label{appendixSTEven}
The explicit forms of the $a_n$ and $b_n$ found in Section \ref{STeven} is given in the Mathematica file `ScalarTensorEvenCoeff' in the public github repository \url{https://github.com/ojtattersall/black-hole-notebooks.git}. They are not reproduced fully here due to the excessive length of some of the expressions.

\section{Vector-Tensor theories}\label{appendixVT}

\subsection{Noether Constraints}\label{appendixVTNC}
The Noether constraints for the $A_{\zeta\,n}$, $B_{\zeta\,n}$, $C_{\zeta\,n}$, and $D_{\zeta\,n}$ are given by
\begin{widetext}
\begin{align}
A_{\zeta5}=&\;-\frac{1}{4 m^2}\left(m \left(4 \left(\frac{d^2D_{\zeta2}}{dr^2}\right) f m-4 \left(\frac{d^2D_{\zeta38}}{dr^2}\right) f m+4 \left(\frac{d^2D_{\zeta7}}{dr^2}\right) f m-3 \left(\frac{dD_{\zeta2}}{dr}\right)
 f^2+2 \left(\frac{dD_{\zeta2}}{dr}\right) f+\left(\frac{dD_{\zeta2}}{dr}\right)\right.\right.\nonumber\\
 &\left.\left.+7 \left(\frac{dD_{\zeta38}}{dr}\right) f^2-6 \left(\frac{dD_{\zeta38}}{dr}\right) f-\left(\frac{dD_{\zeta38}}{dr}\right)-7
 \left(\frac{dD_{\zeta7}}{dr}\right) f^2+6 \left(\frac{dD_{\zeta7}}{dr}\right) f+\left(\frac{dD_{\zeta7}}{dr}\right)\right)+2 (f-1)^2 D_{\zeta7}\right.\nonumber\\
 &\left.-2(f-1)^2 D_{\zeta38}\right),\nonumber\\
A_{\zeta6}=&\;\frac{1}{16 f m^2}\left(-4 f m \left(4 \left(\frac{d^2D_{\zeta38}}{dr^2}\right) f m+\left(\frac{dD_{\zeta2}}{dr}\right) (f-1)^2-\left(\frac{dD_{\zeta38}}{dr}\right) \left(f^2+2
 f-3\right)-\left(\frac{dD_{\zeta60}}{dr}\right) f^2+2 \left(\frac{dD_{\zeta60}}{dr}\right) f\right.\right.\nonumber\\
 &\left.\left.-\left(\frac{dD_{\zeta60}}{dr}\right)+2 \left(\frac{dD_{\zeta7}}{dr}\right) f^2-2
 \left(\frac{dD_{\zeta7}}{dr}\right) f\right)+(f-1)^4 (-D_{\zeta2})+(3 f-1) (f-1)^3
 D_{\zeta60}\right),\nonumber\\ 
 A_{\zeta7}=&\;-\frac{1}{16 f m^2}\left(-4 f \left(2 m \left(f (2 \left(\frac{d^2D_{\zeta11}}{dr^2}\right) m+\left(\frac{dD_{\zeta60}}{dr}\right) (f-1))-\left(\frac{dD_{\zeta11}}{dr}\right)
 f^2+\left(\frac{dD_{\zeta11}}{dr}\right)\right)+(f-1)^3 D_{\zeta60}\right)\right.\nonumber\\
 &\left.+4 f (f-1)^3
 D_{\zeta2}+(9 f-1) (f-1)^3 D_{\zeta11}\right), \nonumber\\
 A_{\zeta8}=&\;\frac{(f-1)}{16
 f m^2}\left( \left(12 \left(\frac{dD_{\zeta15}}{dr}\right) f^2 m+4 \left(\frac{dD_{\zeta15}}{dr}\right) f m+4 \left(\frac{dD_{\zeta19}}{dr}\right) f^2 m-4
 \left(\frac{dD_{\zeta19}}{dr}\right) f m+4 \left(\frac{dD_{\zeta26}}{dr}\right) f^2 m\right.\right.\nonumber\\
 &\left.\left.-4 \left(\frac{dD_{\zeta26}}{dr}\right) f m+4 \left(\frac{dD_{\zeta27}}{dr}\right) f^2 m-4
 \left(\frac{dD_{\zeta27}}{dr}\right) f m-4 f^3 D_{\zeta21}-f^3 D_{\zeta26}-f^3
 D_{\zeta27}+3 f^2 D_{\zeta26}+3 f^2
 D_{\zeta27}-\right.\right.\nonumber\\
 &\left.\left.\left(f^3-11 f^2+11 f-1\right) D_{\zeta19}+4 f
 D_{\zeta21}-3 f D_{\zeta26}-3 f
 D_{\zeta27}+(f-1)^2 (9 f-1)
 D_{\zeta15}+D_{\zeta26}+D_{\zeta27}\right)\right), \nonumber\\
A_{\zeta9}=&\;-\frac{1}{16 f m^2}\left(16 \left(\frac{d^2D_{\zeta15}}{dr^2}\right) f^2 m^2-4 \left(\frac{dD_{\zeta15}}{dr}\right) f^3 m-8 \left(\frac{dD_{\zeta15}}{dr}\right) f^2 m+12 \left(\frac{dD_{\zeta15}}{dr}\right)
 f m+8 \left(\frac{dD_{\zeta19}}{dr}\right) f^3 m\right.\nonumber\\
 &\left.-8 \left(\frac{dD_{\zeta19}}{dr}\right) f^2 m-8 \left(\frac{dD_{\zeta21}}{dr}\right) f^3 m+8 \left(\frac{dD_{\zeta21}}{dr}\right) f^2
 m+4 \left(\frac{dD_{\zeta26}}{dr}\right) f^3 m-8 \left(\frac{dD_{\zeta26}}{dr}\right) f^2 m+4 \left(\frac{dD_{\zeta26}}{dr}\right) f m\right.\nonumber\\
 &\left.+4 \left(\frac{dD_{\zeta27}}{dr}\right) f^3 m-8
 \left(\frac{dD_{\zeta27}}{dr}\right) f^2 m+4 \left(\frac{dD_{\zeta27}}{dr}\right) f m+f^4 D_{\zeta26}+f^4
 D_{\zeta27}-4 f^3 D_{\zeta26}-4 f^3
 D_{\zeta27}-4 (f-1)^2 f^2 D_{\zeta19}\right.\nonumber\\
 &\left.+6 f^2
 D_{\zeta26}+6 f^2 D_{\zeta27}+2 (f-1)^2 (f+1) f
 D_{\zeta21}-4 f D_{\zeta26}-4 f
 D_{\zeta27}+D_{\zeta26}+D_{\zeta27}\right), \nonumber\\
 A_{\zeta10}=&\;-\frac{(f-1)}{4
 m^2} \left(4 \left(\frac{dD_{\zeta15}}{dr}\right) f m+f^2 D_{\zeta26}+f^2
 D_{\zeta27}+(f-1)^2 D_{\zeta15}+2 f (f-1)
 D_{\zeta19}-2 f D_{\zeta21}-2 f D_{\zeta26}-2 f
 D_{\zeta27}\right.\nonumber\\
 &\left.+2
 D_{\zeta21}+D_{\zeta26}+D_{\zeta27}\right), \nonumber\\
 A_{\zeta11}=&\;-\frac{(f-1)}{16 f m^2} \left(-4 \left(\frac{dD_{\zeta2}}{dr}\right) f^2 m+4 \left(\frac{dD_{\zeta2}}{dr}\right) f m-12 \left(\frac{dD_{\zeta38}}{dr}\right) f^2 m-4
 \left(\frac{dD_{\zeta38}}{dr}\right) f m-4 \left(\frac{dD_{\zeta7}}{dr}\right) f^2 m\right.\nonumber\\
 &\left.+4 \left(\frac{dD_{\zeta7}}{dr}\right) f m-4 f^3
 D_{\zeta11}-f^3 D_{\zeta38}+4 f^3 D_{\zeta60}+8
 f^2 D_{\zeta11}+11 f^2 D_{\zeta38}-8 f^2
 D_{\zeta60}+\left(f^3-11 f^2+11 f-1\right)
 D_{\zeta7}\right.\nonumber\\
 &\left.+(f-1)^3 D_{\zeta2}-4 f
 D_{\zeta11}-11 f D_{\zeta38}+4 f
 D_{\zeta60}+D_{\zeta38}\right), \nonumber\\
 A_{\zeta12}=&\;\frac{1}{4 m^2}\left(4 \left(\frac{d^2D_{\zeta15}}{dr^2}\right) f m^2-4 \left(\frac{d^2D_{\zeta19}}{dr^2}\right) f m^2-4 \left(\frac{d^2D_{\zeta26}}{dr^2}\right) f m^2-4 \left(\frac{d^2D_{\zeta27}}{dr^2}\right) f
 m^2-7 \left(\frac{dD_{\zeta15}}{dr}\right) f^2 m\right.\nonumber\\
 &\left.+6 \left(\frac{dD_{\zeta15}}{dr}\right) f m+\left(\frac{dD_{\zeta15}}{dr}\right) m+7 \left(\frac{dD_{\zeta19}}{dr}\right) f^2 m-6
 \left(\frac{dD_{\zeta19}}{dr}\right) f m-\left(\frac{dD_{\zeta19}}{dr}\right) m-4 \left(\frac{dD_{\zeta21}}{dr}\right) f^2 m\right.\nonumber\\
 &\left.+4 \left(\frac{dD_{\zeta21}}{dr}\right) f m+3 \left(\frac{dD_{\zeta26}}{dr}\right)
 f^2 m-2 \left(\frac{dD_{\zeta26}}{dr}\right) f m-\left(\frac{dD_{\zeta26}}{dr}\right) m+3 \left(\frac{dD_{\zeta27}}{dr}\right) f^2 m\right.\nonumber\\
 &\left.-2 \left(\frac{dD_{\zeta27}}{dr}\right) f
 m-\left(\frac{dD_{\zeta27}}{dr}\right) m+3 f^3 D_{\zeta21}+2 f^3 D_{\zeta26}+2 f^3
 D_{\zeta27}-5 f^2 D_{\zeta21}-4 f^2
 D_{\zeta26}-4 f^2 D_{\zeta27}+f D_{\zeta21}\right.\nonumber\\
 &\left.+2 f
 D_{\zeta26}+2 f D_{\zeta27}+2 (f-1)^2
 D_{\zeta15}-2 (f-1)^2
 D_{\zeta19}+D_{\zeta21}\right),\nonumber\\
 A_{\zeta13}=&\;-\frac{(f-1)^2}{8 f
 m^2} \left(4 \left(\frac{dD_{\zeta15}}{dr}\right) f m-4 \left(\frac{dD_{\zeta19}}{dr}\right) f m-4 \left(\frac{dD_{\zeta26}}{dr}\right) f m-4
 \left(\frac{dD_{\zeta27}}{dr}\right) f m-2 f^2 D_{\zeta21}-3 f^2 D_{\zeta26}-3 f^2
 D_{\zeta27}\right.\nonumber\\
 &\left.+(f-1)^2 (-D_{\zeta15})+(f-1)^2
 D_{\zeta19}+2 f D_{\zeta21}+2 f D_{\zeta26}+2 f
 D_{\zeta27}+D_{\zeta26}+D_{\zeta27}\right),\nonumber\\
 A_{\zeta14}=&\;-\frac{(f-1)}{8 f m^2} \left(8 \left(\frac{dD_{\zeta11}}{dr}\right) f^2 m-4 \left(\frac{dD_{\zeta2}}{dr}\right) f^2 m+4 \left(\frac{dD_{\zeta2}}{dr}\right) f m+4
 \left(\frac{dD_{\zeta38}}{dr}\right) f^2 m-4 \left(\frac{dD_{\zeta38}}{dr}\right) f m\right.\nonumber\\
 &\left.-4 \left(\frac{dD_{\zeta7}}{dr}\right) f^2 m+4 \left(\frac{dD_{\zeta7}}{dr}\right) f m-4 f^3
 D_{\zeta11}-5 f^3 D_{\zeta38}+4 f^3
 D_{\zeta60}+4 f^2 D_{\zeta11}+11 f^2
 D_{\zeta38}-4 f^2 D_{\zeta60}\right.\nonumber\\
 &\left.+(f-1)^3
 D_{\zeta2}+(5 f-1) (f-1)^2 D_{\zeta7}-7 f
 D_{\zeta38}+D_{\zeta38}\right),\nonumber\\
 B_{\zeta2}=&\;-\frac{(D_{\zeta21}+D_{\zeta26}+D_{\zeta27}) \sqrt{1-\frac{2 m}{r}}}{r}, \; B_{\zeta3}=\;-\frac{1}{2}B_{\zeta4}=\;-\frac{m (D_{\zeta26}+D_{\zeta27})}{r^2 \sqrt{1-\frac{2 m}{r}}}, \; B_{\zeta5}=\frac{D_{\zeta21} \sqrt{1-\frac{2 m}{r}}}{r}, \nonumber\\
 B_{\zeta6}=&\;\frac{\left(\frac{dD_{\zeta21}}{dr}\right) r (r-2 m)+D_{\zeta19} (2 m-r)+D_{\zeta21}
 (r-m)}{r^2 \sqrt{1-\frac{2 m}{r}}}, \; B_{\zeta7}=\;\frac{\left(\frac{dD_{\zeta21}}{dr}\right) r (2 m-r)+D_{\zeta19} (r-2 m)+D_{\zeta21}
 (m-r)}{r^2 \sqrt{1-\frac{2 m}{r}}}, \nonumber\\
 B_{\zeta8}=&\;\frac{D_{\zeta15} (2 r-5 m)+m
 (D_{\zeta19}+D_{\zeta26}+D_{\zeta27})}{r^2 \sqrt{1-\frac{2 m}{r}}}, \; B_{\zeta9}=\;\frac{(D_{\zeta21}-D_{\zeta15}) \sqrt{1-\frac{2 m}{r}}}{r}, \nonumber\\
 B_{\zeta10}=&\;\frac{2 \left(\frac{dD_{\zeta15}}{dr}\right) m r-\left(\frac{dD_{\zeta15}}{dr}\right) r^2+D_{\zeta15} (3 m-2
 r)+D_{\zeta19} (r-2 m)-m D_{\zeta26}-m
 D_{\zeta27}}{r^2 \sqrt{1-\frac{2 m}{r}}},\nonumber\\
 B_{\zeta11}=&\;\frac{2 (\left(\frac{dD_{\zeta15}}{dr}\right) r (r-2 m)+m D_{\zeta26}+m D_{\zeta27})-2 D_{\zeta15}
 (m-r)+D_{\zeta19} (2 m-r)}{r^2 \sqrt{1-\frac{2 m}{r}}}, \; B_{\zeta12}=-\frac{D_{\zeta21} \sqrt{1-\frac{2 m}{r}}}{r}, \nonumber\\
 B_{\zeta13}=&\frac{1}{r^2 \sqrt{1-\frac{2 m}{r}}}\left(-2 \left(\frac{dD_{\zeta15}}{dr}\right) m r+\left(\frac{dD_{\zeta15}}{dr}\right) r^2+2 \left(\frac{dD_{\zeta19}}{dr}\right) m r-\left(\frac{dD_{\zeta19}}{dr}\right) r^2+2 \left(\frac{dD_{\zeta26}}{dr}\right) m r-\left(\frac{dD_{\zeta26}}{dr}\right)
 r^2\right.\nonumber\\
 &\left.+2 \left(\frac{dD_{\zeta27}}{dr}\right) m r-\left(\frac{dD_{\zeta27}}{dr}\right) r^2+D_{\zeta15} (2 r-5 m)+D_{\zeta19} (5 m-2 r)-4
 m D_{\zeta21}+m D_{\zeta26}+m D_{\zeta27}+2 r
 D_{\zeta21}\right),\nonumber\\
 B_{\zeta14}=&\;-\frac{1}{2}B_{\zeta19}=\;\frac{m (D_{\zeta15}-D_{\zeta19}-D_{\zeta26}-D_{\zeta27})}{r^2
 \sqrt{1-\frac{2 m}{r}}},\nonumber\\
 B_{\zeta16}=&\;\frac{1}{r^2 \sqrt{1-\frac{2 m}{r}}}\left(2 \left(\frac{dD_{\zeta2}}{dr}\right) m r-\left(\frac{dD_{\zeta2}}{dr}\right) r^2-2 \left(\frac{dD_{\zeta38}}{dr}\right) m r+\left(\frac{dD_{\zeta38}}{dr}\right) r^2+2 \left(\frac{dD_{\zeta7}}{dr}\right) m r-\left(\frac{dD_{\zeta7}}{dr}\right)
 r^2\right.\nonumber\\
 &\left.+D_{\zeta7} (3 m-2 r)-m D_{\zeta2}-3 m D_{\zeta38}+2 r
 D_{\zeta38}\right), \nonumber\\
 B_{\zeta17}=&\;\frac{D_{\zeta38} (2 r-3 m)-m D_{\zeta2}-m D_{\zeta7}}{r^2 \sqrt{1-\frac{2
 m}{r}}}, \; B_{\zeta18}=\;\frac{-m D_{\zeta2}-m D_{\zeta7}+4 m D_{\zeta11}+mD_{\zeta38}-2 r D_{\zeta11}}{r^2 \sqrt{1-\frac{2 m}{r}}},\nonumber\\
 B_{\zeta20}=&\;\frac{2 (D_{\zeta11}-D_{\zeta60}) \sqrt{1-\frac{2 m}{r}}}{r}, \; B_{\zeta21}=\;-\frac{2 D_{\zeta21} \sqrt{1-\frac{2 m}{r}}}{r},\nonumber\\
 B_{\zeta22}=&\;\frac{2 \sqrt{1-\frac{2 m}{r}} (\left(\frac{dD_{\zeta2}}{dr}\right) r-\left(\frac{dD_{\zeta38}}{dr}\right) r+\left(\frac{dD_{\zeta7}}{dr}\right) r+2 D_{\zeta2}+2
 D_{\zeta7}-2 D_{\zeta38})}{r}, \; B_{\zeta23}=\;\frac{m (D_{\zeta60}-D_{\zeta2})}{r^2 \sqrt{1-\frac{2 m}{r}}}, \nonumber\\
 B_{\zeta25}=&\;\frac{D_{\zeta60} \sqrt{1-\frac{2 m}{r}}}{r} ,\;  B_{\zeta26}=\;\frac{2 m D_{\zeta2}-r (\left(\frac{dD_{\zeta60}}{dr}\right) (r-2 m)+D_{\zeta60})}{r^2 \sqrt{1-\frac{2 m}{r}}}, \nonumber\\
 B_{\zeta27}=&\;\frac{-4 \left(\frac{dD_{\zeta38}}{dr}\right) m r+2 \left(\frac{dD_{\zeta38}}{dr}\right) r^2+D_{\zeta7} (2 m-r)+2 m D_{\zeta2}-2 m
 D_{\zeta38}-m D_{\zeta60}+2 r D_{\zeta38}}{r^2 \sqrt{1-\frac{2 m}{r}}}, \nonumber\\
 B_{\zeta28}=&\;\left(\frac{dD_{\zeta11}}{dr}\right) \sqrt{1-\frac{2 m}{r}}, \; B_{\zeta29}=\;\frac{D_{\zeta11} (2 r-5 m)}{r^2 \sqrt{1-\frac{2 m}{r}}}, \; D_{\zeta3}=\;-\frac{1}{2}D_{\zeta4}=\;D_{\zeta2}, \;  D_{\zeta8}=-D_{\zeta7}, \; D_{\zeta9}=-D_{\zeta10}=-D_{\zeta12}=D_{\zeta11}, \nonumber\\
 D_{\zeta14}=&-\frac{1}{2}D_{\zeta16}=D_{\zeta15}, \; D_{\zeta20}=-D_{\zeta19}, \; D_{\zeta22}=D_{\zeta23}=-D_{\zeta24}=-D_{\zeta21}, \; D_{\zeta25}=D_{\zeta38}, \; D_{\zeta28}=-D_{\zeta60}, \nonumber\\
 D_{\zeta29}=&-D_{\zeta32}=D_{\zeta33}=-D_{\zeta36}=-D_{\zeta11}, \;  D_{\zeta30}=-D_{\zeta31}=-D_{\zeta34}=D_{\zeta35}=D_{\zeta21}, \;  D_{\zeta37}=D_{\zeta26}+D_{\zeta27},\nonumber\\
 D_{\zeta40}=&-D_{\zeta56}=-D_{\zeta60},\; D_{\zeta41}=-\frac{1}{2}D_{\zeta54}=-(D_{\zeta15}-D_{\zeta19}-D_{\zeta26}-D_{\zeta27}), \; D_{\zeta47}=-\frac{1}{2}D_{\zeta51}=D_{\zeta2}+D_{\zeta7}-D_{\zeta38}, \nonumber\\
 D_{\zeta55}=&-2D_{\zeta38}, \; D_{\zeta59}=-2(D_{\zeta26}+D_{\zeta27}), \; D_{\zeta61}=D_{\zeta2}+D_{\zeta7}-D_{\zeta38}, \; D_{\zeta62}=-D_{\zeta15}+D_{\zeta19}+D_{\zeta26}+D_{\zeta27},
\end{align}
\end{widetext}
with all other remaining coefficients vanishing.

\subsection{Odd Parity Perturbations: Relation between $h_0$ and other fields} \label{appendixVTOdd}
In Section \ref{VTodd} the following relation is found between $h_0$, $h_1$, and $z_0$:
\begin{widetext}
\begin{align}
-i\omega h_0=&\left(1-\frac{2m}{r}\right)\frac{d}{dr}\left[\left(1-\frac{2m}{r}\right)\left(h_1+\frac{\left(2imr\omega\frac{dD_{\zeta11}}{dr}-(l+2)(l-1)D_{\zeta21}\right)}{2C_{41}(l+2)(l-1)}z_0\right)\right]\nonumber\\
&+i\omega\frac{\left(2m\left(\frac{d^2D_{\zeta11}}{dr^2}r(r-2m)^2-\frac{dD_{\zeta11}}{dr}\left(4m^2-5mr+r^2\right)\right)+(l+2)(l-1)r^2D_{\zeta11}\right)}{2C_{41}(l+2)(l-1)r^2\sqrt{1-\frac{2m}{r}}}z_0.
\end{align}
\end{widetext}

\subsection{Odd Parity Perturbations: Equations of motion coefficients}\label{appendixVTOddEq}
The explicit forms of the $c_n$ and $d_n$ found in Section \ref{VTodd} is given in the Mathematica file `VectorTensorOddCoeff' in the public github repository \url{https://github.com/ojtattersall/black-hole-notebooks.git}. They are not reproduced fully here due to the excessive length of some of the expressions.

\subsection{Even Parity Perturbations: Relation between $H_2$ and other fields}\label{appendixVTEven}
The explicit relation between $h_0$ and $H_2$ found in Section \ref{VTeven} is given in the Mathematica file `VectorTensorH2def' in the github folder in the public github repository \url{https://github.com/ojtattersall/black-hole-notebooks.git}. It is not reproduced fully here due to the excessive length of the expression. Schematically:
\begin{align}
H_2=H_0 + L(z_1, z_2, z_3, z_1', z_2', z_3', z_3''),
\end{align} 
where $L$ represents a linear combination of the fields, and a prime represents a derivative with respect to $r$.

\subsection{Even Parity Perturbations: Equations of motion coefficients}\label{appendixVTEvenEq}
The $J_n$ and $K_n$ referred to in eq.~(\ref{VTmatrix}) are given by:
\begin{align}
J_1=&\frac{j_7k_5-j_5k_7}{j_7k_1-j_1k_7},\nonumber\\
J_2=&\frac{-k_7\left(j_6+j_4\left(V_Z-\omega^2\right)\right)+j_7\left(k_6+k_4\left(V_Z-\omega^2\right)\right)}{j_7k_1-j_1k_7},\nonumber\\
J_3=&\frac{j_7k_2-j_2k_7}{j_7k_1-j_1k_7},\; J_4=\frac{j_7k_3-e_2j_7k_4-j_3k_7+e_2j_4k_7}{j_7k_1-j_1k_7},\nonumber\\
J_5=&\frac{j_7k_8-j_8k_7}{j_7k_1-j_1k_7},\; J_6=\frac{j_7k_9-j_9k_7+e_3j_4k_7-e_3j_7k_4}{j_7k_1-j_1k_7},\nonumber\\
J_7=&\frac{j_7k_{10}-e_1j_7k_4-j_{10}k_7+e_1j_4k_7}{j_7k_1-j_1k_7},\nonumber\\
J_8=&\frac{j_7k_{11}-e_4j_7k_4-j_{11}k_7+e_4j_4k_7}{j_7k_1-j_1k_7},\; K_1=\frac{j_5k_1-j_1k_5}{j_7k_1-j_1k_7},\nonumber\\
K_2=&\frac{j_6k_1+j_4k_1\left(V_Z-\omega^2\right)-j_1\left(k_6+k_4\left(V_Z-\omega^2\right)\right)}{j_7k_1-j_1k_7},\nonumber\\
K_3=&\frac{j_2k_1-j_1k_2}{j_7k_1-j_1k_7},\; K_4=\frac{j_3k_1-e_2j_4k_1-j_1k_3+e_2j_1k_4}{j_7k_1-j_1k_7},\nonumber\\
K_5=&\frac{j_8k_1-j_1k_2}{j_7k_1-j_1k_7},\; K_6=\frac{-e_3j_4k_1+j_9k_1+e_3j_1k_4-j_1k_9}{j_7k_1-j_1k_7},\nonumber\\
K_7=&\frac{j_{10}k_1-e_1j_4k_1-j_1k_{10}+e_1j_1k_4}{j_7k_1-j_1k_7},\nonumber\\
K_8=&\frac{j_{11}k_1-e_4j_4k_1-j_1k_{11}+e_4j_1k_4}{j_7k_1-j_1k_7}.
\end{align}

The explicit forms of the $e_n$, $f_n$, $j_n$, and $k_n$ found in Section \ref{VTeven} is given in the Mathematica file `VectorTensorEvenCoeff' in the github folder in the public github repository \url{https://github.com/ojtattersall/black-hole-notebooks.git}. They are not reproduced fully here due to the excessive length of some of the expressions.


\bibliographystyle{apsrev4-1}
\bibliography{blackhole}

\begin{thebibliography}{72}%
\makeatletter
\providecommand \@ifxundefined [1]{%
 \@ifx{#1\undefined}
}%
\providecommand \@ifnum [1]{%
 \ifnum #1\expandafter \@firstoftwo
 \else \expandafter \@secondoftwo
 \fi
}%
\providecommand \@ifx [1]{%
 \ifx #1\expandafter \@firstoftwo
 \else \expandafter \@secondoftwo
 \fi
}%
\providecommand \natexlab [1]{#1}%
\providecommand \enquote  [1]{``#1''}%
\providecommand \bibnamefont  [1]{#1}%
\providecommand \bibfnamefont [1]{#1}%
\providecommand \citenamefont [1]{#1}%
\providecommand \href@noop [0]{\@secondoftwo}%
\providecommand \href [0]{\begingroup \@sanitize@url \@href}%
\providecommand \@href[1]{\@@startlink{#1}\@@href}%
\providecommand \@@href[1]{\endgroup#1\@@endlink}%
\providecommand \@sanitize@url [0]{\catcode `\\12\catcode `\$12\catcode
  `\&12\catcode `\#12\catcode `\^12\catcode `\_12\catcode `\%12\relax}%
\providecommand \@@startlink[1]{}%
\providecommand \@@endlink[0]{}%
\providecommand \url  [0]{\begingroup\@sanitize@url \@url }%
\providecommand \@url [1]{\endgroup\@href {#1}{\urlprefix }}%
\providecommand \urlprefix  [0]{URL }%
\providecommand \Eprint [0]{\href }%
\providecommand \doibase [0]{http://dx.doi.org/}%
\providecommand \selectlanguage [0]{\@gobble}%
\providecommand \bibinfo  [0]{\@secondoftwo}%
\providecommand \bibfield  [0]{\@secondoftwo}%
\providecommand \translation [1]{[#1]}%
\providecommand \BibitemOpen [0]{}%
\providecommand \bibitemStop [0]{}%
\providecommand \bibitemNoStop [0]{.\EOS\space}%
\providecommand \EOS [0]{\spacefactor3000\relax}%
\providecommand \BibitemShut  [1]{\csname bibitem#1\endcsname}%
\let\auto@bib@innerbib\@empty
\bibitem [{\citenamefont {Tattersall}\ \emph {et~al.}(2017)\citenamefont
  {Tattersall}, \citenamefont {Lagos},\ and\ \citenamefont
  {Ferreira}}]{Tattersall:2017eav}%
  \BibitemOpen
  \bibfield  {author} {\bibinfo {author} {\bibfnamefont {O.~J.}\ \bibnamefont
  {Tattersall}}, \bibinfo {author} {\bibfnamefont {M.}~\bibnamefont {Lagos}}, \
  and\ \bibinfo {author} {\bibfnamefont {P.~G.}\ \bibnamefont {Ferreira}},\
  }\href {\doibase 10.1103/PhysRevD.96.064011} {\bibfield  {journal} {\bibinfo
  {journal} {Phys. Rev.}\ }\textbf {\bibinfo {volume} {D96}},\ \bibinfo {pages}
  {064011} (\bibinfo {year} {2017})},\ \Eprint
  {http://arxiv.org/abs/1706.10091} {arXiv:1706.10091 [gr-qc]} \BibitemShut
  {NoStop}%
\bibitem [{\citenamefont {{Abbott}}\ \emph
  {et~al.}(2016{\natexlab{a}})\citenamefont {{Abbott}}, \citenamefont
  {{Abbott}}, \citenamefont {{Abbott}}, \citenamefont {{Abernathy}},
  \citenamefont {{Acernese}}, \citenamefont {{Ackley}}, \citenamefont
  {{Adams}}, \citenamefont {{Adams}}, \citenamefont {{Addesso}}, \citenamefont
  {{Adhikari}},\ and\ \citenamefont {et~al.}}]{2016PhRvL.116f1102A}%
  \BibitemOpen
  \bibfield  {author} {\bibinfo {author} {\bibfnamefont {B.~P.}\ \bibnamefont
  {{Abbott}}}, \bibinfo {author} {\bibfnamefont {R.}~\bibnamefont {{Abbott}}},
  \bibinfo {author} {\bibfnamefont {T.~D.}\ \bibnamefont {{Abbott}}}, \bibinfo
  {author} {\bibfnamefont {M.~R.}\ \bibnamefont {{Abernathy}}}, \bibinfo
  {author} {\bibfnamefont {F.}~\bibnamefont {{Acernese}}}, \bibinfo {author}
  {\bibfnamefont {K.}~\bibnamefont {{Ackley}}}, \bibinfo {author}
  {\bibfnamefont {C.}~\bibnamefont {{Adams}}}, \bibinfo {author} {\bibfnamefont
  {T.}~\bibnamefont {{Adams}}}, \bibinfo {author} {\bibfnamefont
  {P.}~\bibnamefont {{Addesso}}}, \bibinfo {author} {\bibfnamefont {R.~X.}\
  \bibnamefont {{Adhikari}}}, \ and\ \bibinfo {author} {\bibnamefont
  {et~al.}},\ }\href {\doibase 10.1103/PhysRevLett.116.061102} {\bibfield
  {journal} {\bibinfo  {journal} {Physical Review Letters}\ }\textbf {\bibinfo
  {volume} {116}},\ \bibinfo {eid} {061102} (\bibinfo {year}
  {2016}{\natexlab{a}})},\ \Eprint {http://arxiv.org/abs/1602.03837}
  {arXiv:1602.03837 [gr-qc]} \BibitemShut {NoStop}%
\bibitem [{\citenamefont {{Abbott}}\ \emph
  {et~al.}(2016{\natexlab{b}})\citenamefont {{Abbott}}, \citenamefont
  {{Abbott}}, \citenamefont {{Abbott}}, \citenamefont {{Abernathy}},
  \citenamefont {{Acernese}}, \citenamefont {{Ackley}}, \citenamefont
  {{Adams}}, \citenamefont {{Adams}}, \citenamefont {{Addesso}}, \citenamefont
  {{Adhikari}},\ and\ \citenamefont {et~al.}}]{2016PhRvL.116x1103A}%
  \BibitemOpen
  \bibfield  {author} {\bibinfo {author} {\bibfnamefont {B.~P.}\ \bibnamefont
  {{Abbott}}}, \bibinfo {author} {\bibfnamefont {R.}~\bibnamefont {{Abbott}}},
  \bibinfo {author} {\bibfnamefont {T.~D.}\ \bibnamefont {{Abbott}}}, \bibinfo
  {author} {\bibfnamefont {M.~R.}\ \bibnamefont {{Abernathy}}}, \bibinfo
  {author} {\bibfnamefont {F.}~\bibnamefont {{Acernese}}}, \bibinfo {author}
  {\bibfnamefont {K.}~\bibnamefont {{Ackley}}}, \bibinfo {author}
  {\bibfnamefont {C.}~\bibnamefont {{Adams}}}, \bibinfo {author} {\bibfnamefont
  {T.}~\bibnamefont {{Adams}}}, \bibinfo {author} {\bibfnamefont
  {P.}~\bibnamefont {{Addesso}}}, \bibinfo {author} {\bibfnamefont {R.~X.}\
  \bibnamefont {{Adhikari}}}, \ and\ \bibinfo {author} {\bibnamefont
  {et~al.}},\ }\href {\doibase 10.1103/PhysRevLett.116.241103} {\bibfield
  {journal} {\bibinfo  {journal} {Physical Review Letters}\ }\textbf {\bibinfo
  {volume} {116}},\ \bibinfo {eid} {241103} (\bibinfo {year}
  {2016}{\natexlab{b}})},\ \Eprint {http://arxiv.org/abs/1606.04855}
  {arXiv:1606.04855 [gr-qc]} \BibitemShut {NoStop}%
\bibitem [{\citenamefont {{Abbott}}\ \emph {et~al.}(2017)\citenamefont
  {{Abbott}}, \citenamefont {{Abbott}}, \citenamefont {{Abbott}}, \citenamefont
  {{Acernese}}, \citenamefont {{Ackley}}, \citenamefont {{Adams}},
  \citenamefont {{Adams}}, \citenamefont {{Addesso}}, \citenamefont
  {{Adhikari}}, \citenamefont {{Adya}},\ and\ \citenamefont
  {et~al.}}]{2017PhRvL.118v1101A}%
  \BibitemOpen
  \bibfield  {author} {\bibinfo {author} {\bibfnamefont {B.~P.}\ \bibnamefont
  {{Abbott}}}, \bibinfo {author} {\bibfnamefont {R.}~\bibnamefont {{Abbott}}},
  \bibinfo {author} {\bibfnamefont {T.~D.}\ \bibnamefont {{Abbott}}}, \bibinfo
  {author} {\bibfnamefont {F.}~\bibnamefont {{Acernese}}}, \bibinfo {author}
  {\bibfnamefont {K.}~\bibnamefont {{Ackley}}}, \bibinfo {author}
  {\bibfnamefont {C.}~\bibnamefont {{Adams}}}, \bibinfo {author} {\bibfnamefont
  {T.}~\bibnamefont {{Adams}}}, \bibinfo {author} {\bibfnamefont
  {P.}~\bibnamefont {{Addesso}}}, \bibinfo {author} {\bibfnamefont {R.~X.}\
  \bibnamefont {{Adhikari}}}, \bibinfo {author} {\bibfnamefont {V.~B.}\
  \bibnamefont {{Adya}}}, \ and\ \bibinfo {author} {\bibnamefont {et~al.}},\
  }\href {\doibase 10.1103/PhysRevLett.118.221101} {\bibfield  {journal}
  {\bibinfo  {journal} {Physical Review Letters}\ }\textbf {\bibinfo {volume}
  {118}},\ \bibinfo {eid} {221101} (\bibinfo {year} {2017})},\ \Eprint
  {http://arxiv.org/abs/1706.01812} {arXiv:1706.01812 [gr-qc]} \BibitemShut
  {NoStop}%
\bibitem [{\citenamefont {{The LIGO Scientific Collaboration}}\ \emph
  {et~al.}(2017)\citenamefont {{The LIGO Scientific Collaboration}},
  \citenamefont {{the Virgo Collaboration}}, \citenamefont {{Abbott}},
  \citenamefont {{Abbott}}, \citenamefont {{Abbott}}, \citenamefont
  {{Acernese}}, \citenamefont {{Ackley}}, \citenamefont {{Adams}},
  \citenamefont {{Adams}}, \citenamefont {{Addesso}},\ and\ \citenamefont
  {et~al.}}]{2017arXiv170909660T}%
  \BibitemOpen
  \bibfield  {author} {\bibinfo {author} {\bibnamefont {{The LIGO Scientific
  Collaboration}}}, \bibinfo {author} {\bibnamefont {{the Virgo
  Collaboration}}}, \bibinfo {author} {\bibfnamefont {B.~P.}\ \bibnamefont
  {{Abbott}}}, \bibinfo {author} {\bibfnamefont {R.}~\bibnamefont {{Abbott}}},
  \bibinfo {author} {\bibfnamefont {T.~D.}\ \bibnamefont {{Abbott}}}, \bibinfo
  {author} {\bibfnamefont {F.}~\bibnamefont {{Acernese}}}, \bibinfo {author}
  {\bibfnamefont {K.}~\bibnamefont {{Ackley}}}, \bibinfo {author}
  {\bibfnamefont {C.}~\bibnamefont {{Adams}}}, \bibinfo {author} {\bibfnamefont
  {T.}~\bibnamefont {{Adams}}}, \bibinfo {author} {\bibfnamefont
  {P.}~\bibnamefont {{Addesso}}}, \ and\ \bibinfo {author} {\bibnamefont
  {et~al.}},\ }\href@noop {} {\bibfield  {journal} {\bibinfo  {journal} {ArXiv
  e-prints}\ } (\bibinfo {year} {2017})},\ \Eprint
  {http://arxiv.org/abs/1709.09660} {arXiv:1709.09660 [gr-qc]} \BibitemShut
  {NoStop}%
\bibitem [{\citenamefont {{Abbott}}\ \emph
  {et~al.}(2016{\natexlab{c}})\citenamefont {{Abbott}}, \citenamefont
  {{Abbott}}, \citenamefont {{Abbott}}, \citenamefont {{Abernathy}},
  \citenamefont {{Acernese}}, \citenamefont {{Ackley}}, \citenamefont
  {{Adams}}, \citenamefont {{Adams}}, \citenamefont {{Addesso}}, \citenamefont
  {{Adhikari}},\ and\ \citenamefont {et~al.}}]{2016PhRvL.116v1101A}%
  \BibitemOpen
  \bibfield  {author} {\bibinfo {author} {\bibfnamefont {B.~P.}\ \bibnamefont
  {{Abbott}}}, \bibinfo {author} {\bibfnamefont {R.}~\bibnamefont {{Abbott}}},
  \bibinfo {author} {\bibfnamefont {T.~D.}\ \bibnamefont {{Abbott}}}, \bibinfo
  {author} {\bibfnamefont {M.~R.}\ \bibnamefont {{Abernathy}}}, \bibinfo
  {author} {\bibfnamefont {F.}~\bibnamefont {{Acernese}}}, \bibinfo {author}
  {\bibfnamefont {K.}~\bibnamefont {{Ackley}}}, \bibinfo {author}
  {\bibfnamefont {C.}~\bibnamefont {{Adams}}}, \bibinfo {author} {\bibfnamefont
  {T.}~\bibnamefont {{Adams}}}, \bibinfo {author} {\bibfnamefont
  {P.}~\bibnamefont {{Addesso}}}, \bibinfo {author} {\bibfnamefont {R.~X.}\
  \bibnamefont {{Adhikari}}}, \ and\ \bibinfo {author} {\bibnamefont
  {et~al.}},\ }\href {\doibase 10.1103/PhysRevLett.116.221101} {\bibfield
  {journal} {\bibinfo  {journal} {Physical Review Letters}\ }\textbf {\bibinfo
  {volume} {116}},\ \bibinfo {eid} {221101} (\bibinfo {year}
  {2016}{\natexlab{c}})},\ \Eprint {http://arxiv.org/abs/1602.03841}
  {arXiv:1602.03841 [gr-qc]} \BibitemShut {NoStop}%
\bibitem [{\citenamefont {Berti}\ \emph {et~al.}(2015)\citenamefont {Berti}
  \emph {et~al.}}]{Berti:2015itd}%
  \BibitemOpen
  \bibfield  {author} {\bibinfo {author} {\bibfnamefont {E.}~\bibnamefont
  {Berti}} \emph {et~al.},\ }\href {\doibase 10.1088/0264-9381/32/24/243001}
  {\bibfield  {journal} {\bibinfo  {journal} {Class. Quant. Grav.}\ }\textbf
  {\bibinfo {volume} {32}},\ \bibinfo {pages} {243001} (\bibinfo {year}
  {2015})},\ \Eprint {http://arxiv.org/abs/1501.07274} {arXiv:1501.07274
  [gr-qc]} \BibitemShut {NoStop}%
\bibitem [{\citenamefont {Dreyer}\ \emph {et~al.}(2004)\citenamefont {Dreyer},
  \citenamefont {Kelly}, \citenamefont {Krishnan}, \citenamefont {Finn},
  \citenamefont {Garrison},\ and\ \citenamefont
  {Lopez-Aleman}}]{Dreyer:2003bv}%
  \BibitemOpen
  \bibfield  {author} {\bibinfo {author} {\bibfnamefont {O.}~\bibnamefont
  {Dreyer}}, \bibinfo {author} {\bibfnamefont {B.~J.}\ \bibnamefont {Kelly}},
  \bibinfo {author} {\bibfnamefont {B.}~\bibnamefont {Krishnan}}, \bibinfo
  {author} {\bibfnamefont {L.~S.}\ \bibnamefont {Finn}}, \bibinfo {author}
  {\bibfnamefont {D.}~\bibnamefont {Garrison}}, \ and\ \bibinfo {author}
  {\bibfnamefont {R.}~\bibnamefont {Lopez-Aleman}},\ }\href {\doibase
  10.1088/0264-9381/21/4/003} {\bibfield  {journal} {\bibinfo  {journal}
  {Class. Quant. Grav.}\ }\textbf {\bibinfo {volume} {21}},\ \bibinfo {pages}
  {787} (\bibinfo {year} {2004})},\ \Eprint
  {http://arxiv.org/abs/gr-qc/0309007} {arXiv:gr-qc/0309007 [gr-qc]}
  \BibitemShut {NoStop}%
\bibitem [{\citenamefont {Gossan}\ \emph {et~al.}(2012)\citenamefont {Gossan},
  \citenamefont {Veitch},\ and\ \citenamefont {Sathyaprakash}}]{Gossan:2011ha}%
  \BibitemOpen
  \bibfield  {author} {\bibinfo {author} {\bibfnamefont {S.}~\bibnamefont
  {Gossan}}, \bibinfo {author} {\bibfnamefont {J.}~\bibnamefont {Veitch}}, \
  and\ \bibinfo {author} {\bibfnamefont {B.~S.}\ \bibnamefont
  {Sathyaprakash}},\ }\href {\doibase 10.1103/PhysRevD.85.124056} {\bibfield
  {journal} {\bibinfo  {journal} {Phys. Rev.}\ }\textbf {\bibinfo {volume}
  {D85}},\ \bibinfo {pages} {124056} (\bibinfo {year} {2012})},\ \Eprint
  {http://arxiv.org/abs/1111.5819} {arXiv:1111.5819 [gr-qc]} \BibitemShut
  {NoStop}%
\bibitem [{\citenamefont {Kamaretsos}\ \emph {et~al.}(2012)\citenamefont
  {Kamaretsos}, \citenamefont {Hannam}, \citenamefont {Husa},\ and\
  \citenamefont {Sathyaprakash}}]{Kamaretsos:2011um}%
  \BibitemOpen
  \bibfield  {author} {\bibinfo {author} {\bibfnamefont {I.}~\bibnamefont
  {Kamaretsos}}, \bibinfo {author} {\bibfnamefont {M.}~\bibnamefont {Hannam}},
  \bibinfo {author} {\bibfnamefont {S.}~\bibnamefont {Husa}}, \ and\ \bibinfo
  {author} {\bibfnamefont {B.~S.}\ \bibnamefont {Sathyaprakash}},\ }\href
  {\doibase 10.1103/PhysRevD.85.024018} {\bibfield  {journal} {\bibinfo
  {journal} {Phys. Rev.}\ }\textbf {\bibinfo {volume} {D85}},\ \bibinfo {pages}
  {024018} (\bibinfo {year} {2012})},\ \Eprint {http://arxiv.org/abs/1107.0854}
  {arXiv:1107.0854 [gr-qc]} \BibitemShut {NoStop}%
\bibitem [{\citenamefont {Meidam}\ \emph {et~al.}(2014)\citenamefont {Meidam},
  \citenamefont {Agathos}, \citenamefont {Van Den~Broeck}, \citenamefont
  {Veitch},\ and\ \citenamefont {Sathyaprakash}}]{Meidam:2014jpa}%
  \BibitemOpen
  \bibfield  {author} {\bibinfo {author} {\bibfnamefont {J.}~\bibnamefont
  {Meidam}}, \bibinfo {author} {\bibfnamefont {M.}~\bibnamefont {Agathos}},
  \bibinfo {author} {\bibfnamefont {C.}~\bibnamefont {Van Den~Broeck}},
  \bibinfo {author} {\bibfnamefont {J.}~\bibnamefont {Veitch}}, \ and\ \bibinfo
  {author} {\bibfnamefont {B.~S.}\ \bibnamefont {Sathyaprakash}},\ }\href
  {\doibase 10.1103/PhysRevD.90.064009} {\bibfield  {journal} {\bibinfo
  {journal} {Phys. Rev.}\ }\textbf {\bibinfo {volume} {D90}},\ \bibinfo {pages}
  {064009} (\bibinfo {year} {2014})},\ \Eprint {http://arxiv.org/abs/1406.3201}
  {arXiv:1406.3201 [gr-qc]} \BibitemShut {NoStop}%
\bibitem [{\citenamefont {Nakano}\ \emph {et~al.}(2015)\citenamefont {Nakano},
  \citenamefont {Tanaka},\ and\ \citenamefont {Nakamura}}]{Nakano:2015uja}%
  \BibitemOpen
  \bibfield  {author} {\bibinfo {author} {\bibfnamefont {H.}~\bibnamefont
  {Nakano}}, \bibinfo {author} {\bibfnamefont {T.}~\bibnamefont {Tanaka}}, \
  and\ \bibinfo {author} {\bibfnamefont {T.}~\bibnamefont {Nakamura}},\ }\href
  {\doibase 10.1103/PhysRevD.92.064003} {\bibfield  {journal} {\bibinfo
  {journal} {Phys. Rev.}\ }\textbf {\bibinfo {volume} {D92}},\ \bibinfo {pages}
  {064003} (\bibinfo {year} {2015})},\ \Eprint
  {http://arxiv.org/abs/1506.00560} {arXiv:1506.00560 [astro-ph.HE]}
  \BibitemShut {NoStop}%
\bibitem [{\citenamefont {Cardoso}\ and\ \citenamefont
  {Gualtieri}(2016)}]{Cardoso:2016ryw}%
  \BibitemOpen
  \bibfield  {author} {\bibinfo {author} {\bibfnamefont {V.}~\bibnamefont
  {Cardoso}}\ and\ \bibinfo {author} {\bibfnamefont {L.}~\bibnamefont
  {Gualtieri}},\ }\href {\doibase 10.1088/0264-9381/33/17/174001} {\bibfield
  {journal} {\bibinfo  {journal} {Class. Quant. Grav.}\ }\textbf {\bibinfo
  {volume} {33}},\ \bibinfo {pages} {174001} (\bibinfo {year} {2016})},\
  \Eprint {http://arxiv.org/abs/1607.03133} {arXiv:1607.03133 [gr-qc]}
  \BibitemShut {NoStop}%
\bibitem [{\citenamefont {Thrane}\ \emph {et~al.}(2017)\citenamefont {Thrane},
  \citenamefont {Lasky},\ and\ \citenamefont {Levin}}]{Thrane:2017lqn}%
  \BibitemOpen
  \bibfield  {author} {\bibinfo {author} {\bibfnamefont {E.}~\bibnamefont
  {Thrane}}, \bibinfo {author} {\bibfnamefont {P.~D.}\ \bibnamefont {Lasky}}, \
  and\ \bibinfo {author} {\bibfnamefont {Y.}~\bibnamefont {Levin}},\
  }\href@noop {} {\  (\bibinfo {year} {2017})},\ \Eprint
  {http://arxiv.org/abs/1706.05152} {arXiv:1706.05152 [gr-qc]} \BibitemShut
  {NoStop}%
\bibitem [{\citenamefont {Glampedakis}\ \emph {et~al.}(2017)\citenamefont
  {Glampedakis}, \citenamefont {Pappas}, \citenamefont {Silva},\ and\
  \citenamefont {Berti}}]{Glampedakis:2017dvb}%
  \BibitemOpen
  \bibfield  {author} {\bibinfo {author} {\bibfnamefont {K.}~\bibnamefont
  {Glampedakis}}, \bibinfo {author} {\bibfnamefont {G.}~\bibnamefont {Pappas}},
  \bibinfo {author} {\bibfnamefont {H.~O.}\ \bibnamefont {Silva}}, \ and\
  \bibinfo {author} {\bibfnamefont {E.}~\bibnamefont {Berti}},\ }\href
  {\doibase 10.1103/PhysRevD.96.064054} {\bibfield  {journal} {\bibinfo
  {journal} {Phys. Rev.}\ }\textbf {\bibinfo {volume} {D96}},\ \bibinfo {pages}
  {064054} (\bibinfo {year} {2017})},\ \Eprint
  {http://arxiv.org/abs/1706.07658} {arXiv:1706.07658 [gr-qc]} \BibitemShut
  {NoStop}%
\bibitem [{\citenamefont {Barausse}\ and\ \citenamefont
  {Sotiriou}(2008)}]{Barausse:2008xv}%
  \BibitemOpen
  \bibfield  {author} {\bibinfo {author} {\bibfnamefont {E.}~\bibnamefont
  {Barausse}}\ and\ \bibinfo {author} {\bibfnamefont {T.~P.}\ \bibnamefont
  {Sotiriou}},\ }\href {\doibase 10.1103/PhysRevLett.101.099001} {\bibfield
  {journal} {\bibinfo  {journal} {Phys. Rev. Lett.}\ }\textbf {\bibinfo
  {volume} {101}},\ \bibinfo {pages} {099001} (\bibinfo {year} {2008})},\
  \Eprint {http://arxiv.org/abs/0803.3433} {arXiv:0803.3433 [gr-qc]}
  \BibitemShut {NoStop}%
\bibitem [{\citenamefont {Berti}\ \emph {et~al.}(2009)\citenamefont {Berti},
  \citenamefont {Cardoso},\ and\ \citenamefont {Starinets}}]{Berti:2009kk}%
  \BibitemOpen
  \bibfield  {author} {\bibinfo {author} {\bibfnamefont {E.}~\bibnamefont
  {Berti}}, \bibinfo {author} {\bibfnamefont {V.}~\bibnamefont {Cardoso}}, \
  and\ \bibinfo {author} {\bibfnamefont {A.~O.}\ \bibnamefont {Starinets}},\
  }\href {\doibase 10.1088/0264-9381/26/16/163001} {\bibfield  {journal}
  {\bibinfo  {journal} {Class. Quant. Grav.}\ }\textbf {\bibinfo {volume}
  {26}},\ \bibinfo {pages} {163001} (\bibinfo {year} {2009})},\ \Eprint
  {http://arxiv.org/abs/0905.2975} {arXiv:0905.2975 [gr-qc]} \BibitemShut
  {NoStop}%
\bibitem [{\citenamefont {Berti}\ \emph {et~al.}(2006)\citenamefont {Berti},
  \citenamefont {Cardoso},\ and\ \citenamefont {Will}}]{Berti:2005ys}%
  \BibitemOpen
  \bibfield  {author} {\bibinfo {author} {\bibfnamefont {E.}~\bibnamefont
  {Berti}}, \bibinfo {author} {\bibfnamefont {V.}~\bibnamefont {Cardoso}}, \
  and\ \bibinfo {author} {\bibfnamefont {C.~M.}\ \bibnamefont {Will}},\ }\href
  {\doibase 10.1103/PhysRevD.73.064030} {\bibfield  {journal} {\bibinfo
  {journal} {Phys. Rev.}\ }\textbf {\bibinfo {volume} {D73}},\ \bibinfo {pages}
  {064030} (\bibinfo {year} {2006})},\ \Eprint
  {http://arxiv.org/abs/gr-qc/0512160} {arXiv:gr-qc/0512160 [gr-qc]}
  \BibitemShut {NoStop}%
\bibitem [{\citenamefont {Berti}\ \emph {et~al.}(2007)\citenamefont {Berti},
  \citenamefont {Cardoso}, \citenamefont {Cardoso},\ and\ \citenamefont
  {Cavaglia}}]{Berti:2007zu}%
  \BibitemOpen
  \bibfield  {author} {\bibinfo {author} {\bibfnamefont {E.}~\bibnamefont
  {Berti}}, \bibinfo {author} {\bibfnamefont {J.}~\bibnamefont {Cardoso}},
  \bibinfo {author} {\bibfnamefont {V.}~\bibnamefont {Cardoso}}, \ and\
  \bibinfo {author} {\bibfnamefont {M.}~\bibnamefont {Cavaglia}},\ }\href
  {\doibase 10.1103/PhysRevD.76.104044} {\bibfield  {journal} {\bibinfo
  {journal} {Phys. Rev.}\ }\textbf {\bibinfo {volume} {D76}},\ \bibinfo {pages}
  {104044} (\bibinfo {year} {2007})},\ \Eprint {http://arxiv.org/abs/0707.1202}
  {arXiv:0707.1202 [gr-qc]} \BibitemShut {NoStop}%
\bibitem [{\citenamefont {Yang}\ \emph {et~al.}(2017)\citenamefont {Yang},
  \citenamefont {Yagi}, \citenamefont {Blackman}, \citenamefont {Lehner},
  \citenamefont {Paschalidis}, \citenamefont {Pretorius},\ and\ \citenamefont
  {Yunes}}]{Yang:2017zxs}%
  \BibitemOpen
  \bibfield  {author} {\bibinfo {author} {\bibfnamefont {H.}~\bibnamefont
  {Yang}}, \bibinfo {author} {\bibfnamefont {K.}~\bibnamefont {Yagi}}, \bibinfo
  {author} {\bibfnamefont {J.}~\bibnamefont {Blackman}}, \bibinfo {author}
  {\bibfnamefont {L.}~\bibnamefont {Lehner}}, \bibinfo {author} {\bibfnamefont
  {V.}~\bibnamefont {Paschalidis}}, \bibinfo {author} {\bibfnamefont
  {F.}~\bibnamefont {Pretorius}}, \ and\ \bibinfo {author} {\bibfnamefont
  {N.}~\bibnamefont {Yunes}},\ }\href {\doibase 10.1103/PhysRevLett.118.161101}
  {\bibfield  {journal} {\bibinfo  {journal} {Phys. Rev. Lett.}\ }\textbf
  {\bibinfo {volume} {118}},\ \bibinfo {pages} {161101} (\bibinfo {year}
  {2017})},\ \Eprint {http://arxiv.org/abs/1701.05808} {arXiv:1701.05808
  [gr-qc]} \BibitemShut {NoStop}%
\bibitem [{\citenamefont {Baibhav}\ \emph {et~al.}(2017)\citenamefont
  {Baibhav}, \citenamefont {Berti}, \citenamefont {Cardoso},\ and\
  \citenamefont {Khanna}}]{Baibhav:2017jhs}%
  \BibitemOpen
  \bibfield  {author} {\bibinfo {author} {\bibfnamefont {V.}~\bibnamefont
  {Baibhav}}, \bibinfo {author} {\bibfnamefont {E.}~\bibnamefont {Berti}},
  \bibinfo {author} {\bibfnamefont {V.}~\bibnamefont {Cardoso}}, \ and\
  \bibinfo {author} {\bibfnamefont {G.}~\bibnamefont {Khanna}},\ }\href@noop {}
  {\  (\bibinfo {year} {2017})},\ \Eprint {http://arxiv.org/abs/1710.02156}
  {arXiv:1710.02156 [gr-qc]} \BibitemShut {NoStop}%
\bibitem [{\citenamefont {Harada}(1997)}]{Harada:1997mr}%
  \BibitemOpen
  \bibfield  {author} {\bibinfo {author} {\bibfnamefont {T.}~\bibnamefont
  {Harada}},\ }\href {\doibase 10.1143/PTP.98.359} {\bibfield  {journal}
  {\bibinfo  {journal} {Prog. Theor. Phys.}\ }\textbf {\bibinfo {volume}
  {98}},\ \bibinfo {pages} {359} (\bibinfo {year} {1997})},\ \Eprint
  {http://arxiv.org/abs/gr-qc/9706014} {arXiv:gr-qc/9706014 [gr-qc]}
  \BibitemShut {NoStop}%
\bibitem [{\citenamefont {Kwon}\ \emph {et~al.}(1986)\citenamefont {Kwon},
  \citenamefont {Kim}, \citenamefont {Myung}, \citenamefont {Cho},\ and\
  \citenamefont {Park}}]{PhysRevD.34.333}%
  \BibitemOpen
  \bibfield  {author} {\bibinfo {author} {\bibfnamefont {O.~J.}\ \bibnamefont
  {Kwon}}, \bibinfo {author} {\bibfnamefont {Y.~D.}\ \bibnamefont {Kim}},
  \bibinfo {author} {\bibfnamefont {Y.~S.}\ \bibnamefont {Myung}}, \bibinfo
  {author} {\bibfnamefont {B.~H.}\ \bibnamefont {Cho}}, \ and\ \bibinfo
  {author} {\bibfnamefont {Y.~J.}\ \bibnamefont {Park}},\ }\href {\doibase
  10.1103/PhysRevD.34.333} {\bibfield  {journal} {\bibinfo  {journal} {Phys.
  Rev. D}\ }\textbf {\bibinfo {volume} {34}},\ \bibinfo {pages} {333} (\bibinfo
  {year} {1986})}\BibitemShut {NoStop}%
\bibitem [{\citenamefont {Dong}\ \emph {et~al.}(2017)\citenamefont {Dong},
  \citenamefont {Sakstein},\ and\ \citenamefont {Stojkovic}}]{Dong:2017toi}%
  \BibitemOpen
  \bibfield  {author} {\bibinfo {author} {\bibfnamefont {R.}~\bibnamefont
  {Dong}}, \bibinfo {author} {\bibfnamefont {J.}~\bibnamefont {Sakstein}}, \
  and\ \bibinfo {author} {\bibfnamefont {D.}~\bibnamefont {Stojkovic}},\ }\href
  {\doibase 10.1103/PhysRevD.96.064048} {\bibfield  {journal} {\bibinfo
  {journal} {Phys. Rev.}\ }\textbf {\bibinfo {volume} {D96}},\ \bibinfo {pages}
  {064048} (\bibinfo {year} {2017})},\ \Eprint
  {http://arxiv.org/abs/1709.01641} {arXiv:1709.01641 [gr-qc]} \BibitemShut
  {NoStop}%
\bibitem [{\citenamefont {Blazquez-Salcedo}\ \emph {et~al.}(2016)\citenamefont
  {Blazquez-Salcedo}, \citenamefont {Macedo}, \citenamefont {Cardoso},
  \citenamefont {Ferrari}, \citenamefont {Gualtieri}, \citenamefont {Khoo},
  \citenamefont {Kunz},\ and\ \citenamefont {Pani}}]{Blazquez-Salcedo:2016enn}%
  \BibitemOpen
  \bibfield  {author} {\bibinfo {author} {\bibfnamefont {J.~L.}\ \bibnamefont
  {Blazquez-Salcedo}}, \bibinfo {author} {\bibfnamefont {C.~F.~B.}\
  \bibnamefont {Macedo}}, \bibinfo {author} {\bibfnamefont {V.}~\bibnamefont
  {Cardoso}}, \bibinfo {author} {\bibfnamefont {V.}~\bibnamefont {Ferrari}},
  \bibinfo {author} {\bibfnamefont {L.}~\bibnamefont {Gualtieri}}, \bibinfo
  {author} {\bibfnamefont {F.~S.}\ \bibnamefont {Khoo}}, \bibinfo {author}
  {\bibfnamefont {J.}~\bibnamefont {Kunz}}, \ and\ \bibinfo {author}
  {\bibfnamefont {P.}~\bibnamefont {Pani}},\ }\href {\doibase
  10.1103/PhysRevD.94.104024} {\bibfield  {journal} {\bibinfo  {journal} {Phys.
  Rev.}\ }\textbf {\bibinfo {volume} {D94}},\ \bibinfo {pages} {104024}
  (\bibinfo {year} {2016})},\ \Eprint {http://arxiv.org/abs/1609.01286}
  {arXiv:1609.01286 [gr-qc]} \BibitemShut {NoStop}%
\bibitem [{\citenamefont {Lasky}\ and\ \citenamefont
  {Doneva}(2010)}]{Lasky:2010bd}%
  \BibitemOpen
  \bibfield  {author} {\bibinfo {author} {\bibfnamefont {P.~D.}\ \bibnamefont
  {Lasky}}\ and\ \bibinfo {author} {\bibfnamefont {D.~D.}\ \bibnamefont
  {Doneva}},\ }\href {\doibase 10.1103/PhysRevD.82.124068} {\bibfield
  {journal} {\bibinfo  {journal} {Phys. Rev.}\ }\textbf {\bibinfo {volume}
  {D82}},\ \bibinfo {pages} {124068} (\bibinfo {year} {2010})},\ \Eprint
  {http://arxiv.org/abs/1011.0747} {arXiv:1011.0747 [gr-qc]} \BibitemShut
  {NoStop}%
\bibitem [{\citenamefont {Cardoso}\ and\ \citenamefont
  {Gualtieri}(2009)}]{Cardoso:2009pk}%
  \BibitemOpen
  \bibfield  {author} {\bibinfo {author} {\bibfnamefont {V.}~\bibnamefont
  {Cardoso}}\ and\ \bibinfo {author} {\bibfnamefont {L.}~\bibnamefont
  {Gualtieri}},\ }\href {\doibase 10.1103/PhysRevD.81.089903,
  10.1103/PhysRevD.80.064008} {\bibfield  {journal} {\bibinfo  {journal} {Phys.
  Rev.}\ }\textbf {\bibinfo {volume} {D80}},\ \bibinfo {pages} {064008}
  (\bibinfo {year} {2009})},\ \bibinfo {note} {[Erratum: Phys.
  Rev.D81,089903(2010)]},\ \Eprint {http://arxiv.org/abs/0907.5008}
  {arXiv:0907.5008 [gr-qc]} \BibitemShut {NoStop}%
\bibitem [{\citenamefont {Molina}\ \emph {et~al.}(2010)\citenamefont {Molina},
  \citenamefont {Pani}, \citenamefont {Cardoso},\ and\ \citenamefont
  {Gualtieri}}]{Molina:2010fb}%
  \BibitemOpen
  \bibfield  {author} {\bibinfo {author} {\bibfnamefont {C.}~\bibnamefont
  {Molina}}, \bibinfo {author} {\bibfnamefont {P.}~\bibnamefont {Pani}},
  \bibinfo {author} {\bibfnamefont {V.}~\bibnamefont {Cardoso}}, \ and\
  \bibinfo {author} {\bibfnamefont {L.}~\bibnamefont {Gualtieri}},\ }\href
  {\doibase 10.1103/PhysRevD.81.124021} {\bibfield  {journal} {\bibinfo
  {journal} {Phys. Rev.}\ }\textbf {\bibinfo {volume} {D81}},\ \bibinfo {pages}
  {124021} (\bibinfo {year} {2010})},\ \Eprint {http://arxiv.org/abs/1004.4007}
  {arXiv:1004.4007 [gr-qc]} \BibitemShut {NoStop}%
\bibitem [{\citenamefont {Dodelson}(2003)}]{Dodelson:2003ft}%
  \BibitemOpen
  \bibfield  {author} {\bibinfo {author} {\bibfnamefont {S.}~\bibnamefont
  {Dodelson}},\ }\href
  {http://www.slac.stanford.edu/spires/find/books/www?cl=QB981:D62:2003} {\emph
  {\bibinfo {title} {{Modern Cosmology}}}}\ (\bibinfo  {publisher} {Academic
  Press},\ \bibinfo {address} {Amsterdam},\ \bibinfo {year} {2003})\BibitemShut
  {NoStop}%
\bibitem [{\citenamefont {Lagos}\ \emph {et~al.}(2016)\citenamefont {Lagos},
  \citenamefont {Baker}, \citenamefont {Ferreira},\ and\ \citenamefont
  {Noller}}]{Lagos:2016wyv}%
  \BibitemOpen
  \bibfield  {author} {\bibinfo {author} {\bibfnamefont {M.}~\bibnamefont
  {Lagos}}, \bibinfo {author} {\bibfnamefont {T.}~\bibnamefont {Baker}},
  \bibinfo {author} {\bibfnamefont {P.~G.}\ \bibnamefont {Ferreira}}, \ and\
  \bibinfo {author} {\bibfnamefont {J.}~\bibnamefont {Noller}},\ }\href
  {\doibase 10.1088/1475-7516/2016/08/007} {\bibfield  {journal} {\bibinfo
  {journal} {JCAP}\ }\textbf {\bibinfo {volume} {1608}},\ \bibinfo {pages}
  {007} (\bibinfo {year} {2016})},\ \Eprint {http://arxiv.org/abs/1604.01396}
  {arXiv:1604.01396 [gr-qc]} \BibitemShut {NoStop}%
\bibitem [{\citenamefont {{Lagos}}\ and\ \citenamefont
  {{Ferreira}}(2017)}]{2017JCAP...01..047L}%
  \BibitemOpen
  \bibfield  {author} {\bibinfo {author} {\bibfnamefont {M.}~\bibnamefont
  {{Lagos}}}\ and\ \bibinfo {author} {\bibfnamefont {P.~G.}\ \bibnamefont
  {{Ferreira}}},\ }\href {\doibase 10.1088/1475-7516/2017/01/047} {\bibfield
  {journal} {\bibinfo  {journal} {Journal of Cosmology and Astroparticle
  Physics}\ }\textbf {\bibinfo {volume} {1}},\ \bibinfo {eid} {047} (\bibinfo
  {year} {2017})},\ \Eprint {http://arxiv.org/abs/1610.00553} {arXiv:1610.00553
  [gr-qc]} \BibitemShut {NoStop}%
\bibitem [{\citenamefont {{Sotiriou}}\ and\ \citenamefont
  {{Faraoni}}(2012)}]{2012PhRvL.108h1103S}%
  \BibitemOpen
  \bibfield  {author} {\bibinfo {author} {\bibfnamefont {T.~P.}\ \bibnamefont
  {{Sotiriou}}}\ and\ \bibinfo {author} {\bibfnamefont {V.}~\bibnamefont
  {{Faraoni}}},\ }\href {\doibase 10.1103/PhysRevLett.108.081103} {\bibfield
  {journal} {\bibinfo  {journal} {Physical Review Letters}\ }\textbf {\bibinfo
  {volume} {108}},\ \bibinfo {eid} {081103} (\bibinfo {year} {2012})},\ \Eprint
  {http://arxiv.org/abs/1109.6324} {arXiv:1109.6324 [gr-qc]} \BibitemShut
  {NoStop}%
\bibitem [{\citenamefont {Sotiriou}\ and\ \citenamefont
  {Zhou}(2014)}]{Sotiriou:2013qea}%
  \BibitemOpen
  \bibfield  {author} {\bibinfo {author} {\bibfnamefont {T.~P.}\ \bibnamefont
  {Sotiriou}}\ and\ \bibinfo {author} {\bibfnamefont {S.-Y.}\ \bibnamefont
  {Zhou}},\ }\href {\doibase 10.1103/PhysRevLett.112.251102} {\bibfield
  {journal} {\bibinfo  {journal} {Phys. Rev. Lett.}\ }\textbf {\bibinfo
  {volume} {112}},\ \bibinfo {pages} {251102} (\bibinfo {year} {2014})},\
  \Eprint {http://arxiv.org/abs/1312.3622} {arXiv:1312.3622 [gr-qc]}
  \BibitemShut {NoStop}%
\bibitem [{\citenamefont {Chagoya}\ \emph {et~al.}(2016)\citenamefont
  {Chagoya}, \citenamefont {Niz},\ and\ \citenamefont
  {Tasinato}}]{Chagoya:2016aar}%
  \BibitemOpen
  \bibfield  {author} {\bibinfo {author} {\bibfnamefont {J.}~\bibnamefont
  {Chagoya}}, \bibinfo {author} {\bibfnamefont {G.}~\bibnamefont {Niz}}, \ and\
  \bibinfo {author} {\bibfnamefont {G.}~\bibnamefont {Tasinato}},\ }\href
  {\doibase 10.1088/0264-9381/33/17/175007} {\bibfield  {journal} {\bibinfo
  {journal} {Class. Quant. Grav.}\ }\textbf {\bibinfo {volume} {33}},\ \bibinfo
  {pages} {175007} (\bibinfo {year} {2016})},\ \Eprint
  {http://arxiv.org/abs/1602.08697} {arXiv:1602.08697 [hep-th]} \BibitemShut
  {NoStop}%
\bibitem [{\citenamefont {{Heisenberg}}\ \emph {et~al.}(2017)\citenamefont
  {{Heisenberg}}, \citenamefont {{Kase}}, \citenamefont {{Minamitsuji}},\ and\
  \citenamefont {{Tsujikawa}}}]{2017JCAP...08..024H}%
  \BibitemOpen
  \bibfield  {author} {\bibinfo {author} {\bibfnamefont {L.}~\bibnamefont
  {{Heisenberg}}}, \bibinfo {author} {\bibfnamefont {R.}~\bibnamefont
  {{Kase}}}, \bibinfo {author} {\bibfnamefont {M.}~\bibnamefont
  {{Minamitsuji}}}, \ and\ \bibinfo {author} {\bibfnamefont {S.}~\bibnamefont
  {{Tsujikawa}}},\ }\href {\doibase 10.1088/1475-7516/2017/08/024} {\bibfield
  {journal} {\bibinfo  {journal} {JCAP}\ }\textbf {\bibinfo {volume} {8}},\
  \bibinfo {eid} {024} (\bibinfo {year} {2017})},\ \Eprint
  {http://arxiv.org/abs/1706.05115} {arXiv:1706.05115 [gr-qc]} \BibitemShut
  {NoStop}%
\bibitem [{\citenamefont {{Stein}}\ and\ \citenamefont
  {{Yunes}}(2011)}]{2011PhRvD..83f4038S}%
  \BibitemOpen
  \bibfield  {author} {\bibinfo {author} {\bibfnamefont {L.~C.}\ \bibnamefont
  {{Stein}}}\ and\ \bibinfo {author} {\bibfnamefont {N.}~\bibnamefont
  {{Yunes}}},\ }\href {\doibase 10.1103/PhysRevD.83.064038} {\bibfield
  {journal} {\bibinfo  {journal} {Physicsl Review D}\ }\textbf {\bibinfo
  {volume} {83}},\ \bibinfo {eid} {064038} (\bibinfo {year} {2011})},\ \Eprint
  {http://arxiv.org/abs/1012.3144} {arXiv:1012.3144 [gr-qc]} \BibitemShut
  {NoStop}%
\bibitem [{\citenamefont {Alexander}\ and\ \citenamefont
  {Yunes}(2009)}]{Alexander:2009tp}%
  \BibitemOpen
  \bibfield  {author} {\bibinfo {author} {\bibfnamefont {S.}~\bibnamefont
  {Alexander}}\ and\ \bibinfo {author} {\bibfnamefont {N.}~\bibnamefont
  {Yunes}},\ }\href {\doibase 10.1016/j.physrep.2009.07.002} {\bibfield
  {journal} {\bibinfo  {journal} {Phys. Rept.}\ }\textbf {\bibinfo {volume}
  {480}},\ \bibinfo {pages} {1} (\bibinfo {year} {2009})},\ \Eprint
  {http://arxiv.org/abs/0907.2562} {arXiv:0907.2562 [hep-th]} \BibitemShut
  {NoStop}%
\bibitem [{\citenamefont {Regge}\ and\ \citenamefont
  {Wheeler}(1957)}]{Regge:1957td}%
  \BibitemOpen
  \bibfield  {author} {\bibinfo {author} {\bibfnamefont {T.}~\bibnamefont
  {Regge}}\ and\ \bibinfo {author} {\bibfnamefont {J.~A.}\ \bibnamefont
  {Wheeler}},\ }\href {\doibase 10.1103/PhysRev.108.1063} {\bibfield  {journal}
  {\bibinfo  {journal} {Phys. Rev.}\ }\textbf {\bibinfo {volume} {108}},\
  \bibinfo {pages} {1063} (\bibinfo {year} {1957})}\BibitemShut {NoStop}%
\bibitem [{\citenamefont {Rezzolla}(2003)}]{Rezzolla:2003ua}%
  \BibitemOpen
  \bibfield  {author} {\bibinfo {author} {\bibfnamefont {L.}~\bibnamefont
  {Rezzolla}},\ }\bibfield  {booktitle} {\emph {\bibinfo {booktitle}
  {{Astroparticle physics and cosmology. Proceedings: Summer School, Trieste,
  Italy, Jun 17-Jul 5 2002}}},\ }\href@noop {} {\bibfield  {journal} {\bibinfo
  {journal} {ICTP Lect. Notes Ser.}\ }\textbf {\bibinfo {volume} {14}},\
  \bibinfo {pages} {255} (\bibinfo {year} {2003})},\ \Eprint
  {http://arxiv.org/abs/gr-qc/0302025} {arXiv:gr-qc/0302025 [gr-qc]}
  \BibitemShut {NoStop}%
\bibitem [{\citenamefont {Martel}\ and\ \citenamefont
  {Poisson}(2005)}]{Martel:2005ir}%
  \BibitemOpen
  \bibfield  {author} {\bibinfo {author} {\bibfnamefont {K.}~\bibnamefont
  {Martel}}\ and\ \bibinfo {author} {\bibfnamefont {E.}~\bibnamefont
  {Poisson}},\ }\href {\doibase 10.1103/PhysRevD.71.104003} {\bibfield
  {journal} {\bibinfo  {journal} {Phys. Rev.}\ }\textbf {\bibinfo {volume}
  {D71}},\ \bibinfo {pages} {104003} (\bibinfo {year} {2005})},\ \Eprint
  {http://arxiv.org/abs/gr-qc/0502028} {arXiv:gr-qc/0502028 [gr-qc]}
  \BibitemShut {NoStop}%
\bibitem [{\citenamefont {Ripley}\ and\ \citenamefont
  {Yagi}(2017)}]{Ripley:2017kqg}%
  \BibitemOpen
  \bibfield  {author} {\bibinfo {author} {\bibfnamefont {J.~L.}\ \bibnamefont
  {Ripley}}\ and\ \bibinfo {author} {\bibfnamefont {K.}~\bibnamefont {Yagi}},\
  }\href@noop {} {\  (\bibinfo {year} {2017})},\ \Eprint
  {http://arxiv.org/abs/1705.03068} {arXiv:1705.03068 [gr-qc]} \BibitemShut
  {NoStop}%
\bibitem [{\citenamefont {Gleiser}\ and\ \citenamefont
  {Dotti}(2006)}]{Gleiser:2006yz}%
  \BibitemOpen
  \bibfield  {author} {\bibinfo {author} {\bibfnamefont {R.~J.}\ \bibnamefont
  {Gleiser}}\ and\ \bibinfo {author} {\bibfnamefont {G.}~\bibnamefont
  {Dotti}},\ }\href {\doibase 10.1088/0264-9381/23/15/021} {\bibfield
  {journal} {\bibinfo  {journal} {Class. Quant. Grav.}\ }\textbf {\bibinfo
  {volume} {23}},\ \bibinfo {pages} {5063} (\bibinfo {year} {2006})},\ \Eprint
  {http://arxiv.org/abs/gr-qc/0604021} {arXiv:gr-qc/0604021 [gr-qc]}
  \BibitemShut {NoStop}%
\bibitem [{\citenamefont {Zerilli}(1970)}]{Zerilli:1970se}%
  \BibitemOpen
  \bibfield  {author} {\bibinfo {author} {\bibfnamefont {F.~J.}\ \bibnamefont
  {Zerilli}},\ }\href {\doibase 10.1103/PhysRevLett.24.737} {\bibfield
  {journal} {\bibinfo  {journal} {Phys. Rev. Lett.}\ }\textbf {\bibinfo
  {volume} {24}},\ \bibinfo {pages} {737} (\bibinfo {year} {1970})}\BibitemShut
  {NoStop}%
\bibitem [{\citenamefont {Ganguly}\ \emph {et~al.}(2017)\citenamefont
  {Ganguly}, \citenamefont {Gannouji}, \citenamefont {Gonzalez-Espinoza},\ and\
  \citenamefont {Pizarro-Moya}}]{Ganguly:2017ort}%
  \BibitemOpen
  \bibfield  {author} {\bibinfo {author} {\bibfnamefont {A.}~\bibnamefont
  {Ganguly}}, \bibinfo {author} {\bibfnamefont {R.}~\bibnamefont {Gannouji}},
  \bibinfo {author} {\bibfnamefont {M.}~\bibnamefont {Gonzalez-Espinoza}}, \
  and\ \bibinfo {author} {\bibfnamefont {C.}~\bibnamefont {Pizarro-Moya}},\
  }\href@noop {} {\  (\bibinfo {year} {2017})},\ \Eprint
  {http://arxiv.org/abs/1710.07669} {arXiv:1710.07669 [gr-qc]} \BibitemShut
  {NoStop}%
\bibitem [{\citenamefont {Kobayashi}\ \emph {et~al.}(2012)\citenamefont
  {Kobayashi}, \citenamefont {Motohashi},\ and\ \citenamefont
  {Suyama}}]{Kobayashi:2012kh}%
  \BibitemOpen
  \bibfield  {author} {\bibinfo {author} {\bibfnamefont {T.}~\bibnamefont
  {Kobayashi}}, \bibinfo {author} {\bibfnamefont {H.}~\bibnamefont
  {Motohashi}}, \ and\ \bibinfo {author} {\bibfnamefont {T.}~\bibnamefont
  {Suyama}},\ }\href {\doibase 10.1103/PhysRevD.85.084025} {\bibfield
  {journal} {\bibinfo  {journal} {Phys. Rev.}\ }\textbf {\bibinfo {volume}
  {D85}},\ \bibinfo {pages} {084025} (\bibinfo {year} {2012})},\ \Eprint
  {http://arxiv.org/abs/1202.4893} {arXiv:1202.4893 [gr-qc]} \BibitemShut
  {NoStop}%
\bibitem [{\citenamefont {Kobayashi}\ \emph {et~al.}(2014)\citenamefont
  {Kobayashi}, \citenamefont {Motohashi},\ and\ \citenamefont
  {Suyama}}]{Kobayashi:2014wsa}%
  \BibitemOpen
  \bibfield  {author} {\bibinfo {author} {\bibfnamefont {T.}~\bibnamefont
  {Kobayashi}}, \bibinfo {author} {\bibfnamefont {H.}~\bibnamefont
  {Motohashi}}, \ and\ \bibinfo {author} {\bibfnamefont {T.}~\bibnamefont
  {Suyama}},\ }\href {\doibase 10.1103/PhysRevD.89.084042} {\bibfield
  {journal} {\bibinfo  {journal} {Phys. Rev.}\ }\textbf {\bibinfo {volume}
  {D89}},\ \bibinfo {pages} {084042} (\bibinfo {year} {2014})},\ \Eprint
  {http://arxiv.org/abs/1402.6740} {arXiv:1402.6740 [gr-qc]} \BibitemShut
  {NoStop}%
\bibitem [{\citenamefont {Creminelli}\ and\ \citenamefont
  {Vernizzi}(2017)}]{Creminelli:2017sry}%
  \BibitemOpen
  \bibfield  {author} {\bibinfo {author} {\bibfnamefont {P.}~\bibnamefont
  {Creminelli}}\ and\ \bibinfo {author} {\bibfnamefont {F.}~\bibnamefont
  {Vernizzi}},\ }\href@noop {} {\  (\bibinfo {year} {2017})},\ \Eprint
  {http://arxiv.org/abs/1710.05877} {arXiv:1710.05877 [astro-ph.CO]}
  \BibitemShut {NoStop}%
\bibitem [{\citenamefont {Baker}\ \emph {et~al.}(2017)\citenamefont {Baker},
  \citenamefont {Bellini}, \citenamefont {Ferreira}, \citenamefont {Lagos},
  \citenamefont {Noller},\ and\ \citenamefont {Sawicki}}]{Baker:2017hug}%
  \BibitemOpen
  \bibfield  {author} {\bibinfo {author} {\bibfnamefont {T.}~\bibnamefont
  {Baker}}, \bibinfo {author} {\bibfnamefont {E.}~\bibnamefont {Bellini}},
  \bibinfo {author} {\bibfnamefont {P.~G.}\ \bibnamefont {Ferreira}}, \bibinfo
  {author} {\bibfnamefont {M.}~\bibnamefont {Lagos}}, \bibinfo {author}
  {\bibfnamefont {J.}~\bibnamefont {Noller}}, \ and\ \bibinfo {author}
  {\bibfnamefont {I.}~\bibnamefont {Sawicki}},\ }\href@noop {} {\  (\bibinfo
  {year} {2017})},\ \Eprint {http://arxiv.org/abs/1710.06394} {arXiv:1710.06394
  [astro-ph.CO]} \BibitemShut {NoStop}%
\bibitem [{\citenamefont {{Mar{\'{\i}}a Ezquiaga}}\ and\ \citenamefont
  {{Zumalac{\'a}rregui}}(2017)}]{2017arXiv171005901M}%
  \BibitemOpen
  \bibfield  {author} {\bibinfo {author} {\bibfnamefont {J.}~\bibnamefont
  {{Mar{\'{\i}}a Ezquiaga}}}\ and\ \bibinfo {author} {\bibfnamefont
  {M.}~\bibnamefont {{Zumalac{\'a}rregui}}},\ }\href@noop {} {\bibfield
  {journal} {\bibinfo  {journal} {ArXiv e-prints}\ } (\bibinfo {year}
  {2017})},\ \Eprint {http://arxiv.org/abs/1710.05901} {arXiv:1710.05901}
  \BibitemShut {NoStop}%
\bibitem [{\citenamefont {{Sakstein}}\ and\ \citenamefont
  {{Jain}}(2017)}]{2017arXiv171005893S}%
  \BibitemOpen
  \bibfield  {author} {\bibinfo {author} {\bibfnamefont {J.}~\bibnamefont
  {{Sakstein}}}\ and\ \bibinfo {author} {\bibfnamefont {B.}~\bibnamefont
  {{Jain}}},\ }\href@noop {} {\bibfield  {journal} {\bibinfo  {journal} {ArXiv
  e-prints}\ } (\bibinfo {year} {2017})},\ \Eprint
  {http://arxiv.org/abs/1710.05893} {arXiv:1710.05893} \BibitemShut {NoStop}%
\bibitem [{\citenamefont {{The LIGO Scientific Collaboration}}\ and\
  \citenamefont {{The Virgo Collaboration}}(2017)}]{2017arXiv171005832T}%
  \BibitemOpen
  \bibfield  {author} {\bibinfo {author} {\bibnamefont {{The LIGO Scientific
  Collaboration}}}\ and\ \bibinfo {author} {\bibnamefont {{The Virgo
  Collaboration}}},\ }\href@noop {} {\bibfield  {journal} {\bibinfo  {journal}
  {ArXiv e-prints}\ } (\bibinfo {year} {2017})},\ \Eprint
  {http://arxiv.org/abs/1710.05832} {arXiv:1710.05832 [gr-qc]} \BibitemShut
  {NoStop}%
\bibitem [{\citenamefont {{LIGO Scientific Collaboration}}\ \emph
  {et~al.}(2017{\natexlab{a}})\citenamefont {{LIGO Scientific Collaboration}},
  \citenamefont {{Virgo Collaboration}}, \citenamefont {{Gamma-Ray Burst
  Monitor}},\ and\ \citenamefont {{INTEGRAL}}}]{2017arXiv171005834L}%
  \BibitemOpen
  \bibfield  {author} {\bibinfo {author} {\bibnamefont {{LIGO Scientific
  Collaboration}}}, \bibinfo {author} {\bibnamefont {{Virgo Collaboration}}},
  \bibinfo {author} {\bibfnamefont {F.}~\bibnamefont {{Gamma-Ray Burst
  Monitor}}}, \ and\ \bibinfo {author} {\bibnamefont {{INTEGRAL}}},\
  }\href@noop {} {\bibfield  {journal} {\bibinfo  {journal} {ArXiv e-prints}\ }
  (\bibinfo {year} {2017}{\natexlab{a}})},\ \Eprint
  {http://arxiv.org/abs/1710.05834} {arXiv:1710.05834 [astro-ph.HE]}
  \BibitemShut {NoStop}%
\bibitem [{\citenamefont {{LIGO Scientific Collaboration}}\ \emph
  {et~al.}(2017{\natexlab{b}})\citenamefont {{LIGO Scientific Collaboration}},
  \citenamefont {{Virgo Collaboration}}, \citenamefont {{GBM}}, \citenamefont
  {{INTEGRAL}}, \citenamefont {{IceCube Collaboration}}, \citenamefont
  {{AstroSat Cadmium Zinc Telluride Imager Team}}, \citenamefont {{IPN
  Collaboration}}, \citenamefont {{The Insight-Hxmt Collaboration}},
  \citenamefont {{ANTARES Collaboration}}, \citenamefont {{The Swift
  Collaboration}}, \citenamefont {{AGILE Team}}, \citenamefont {{The 1M2H
  Team}}, \citenamefont {{The Dark Energy Camera GW-EM Collaboration}},
  \citenamefont {{the DES Collaboration}}, \citenamefont {{The DLT40
  Collaboration}}, \citenamefont {{GRAWITA}}, \citenamefont {{:}},
  \citenamefont {{GRAvitational Wave Inaf TeAm}}, \citenamefont {{The Fermi
  Large Area Telescope Collaboration}}, \citenamefont {{ATCA}}, \citenamefont
  {{:}}, \citenamefont {{Telescope Compact Array}}, \citenamefont {{ASKAP}},
  \citenamefont {{:}}, \citenamefont {{SKA Pathfinder}}, \citenamefont {{Las
  Cumbres Observatory Group}}, \citenamefont {{OzGrav}}, \citenamefont {{DWF}},
  \citenamefont {{AST3}}, \citenamefont {{CAASTRO Collaborations}},
  \citenamefont {{The VINROUGE Collaboration}}, \citenamefont {{MASTER
  Collaboration}}, \citenamefont {{J-GEM}}, \citenamefont {{GROWTH}},
  \citenamefont {{JAGWAR}}, \citenamefont {{NRAO}}, \citenamefont {{TTU-NRAO}},
  \citenamefont {{NuSTAR Collaborations}}, \citenamefont {{Pan-STARRS}},
  \citenamefont {{The MAXI Team}}, \citenamefont {{Consortium}}, \citenamefont
  {{KU Collaboration}}, \citenamefont {{Optical Telescope}}, \citenamefont
  {{ePESSTO}}, \citenamefont {{GROND}}, \citenamefont {{Tech University}},
  \citenamefont {{SALT Group}}, \citenamefont {{TOROS}}, \citenamefont {{:}},
  \citenamefont {{Transient Robotic Observatory of the South Collaboration}},
  \citenamefont {{The BOOTES Collaboration}}, \citenamefont {{MWA}},
  \citenamefont {{:}}, \citenamefont {{Widefield Array}}, \citenamefont {{The
  CALET Collaboration}}, \citenamefont {{IKI-GW Follow-up Collaboration}},
  \citenamefont {{H.~E.~S.~S.~Collaboration}}, \citenamefont {{LOFAR
  Collaboration}}, \citenamefont {{LWA}}, \citenamefont {{:}}, \citenamefont
  {{Wavelength Array}}, \citenamefont {{HAWC Collaboration}}, \citenamefont
  {{The Pierre Auger Collaboration}}, \citenamefont {{ALMA Collaboration}},
  \citenamefont {{Euro VLBI Team}}, \citenamefont {{Pi of the Sky
  Collaboration}}, \citenamefont {{The Chandra Team at McGill University}},
  \citenamefont {{DFN}}, \citenamefont {{:}}, \citenamefont {{Fireball
  Network}}, \citenamefont {{ATLAS}}, \citenamefont {{Time Resolution Universe
  Survey}}, \citenamefont {{RIMAS}}, \citenamefont {{RATIR}},\ and\
  \citenamefont {{South Africa/MeerKAT}}}]{2017arXiv171005833L}%
  \BibitemOpen
  \bibfield  {author} {\bibinfo {author} {\bibnamefont {{LIGO Scientific
  Collaboration}}}, \bibinfo {author} {\bibnamefont {{Virgo Collaboration}}},
  \bibinfo {author} {\bibfnamefont {F.}~\bibnamefont {{GBM}}}, \bibinfo
  {author} {\bibnamefont {{INTEGRAL}}}, \bibinfo {author} {\bibnamefont
  {{IceCube Collaboration}}}, \bibinfo {author} {\bibnamefont {{AstroSat
  Cadmium Zinc Telluride Imager Team}}}, \bibinfo {author} {\bibnamefont {{IPN
  Collaboration}}}, \bibinfo {author} {\bibnamefont {{The Insight-Hxmt
  Collaboration}}}, \bibinfo {author} {\bibnamefont {{ANTARES Collaboration}}},
  \bibinfo {author} {\bibnamefont {{The Swift Collaboration}}}, \bibinfo
  {author} {\bibnamefont {{AGILE Team}}}, \bibinfo {author} {\bibnamefont {{The
  1M2H Team}}}, \bibinfo {author} {\bibnamefont {{The Dark Energy Camera GW-EM
  Collaboration}}}, \bibinfo {author} {\bibnamefont {{the DES Collaboration}}},
  \bibinfo {author} {\bibnamefont {{The DLT40 Collaboration}}}, \bibinfo
  {author} {\bibnamefont {{GRAWITA}}}, \bibinfo {author} {\bibnamefont {{:}}},
  \bibinfo {author} {\bibnamefont {{GRAvitational Wave Inaf TeAm}}}, \bibinfo
  {author} {\bibnamefont {{The Fermi Large Area Telescope Collaboration}}},
  \bibinfo {author} {\bibnamefont {{ATCA}}}, \bibinfo {author} {\bibnamefont
  {{:}}}, \bibinfo {author} {\bibfnamefont {A.}~\bibnamefont {{Telescope
  Compact Array}}}, \bibinfo {author} {\bibnamefont {{ASKAP}}}, \bibinfo
  {author} {\bibnamefont {{:}}}, \bibinfo {author} {\bibfnamefont
  {A.}~\bibnamefont {{SKA Pathfinder}}}, \bibinfo {author} {\bibnamefont {{Las
  Cumbres Observatory Group}}}, \bibinfo {author} {\bibnamefont {{OzGrav}}},
  \bibinfo {author} {\bibnamefont {{DWF}}}, \bibinfo {author} {\bibnamefont
  {{AST3}}}, \bibinfo {author} {\bibnamefont {{CAASTRO Collaborations}}},
  \bibinfo {author} {\bibnamefont {{The VINROUGE Collaboration}}}, \bibinfo
  {author} {\bibnamefont {{MASTER Collaboration}}}, \bibinfo {author}
  {\bibnamefont {{J-GEM}}}, \bibinfo {author} {\bibnamefont {{GROWTH}}},
  \bibinfo {author} {\bibnamefont {{JAGWAR}}}, \bibinfo {author} {\bibfnamefont
  {C.}~\bibnamefont {{NRAO}}}, \bibinfo {author} {\bibnamefont {{TTU-NRAO}}},
  \bibinfo {author} {\bibnamefont {{NuSTAR Collaborations}}}, \bibinfo {author}
  {\bibnamefont {{Pan-STARRS}}}, \bibinfo {author} {\bibnamefont {{The MAXI
  Team}}}, \bibinfo {author} {\bibfnamefont {T.}~\bibnamefont {{Consortium}}},
  \bibinfo {author} {\bibnamefont {{KU Collaboration}}}, \bibinfo {author}
  {\bibfnamefont {N.}~\bibnamefont {{Optical Telescope}}}, \bibinfo {author}
  {\bibnamefont {{ePESSTO}}}, \bibinfo {author} {\bibnamefont {{GROND}}},
  \bibinfo {author} {\bibfnamefont {T.}~\bibnamefont {{Tech University}}},
  \bibinfo {author} {\bibnamefont {{SALT Group}}}, \bibinfo {author}
  {\bibnamefont {{TOROS}}}, \bibinfo {author} {\bibnamefont {{:}}}, \bibinfo
  {author} {\bibnamefont {{Transient Robotic Observatory of the South
  Collaboration}}}, \bibinfo {author} {\bibnamefont {{The BOOTES
  Collaboration}}}, \bibinfo {author} {\bibnamefont {{MWA}}}, \bibinfo {author}
  {\bibnamefont {{:}}}, \bibinfo {author} {\bibfnamefont {M.}~\bibnamefont
  {{Widefield Array}}}, \bibinfo {author} {\bibnamefont {{The CALET
  Collaboration}}}, \bibinfo {author} {\bibnamefont {{IKI-GW Follow-up
  Collaboration}}}, \bibinfo {author} {\bibnamefont
  {{H.~E.~S.~S.~Collaboration}}}, \bibinfo {author} {\bibnamefont {{LOFAR
  Collaboration}}}, \bibinfo {author} {\bibnamefont {{LWA}}}, \bibinfo {author}
  {\bibnamefont {{:}}}, \bibinfo {author} {\bibfnamefont {L.}~\bibnamefont
  {{Wavelength Array}}}, \bibinfo {author} {\bibnamefont {{HAWC
  Collaboration}}}, \bibinfo {author} {\bibnamefont {{The Pierre Auger
  Collaboration}}}, \bibinfo {author} {\bibnamefont {{ALMA Collaboration}}},
  \bibinfo {author} {\bibnamefont {{Euro VLBI Team}}}, \bibinfo {author}
  {\bibnamefont {{Pi of the Sky Collaboration}}}, \bibinfo {author}
  {\bibnamefont {{The Chandra Team at McGill University}}}, \bibinfo {author}
  {\bibnamefont {{DFN}}}, \bibinfo {author} {\bibnamefont {{:}}}, \bibinfo
  {author} {\bibfnamefont {D.}~\bibnamefont {{Fireball Network}}}, \bibinfo
  {author} {\bibnamefont {{ATLAS}}}, \bibinfo {author} {\bibfnamefont
  {H.}~\bibnamefont {{Time Resolution Universe Survey}}}, \bibinfo {author}
  {\bibnamefont {{RIMAS}}}, \bibinfo {author} {\bibnamefont {{RATIR}}}, \ and\
  \bibinfo {author} {\bibfnamefont {S.}~\bibnamefont {{South
  Africa/MeerKAT}}},\ }\href@noop {} {\bibfield  {journal} {\bibinfo  {journal}
  {ArXiv e-prints}\ } (\bibinfo {year} {2017}{\natexlab{b}})},\ \Eprint
  {http://arxiv.org/abs/1710.05833} {arXiv:1710.05833 [astro-ph.HE]}
  \BibitemShut {NoStop}%
\bibitem [{\citenamefont {Saijo}\ \emph {et~al.}(1997)\citenamefont {Saijo},
  \citenamefont {Shinkai},\ and\ \citenamefont {Maeda}}]{Saijo:1996iz}%
  \BibitemOpen
  \bibfield  {author} {\bibinfo {author} {\bibfnamefont {M.}~\bibnamefont
  {Saijo}}, \bibinfo {author} {\bibfnamefont {H.-a.}\ \bibnamefont {Shinkai}},
  \ and\ \bibinfo {author} {\bibfnamefont {K.-i.}\ \bibnamefont {Maeda}},\
  }\href {\doibase 10.1103/PhysRevD.56.785} {\bibfield  {journal} {\bibinfo
  {journal} {Phys. Rev.}\ }\textbf {\bibinfo {volume} {D56}},\ \bibinfo {pages}
  {785} (\bibinfo {year} {1997})},\ \Eprint
  {http://arxiv.org/abs/gr-qc/9701001} {arXiv:gr-qc/9701001 [gr-qc]}
  \BibitemShut {NoStop}%
\bibitem [{\citenamefont {Barreira}\ \emph {et~al.}(2013)\citenamefont
  {Barreira}, \citenamefont {Li}, \citenamefont {Hellwing}, \citenamefont
  {Baugh},\ and\ \citenamefont {Pascoli}}]{Barreira:2013eea}%
  \BibitemOpen
  \bibfield  {author} {\bibinfo {author} {\bibfnamefont {A.}~\bibnamefont
  {Barreira}}, \bibinfo {author} {\bibfnamefont {B.}~\bibnamefont {Li}},
  \bibinfo {author} {\bibfnamefont {W.~A.}\ \bibnamefont {Hellwing}}, \bibinfo
  {author} {\bibfnamefont {C.~M.}\ \bibnamefont {Baugh}}, \ and\ \bibinfo
  {author} {\bibfnamefont {S.}~\bibnamefont {Pascoli}},\ }\href {\doibase
  10.1088/1475-7516/2013/10/027} {\bibfield  {journal} {\bibinfo  {journal}
  {JCAP}\ }\textbf {\bibinfo {volume} {1310}},\ \bibinfo {pages} {027}
  (\bibinfo {year} {2013})},\ \Eprint {http://arxiv.org/abs/1306.3219}
  {arXiv:1306.3219 [astro-ph.CO]} \BibitemShut {NoStop}%
\bibitem [{\citenamefont {{Chandrasekhar}}(1975)}]{1975RSPSA.343..289C}%
  \BibitemOpen
  \bibfield  {author} {\bibinfo {author} {\bibfnamefont {S.}~\bibnamefont
  {{Chandrasekhar}}},\ }\href {\doibase 10.1098/rspa.1975.0066} {\bibfield
  {journal} {\bibinfo  {journal} {Proceedings of the Royal Society of London
  Series A}\ }\textbf {\bibinfo {volume} {343}},\ \bibinfo {pages} {289}
  (\bibinfo {year} {1975})}\BibitemShut {NoStop}%
\bibitem [{\citenamefont {{Chandrasekhar}}\ and\ \citenamefont
  {{Detweiler}}(1975)}]{1975RSPSA.344..441C}%
  \BibitemOpen
  \bibfield  {author} {\bibinfo {author} {\bibfnamefont {S.}~\bibnamefont
  {{Chandrasekhar}}}\ and\ \bibinfo {author} {\bibfnamefont {S.}~\bibnamefont
  {{Detweiler}}},\ }\href {\doibase 10.1098/rspa.1975.0112} {\bibfield
  {journal} {\bibinfo  {journal} {Proceedings of the Royal Society of London
  Series A}\ }\textbf {\bibinfo {volume} {344}},\ \bibinfo {pages} {441}
  (\bibinfo {year} {1975})}\BibitemShut {NoStop}%
\bibitem [{\citenamefont {Zhang}\ \emph {et~al.}(2007)\citenamefont {Zhang},
  \citenamefont {Gui},\ and\ \citenamefont {Li}}]{Zhang:2006hh}%
  \BibitemOpen
  \bibfield  {author} {\bibinfo {author} {\bibfnamefont {Y.}~\bibnamefont
  {Zhang}}, \bibinfo {author} {\bibfnamefont {Y.~X.}\ \bibnamefont {Gui}}, \
  and\ \bibinfo {author} {\bibfnamefont {F.}~\bibnamefont {Li}},\ }\href
  {\doibase 10.1007/s10714-007-0434-2} {\bibfield  {journal} {\bibinfo
  {journal} {Gen. Rel. Grav.}\ }\textbf {\bibinfo {volume} {39}},\ \bibinfo
  {pages} {1003} (\bibinfo {year} {2007})},\ \Eprint
  {http://arxiv.org/abs/gr-qc/0612010} {arXiv:gr-qc/0612010 [gr-qc]}
  \BibitemShut {NoStop}%
\bibitem [{\citenamefont {Toshmatov}\ \emph {et~al.}(2015)\citenamefont
  {Toshmatov}, \citenamefont {Abdujabbarov}, \citenamefont {Stuchlik},\ and\
  \citenamefont {Ahmedov}}]{Toshmatov:2015wga}%
  \BibitemOpen
  \bibfield  {author} {\bibinfo {author} {\bibfnamefont {B.}~\bibnamefont
  {Toshmatov}}, \bibinfo {author} {\bibfnamefont {A.}~\bibnamefont
  {Abdujabbarov}}, \bibinfo {author} {\bibfnamefont {Z.}~\bibnamefont
  {Stuchlik}}, \ and\ \bibinfo {author} {\bibfnamefont {B.}~\bibnamefont
  {Ahmedov}},\ }\href {\doibase 10.1103/PhysRevD.91.083008} {\bibfield
  {journal} {\bibinfo  {journal} {Phys. Rev.}\ }\textbf {\bibinfo {volume}
  {D91}},\ \bibinfo {pages} {083008} (\bibinfo {year} {2015})},\ \Eprint
  {http://arxiv.org/abs/1503.05737} {arXiv:1503.05737 [gr-qc]} \BibitemShut
  {NoStop}%
\bibitem [{\citenamefont {Medved}\ \emph {et~al.}(2004)\citenamefont {Medved},
  \citenamefont {Martin},\ and\ \citenamefont {Visser}}]{Medved:2003rga}%
  \BibitemOpen
  \bibfield  {author} {\bibinfo {author} {\bibfnamefont {A.~J.~M.}\
  \bibnamefont {Medved}}, \bibinfo {author} {\bibfnamefont {D.}~\bibnamefont
  {Martin}}, \ and\ \bibinfo {author} {\bibfnamefont {M.}~\bibnamefont
  {Visser}},\ }\href {\doibase 10.1088/0264-9381/21/6/008} {\bibfield
  {journal} {\bibinfo  {journal} {Class. Quant. Grav.}\ }\textbf {\bibinfo
  {volume} {21}},\ \bibinfo {pages} {1393} (\bibinfo {year} {2004})},\ \Eprint
  {http://arxiv.org/abs/gr-qc/0310009} {arXiv:gr-qc/0310009 [gr-qc]}
  \BibitemShut {NoStop}%
\bibitem [{\citenamefont {Eling}\ and\ \citenamefont
  {Jacobson}(2006)}]{Eling:2006ec}%
  \BibitemOpen
  \bibfield  {author} {\bibinfo {author} {\bibfnamefont {C.}~\bibnamefont
  {Eling}}\ and\ \bibinfo {author} {\bibfnamefont {T.}~\bibnamefont
  {Jacobson}},\ }\href {\doibase 10.1088/0264-9381/23/18/009,
  10.1088/0264-9381/27/4/049802} {\bibfield  {journal} {\bibinfo  {journal}
  {Class. Quant. Grav.}\ }\textbf {\bibinfo {volume} {23}},\ \bibinfo {pages}
  {5643} (\bibinfo {year} {2006})},\ \bibinfo {note} {[Erratum: Class. Quant.
  Grav.27,049802(2010)]},\ \Eprint {http://arxiv.org/abs/gr-qc/0604088}
  {arXiv:gr-qc/0604088 [gr-qc]} \BibitemShut {NoStop}%
\bibitem [{\citenamefont {Bekenstein}(1972)}]{PhysRevD.5.2403}%
  \BibitemOpen
  \bibfield  {author} {\bibinfo {author} {\bibfnamefont {J.~D.}\ \bibnamefont
  {Bekenstein}},\ }\href {\doibase 10.1103/PhysRevD.5.2403} {\bibfield
  {journal} {\bibinfo  {journal} {Phys. Rev. D}\ }\textbf {\bibinfo {volume}
  {5}},\ \bibinfo {pages} {2403} (\bibinfo {year} {1972})}\BibitemShut
  {NoStop}%
\bibitem [{\citenamefont {Teukolsky}(2015)}]{Teukolsky:2014vca}%
  \BibitemOpen
  \bibfield  {author} {\bibinfo {author} {\bibfnamefont {S.~A.}\ \bibnamefont
  {Teukolsky}},\ }\href {\doibase 10.1088/0264-9381/32/12/124006} {\bibfield
  {journal} {\bibinfo  {journal} {Class. Quant. Grav.}\ }\textbf {\bibinfo
  {volume} {32}},\ \bibinfo {pages} {124006} (\bibinfo {year} {2015})},\
  \Eprint {http://arxiv.org/abs/1410.2130} {arXiv:1410.2130 [gr-qc]}
  \BibitemShut {NoStop}%
\bibitem [{\citenamefont {Pani}(2012)}]{Pani:2012zz}%
  \BibitemOpen
  \bibfield  {author} {\bibinfo {author} {\bibfnamefont {P.}~\bibnamefont
  {Pani}},\ }\href {\doibase 10.1140/epjp/i2012-12067-1} {\bibfield  {journal}
  {\bibinfo  {journal} {Eur. Phys. J. Plus}\ }\textbf {\bibinfo {volume}
  {127}},\ \bibinfo {pages} {67} (\bibinfo {year} {2012})}\BibitemShut
  {NoStop}%
\bibitem [{\citenamefont {et.
  al.}(2017{\natexlab{a}})}]{PhysRevLett.119.161101}%
  \BibitemOpen
  \bibfield  {author} {\bibinfo {author} {\bibfnamefont {B.~P.~A.}\
  \bibnamefont {et. al.}} (\bibinfo {collaboration} {LIGO Scientific
  Collaboration and Virgo Collaboration}),\ }\href {\doibase
  10.1103/PhysRevLett.119.161101} {\bibfield  {journal} {\bibinfo  {journal}
  {Phys. Rev. Lett.}\ }\textbf {\bibinfo {volume} {119}},\ \bibinfo {pages}
  {161101} (\bibinfo {year} {2017}{\natexlab{a}})}\BibitemShut {NoStop}%
\bibitem [{\citenamefont {et. al.}(2017{\natexlab{b}})}]{2041-8205-848-2-L12}%
  \BibitemOpen
  \bibfield  {author} {\bibinfo {author} {\bibfnamefont {B.~P.~A.}\
  \bibnamefont {et. al.}},\ }\href
  {http://stacks.iop.org/2041-8205/848/i=2/a=L12} {\bibfield  {journal}
  {\bibinfo  {journal} {The Astrophysical Journal Letters}\ }\textbf {\bibinfo
  {volume} {848}},\ \bibinfo {pages} {L12} (\bibinfo {year}
  {2017}{\natexlab{b}})}\BibitemShut {NoStop}%
\bibitem [{\citenamefont {et. al.}(2017{\natexlab{c}})}]{2041-8205-848-2-L13}%
  \BibitemOpen
  \bibfield  {author} {\bibinfo {author} {\bibfnamefont {B.~P.~A.}\
  \bibnamefont {et. al.}},\ }\href
  {http://stacks.iop.org/2041-8205/848/i=2/a=L13} {\bibfield  {journal}
  {\bibinfo  {journal} {The Astrophysical Journal Letters}\ }\textbf {\bibinfo
  {volume} {848}},\ \bibinfo {pages} {L13} (\bibinfo {year}
  {2017}{\natexlab{c}})}\BibitemShut {NoStop}%
\bibitem [{\citenamefont {et. al.}(2017{\natexlab{d}})}]{2041-8205-848-2-L14}%
  \BibitemOpen
  \bibfield  {author} {\bibinfo {author} {\bibfnamefont {A.~G.}\ \bibnamefont
  {et. al.}},\ }\href {http://stacks.iop.org/2041-8205/848/i=2/a=L14}
  {\bibfield  {journal} {\bibinfo  {journal} {The Astrophysical Journal
  Letters}\ }\textbf {\bibinfo {volume} {848}},\ \bibinfo {pages} {L14}
  (\bibinfo {year} {2017}{\natexlab{d}})}\BibitemShut {NoStop}%
\bibitem [{\citenamefont {et. al.}(2017{\natexlab{e}})}]{2041-8205-848-2-L15}%
  \BibitemOpen
  \bibfield  {author} {\bibinfo {author} {\bibfnamefont {V.~S.}\ \bibnamefont
  {et. al.}},\ }\href {http://stacks.iop.org/2041-8205/848/i=2/a=L15}
  {\bibfield  {journal} {\bibinfo  {journal} {The Astrophysical Journal
  Letters}\ }\textbf {\bibinfo {volume} {848}},\ \bibinfo {pages} {L15}
  (\bibinfo {year} {2017}{\natexlab{e}})}\BibitemShut {NoStop}%
\bibitem [{\citenamefont {Sakstein}\ and\ \citenamefont
  {Jain}(2017)}]{Sakstein:2017xjx}%
  \BibitemOpen
  \bibfield  {author} {\bibinfo {author} {\bibfnamefont {J.}~\bibnamefont
  {Sakstein}}\ and\ \bibinfo {author} {\bibfnamefont {B.}~\bibnamefont
  {Jain}},\ }\href@noop {} {\  (\bibinfo {year} {2017})},\ \Eprint
  {http://arxiv.org/abs/1710.05893} {arXiv:1710.05893 [astro-ph.CO]}
  \BibitemShut {NoStop}%
\bibitem [{\citenamefont {Ezquiaga}\ and\ \citenamefont
  {Zumalacárregui}(2017)}]{Ezquiaga:2017ekz}%
  \BibitemOpen
  \bibfield  {author} {\bibinfo {author} {\bibfnamefont {J.~M.}\ \bibnamefont
  {Ezquiaga}}\ and\ \bibinfo {author} {\bibfnamefont {M.}~\bibnamefont
  {Zumalacárregui}},\ }\href@noop {} {\  (\bibinfo {year} {2017})},\ \Eprint
  {http://arxiv.org/abs/1710.05901} {arXiv:1710.05901 [astro-ph.CO]}
  \BibitemShut {NoStop}%
\bibitem [{\citenamefont {Nutma}(2014)}]{Nutma:2013zea}%
  \BibitemOpen
  \bibfield  {author} {\bibinfo {author} {\bibfnamefont {T.}~\bibnamefont
  {Nutma}},\ }\href {\doibase 10.1016/j.cpc.2014.02.006} {\bibfield  {journal}
  {\bibinfo  {journal} {Comput. Phys. Commun.}\ }\textbf {\bibinfo {volume}
  {185}},\ \bibinfo {pages} {1719} (\bibinfo {year} {2014})},\ \Eprint
  {http://arxiv.org/abs/1308.3493} {arXiv:1308.3493 [cs.SC]} \BibitemShut
  {NoStop}%
\end{thebibliography}%

\end{document}